\newdimen\nude\newbox\chek
\def\slash#1{\setbox\chek=\hbox{$#1$}\nude=\wd\chek#1{\kern-\nude/}}

\def\to{\rightarrow}

\def\llgm{\left\lgroup\matrix}
\def\rrgm{\right\rgroup}
\def\vectrl #1{\buildrel\leftrightarrow \over #1}
\def\partrl{\vectrl{\partial}}
\magnification=1095
\parskip=-1pt plus 0.07 cm
\baselineskip= 14pt

\rightline{ KEK CP-080}

\vskip 2 cm
\centerline{{\bf Complete Lagrangian of MSSM}\footnote{*}
{This work is partially supported by the Ministry of Education,
Science and Culture,
Japan, under the Grant-in-Aid for basic research program C (no.08640391).}}
\vskip 2cm
\centerline{Masaaki Kuroda}
\centerline{Institute of Physics, Meiji-Gakuin University}
\centerline{Yokohama, Japan}
\vskip 2 true cm

\noindent{\bf Abstract}\par
    A brief derivation of the lagrangian of the minimal supersymmetric 
theory is given and  the complete expression of the lagrangian in terms of mass 
eigenstates is presented.
\medskip
\vskip 1 true cm
     
\noindent{\bf 1. Introduction}\par
     Theories of supersymmetry (SUSY) draw much attentions in the past decades
as a realistic possibility of the physics beyond the Standard Model
(SM)[1].  Due to a large number
of particles and interactions involved in any of SUSY theories,
the actual calculation of the production rates or decay rates of
supersymmetric particles,  which are predicted 
but not yet discovered, is very complicated.
Consequently, it is highly desirable to have a system of  automatic
computation of such processes.  The Minami-Tateya collaboration at 
KEK has already developed a system called {\tt GRACE} [2] which  generates
automatically the tree amplitudes of the Standard Model.  Implementing
SUSY interactions in {\tt GRACE} with some additional modifications,
necessary for the treatment of Majorana particles and  
fermion number clashing vertices, they are  upgrading {\tt GRACE} so that
they can use it to compute SUSY processes automatically. 
As a prototype of SUSY theories, this group has chosen
MSSM, the minimal supersymmetric extension of the standard model,
which is the smallest but the most basic model of SUSY that includes
the SM, and they have coded its interactions in the model file of {\tt GRACE}.  
\par
    Although the MSSM lagrangian is given in several papers [3,4], 
we need the self-contained and full expression including ghost interactions
and gauge-fixing terms.
In this paper I will present 
the complete MSSM lagrangian which is  written
in terms of mass eigenstates.  
The model files of {\tt GRACE} for SUSY are coded based on this
lagrangian, which  also defines the phase convention of physical particles.  
A package of MSSM processes, though limited to twenty-three processes,
was  generated by {\tt GRACE}  and 
is already released as {\tt SUSY23}[5].  \par
    The paper is organized as follows.  The superfields and  their component
fields used in MSSM are introduced  in section 2, where  the physical 
states such
as fermions, gauge bosons etc. are defined in terms of the component fields.
In section 3, the lagrangian is represented by superfields
and then decomposed into the  component fields.  The complete expression 
of MSSM lagrangian is given in Appendix C in terms of mass eigenstates,
except for the case of sfermion self-interactions, in which
the lagrangian is given in a form with  $\tilde f_L$ and $\tilde f_R$. 
For simplicity we consider fermions only in the first generation.
Inclusion of the second and the third generations 
is trivial except for the four-sfermion interactions, in which 
the inter-generation interactions occur.\par
   Notations, conventions and several important formulae are 
compiled in Appendix A.   In Appendix B, the structure of the $F$-term
and $D$-term is discussed.  In Appendix D,  the products of sfermion,
which are expressed in Appendix C by $\tilde f_L$ and $\tilde f_R$
are converted to the  mass eigenstates $\tilde f_1$ and $\tilde f_2$.
In Appendix E, for the sake of those  who
are consulting the Hikasa's manuscript [4], his notation is compared
with mine.   It is confirmed that upon adjusting the conventions
the lagrangian presented 
in this paper fully agrees with
ref.[4] up to some trivial misprints in it.
\vskip 2 cm

\noindent{\bf 2.  Superfields and physical states}\par

  The MSSM, the minimal supersymmetric extension of the standard model
which is based on the gauge group $SU(2)_L\times U(1)\times SU(3)_c$,
consists of three gauge-superfields, $ V^a$, $V$ and $V_s^\alpha$ 
for gauge bosons,
five left-handed chiral-superfields\footnote{*}{Assume that the
neutrinos are massless, we don't introduce the superfields 
corresponding to the right-handed singlet neutrino fields.}, ${\bf \Phi}_\ell$, $\Phi_e$, 
${\bf \Phi}_q$, $\Phi_u$, $\Phi_d$, for spinors of each generation, 
and two left-handed chiral superfields, ${\bf \Phi_{H1}}$ and 
${\bf \Phi_{H2}}$ for Higgs particles.  
The model contains  three gauge coupling constants,
$g$, $g^\prime$ and $g_s$, and one Higgs coupling constant, $\mu$, and in 
addition to the fermion masses which I don't count as  free parameters,
$6+8N_G$ free parameters of the mass dimension for soft SUSY-breaking 
terms. \par
   The chiral superfields represented by the bold letters are $SU(2)_L$ 
doublet, while non-bold superfields are $SU(2)_L$ singlet.
The gauge superfield $V^a$ is $SU(2)_L$ triplet,  while the gauge superfield
$V$ is $SU(2)_L$ singlet. The gauge superfield $V_s^\alpha$ is $SU(3)_c$
octet. We use the index $a, b \cdots$ for $SU(2)_L$ and the index 
$\alpha, \beta \cdots$ for $SU(3)_c$.\par
     Since we use only the left-handed chiral
superfields subject to the condition
$$
 \bar{\cal D}_L {\bf \Phi} \equiv -({{\partial}\over{\partial \bar\theta}}
              +i\theta\sigma^\mu\partial_\mu){\bf \Phi}=0,
\eqno(2.1)
$$
with the four-dimensional $\sigma^\mu$ being defined  by (A.3),
we consider $\Phi_e$, $\Phi_u$ and $\Phi_d$ 
as the superfields for the left-handed antifermions
which contain right-handed fermions; for example, 
the superfield $\Phi_e$ has a positive charge, 
and its hypercharge is given by  $Y=2$.  For gauge-superfields, 
we work in the so-called Wess-Zumino gauge, in which several component
fields contained in the gauge-superfield are gauged away and gauge-superfields
consist of three component fields.  The content of component fields and
the quantum number of each superfield appearing in MSSM are listed 
in Table 1.  

\topinsert
{\offinterlineskip \tabskip=0pt
\halign{ \strut
  \vrule#& \quad # \quad &
  \vrule#& \quad # \quad &
  \vrule#& \quad # \quad &
  \vrule#& \quad # \quad &
  \vrule#  \cr
\noalign{\hrule}
 &             && $SU(2)_L$ &&  $U(1)$ && particle content ~~~~~~~~~~&\cr
\noalign{\hrule}
 & $V^a$ &&  $\bf 3$ &&  1 &&  $ W^a_\mu$,~$\lambda^a$, $ D^a$ & \cr
\noalign{\hrule}
 & $V$     &&  $\bf 1$ &&  0 &&  $B_\mu$,~$\lambda$, $ D$ & \cr
\noalign{\hrule}
\noalign{\hrule}
 & $\bf\Phi_\ell$ && $\bf 2$ && $-1$  && $\llgm{ \nu_L\cr e_L}\rrgm$, 
    $\llgm{ A(\nu_L)\cr A(e_L)}\rrgm$, $\llgm{ F(\nu_L)\cr F(e_L)}\rrgm$ &\cr
\noalign{\hrule}
 & $\Phi_e$ && $\bf 1$ && 2 &&  $e_R$,~~$A(e_R)$,~~ $ F(e_R)$ & \cr
\noalign{\hrule}
 & $\bf\Phi_q$ && $\bf 2$ && ${1\over 3}$ && $\llgm{ u_L\cr d_L}\rrgm$, 
   $\llgm{ A(u_L)\cr A(d_L)}\rrgm$, $\llgm{ F(u_L)\cr F(d_L)}\rrgm$ & \cr
\noalign{\hrule}
 & $\Phi_u$ &&  $\bf 1$ &&$-{4\over 3}$&&  $u_R$,~~$A(u_R)$,~~ $F(u_R)$ & \cr
\noalign{\hrule}
 & $\Phi_d$ &&  $\bf 1$ &&$ {2\over 3}$&&  $d_R$,~~$A(d_R)$,~~ $F(d_R)$ & \cr
\noalign{\hrule}
\noalign{\hrule}
 & $\bf\Phi_{H_1}$ && $\bf 2$ && $-1$ && $\llgm{ H^0_1\cr H^-_1}\rrgm$, 
   $\llgm{ \tilde{H^0_1}\cr \tilde{H^-_1}}\rrgm$, 
   $\llgm{ F(H^0_1)\cr F(H^-_1)}\rrgm$ & \cr
\noalign{\hrule}
 & $\bf \Phi_{H_2}$ && $\bf 2$  &&  1  && $\llgm{ H^+_2\cr H^0_2}\rrgm$, 
   $\llgm{\tilde{H^+_2}\cr \tilde{H^0_2} }\rrgm$, 
   $\llgm{ F(H^+_2)\cr F(H^0_2)}\rrgm$ & \cr 
\noalign{\hrule} }}

\medskip

{\offinterlineskip \tabskip=0pt
\halign{ \strut
  \vrule#& \quad # \quad &
  \vrule#& \quad # \quad &
  \vrule#& \quad # \quad &
  \vrule#& \quad # \quad &
  \vrule#  \cr
\noalign{\hrule}
 &             && $SU(3)_c$ && $SU(2)_L\times U(1)$&& particle content &\cr
\noalign{\hrule}
 & $V_s^\alpha$~~ &&  $\bf 8$ && singlet                && 
                 $ g_\mu^\alpha$,~$\tilde g^\alpha$, $ D^\alpha$  & \cr
\noalign{\hrule} }}

\medskip
\noindent{Table 1. Quantum numbers and component fields contained in each
superfield appearing in MSSM}\par
\vskip 1 cm
\endinsert
\par
    Upon shifting the vacuum expectation values, 
we define the Higgs scalar fields as follows,
$$
\eqalign{
  {\bf H}_1\equiv\llgm{ H^0_1\cr H^-_1}\rrgm=& 
       \llgm{(v_1+\phi_1^0-i\chi_1^0)/\sqrt 2 \cr
             -\phi_1^-}\rrgm,\cr
  {\bf H}_2\equiv\llgm{ H^+_2\cr H^0_2}\rrgm=& 
       \llgm{\phi_2^+ \cr 
          (v_2+\phi_2^0 + i \chi_2^0)/\sqrt 2}\rrgm.
}\eqno(2.2)
$$\par

  In terms of component fields the conventional spinors are expressed as
$$
\eqalign{
     \Psi(e) =& \left\lgroup\matrix{\overline e_R\cr e_L}\right\rgroup,~~~
   \Psi(\nu) = \left\lgroup\matrix{0 \cr \nu_L}\right\rgroup.
                          \cr
     \Psi(u) =& \left\lgroup\matrix{\overline u_R\cr u_L}\right\rgroup,~~~
     \Psi(d) = \left\lgroup\matrix{\overline d_R\cr d_L}\right\rgroup.
}\eqno(2.3)
$$
Here, the bar is on the upper component, since $\Phi_e$ is
the left-handed chiral superfield for positron and the complex 
conjugate corresponds to the charge conjugated state.\par
   Charged gauge bosons are defined as
$$
      W_\mu^\pm \equiv {{W_\mu^1 \mp i W_\mu^2}\over {\sqrt 2}},
\eqno(2.4a)
$$
while neutral gauge bosons are 
$$
     \left\lgroup\matrix{Z_\mu\cr A_\mu}\right\rgroup
    = \left\lgroup\matrix{ \cos\theta_W & -\sin\theta_W \cr
                           \sin\theta_W &  \cos\theta_W }\right\rgroup
      \left\lgroup\matrix{W_\mu^3 \cr B_\mu}\right\rgroup.
\eqno(2.4b)
$$
The gauge boson masses are given in terms of vacuum expectation values 
$v_1$ and $v_2$;
$$
     M_W^2 = {1\over 4} g^2 (v_1^2+v_2^2),~~~~~
     M_Z^2 = {1\over 4} (g^2+g^{\prime 2})(v_1^2+v_2^2).
\eqno(2.5)
$$\par
     Sfermions are defined as
$$
    \llgm{\tilde f_1 \cr \tilde f_2} \rrgm =
    \llgm{\cos\theta_f & \sin\theta_f \cr -\sin\theta_f & \cos\theta_f}\rrgm
    \llgm{\tilde f_L \cr \tilde f_R}\rrgm,~~~~f=e,u,d,
\eqno(2.6)
$$
where $\tilde f_L$ and $\tilde f_R$ are given in terms of the 
scalar component fields, $A(f_L)$, $A(f_R)$,  listed in Table 1 as
$$
\eqalign{
     \tilde e_L =& A(e_L), ~~~\tilde e_R = A(e_R)^*, \cr
     \tilde \nu_L =& A(\nu_L),\cr 
     \tilde u_L =& A(u_L), ~~~\tilde u_R = A(u_R)^*, \cr
     \tilde d_L =& A(d_L), ~~~\tilde d_R = A(d_R)^*, \cr
}\eqno(2.7)
$$
and the mixing angle is defined such that the mass matrix becomes 
diagonal with eigenvalues $m^2_{\tilde f_1}$ and $m^2_{\tilde f_2}$ 
($m_{\tilde f_1}<m_{\tilde f_2}$):
$$
\eqalign{
    \llgm{\cos\theta_f & \sin\theta_f \cr -\sin\theta_f & \cos\theta_f}\rrgm
   & \llgm{m^2_{\tilde f_L} & m^2_{\tilde f_{LR}} \cr 
          m^{2*}_{\tilde f_{LR}} & m^2_{\tilde f_R} }\rrgm
    \llgm{\cos\theta_f & -\sin\theta_f \cr \sin\theta_f & \cos\theta_f}\rrgm\cr
  & = \llgm{m^2_{\tilde f_1} & 0 \cr 0 & m^2_{\tilde f_2} }\rrgm. 
}\eqno(2.8)
$$
The explicit form of the mass matrix is given in section 4.
\par

     As we see in section 3, charginos don't conserve fermion number.
Therefore, the fermion number of charginos is not determined by interactions, 
but it is a matter of convention.  We adopt the convention that the positively 
charged charginos are  Dirac-particles.  This convention is 
recommended by the LEP2 working group [6] and it is used also 
in the generator {\tt SUSYGEN} [7].  There are two charginos which
are made of four Weyl spinors, $\lambda^+$, $\lambda^-$, $\tilde H^-_1$
and $\tilde H^+_2$.  The physical states with mass $m_{\tilde \chi^\pm_1}$
and $m_{\tilde \chi^\pm_2}$ are given by
$$
   {\rm chargino}:~~~~~~~~
   \Psi( \tilde \chi^+_i) = 
      \llgm{\overline{\lambda^-_{iR}} \cr \lambda^+_{iL}}
                      \rrgm,~~~~~~~~~~
   \Psi(\tilde \chi^-_i) \equiv \Psi( \tilde \chi^+_i)^c  = 
      \llgm{\overline{\lambda^+_{iL}} \cr \lambda^-_{iR}} \rrgm,~~~~
       i=1,2
\eqno(2.9)
$$
where, 
$$
\eqalign{
   \llgm{\lambda^-_{1R} \cr \lambda^-_{2R}}\rrgm =&
   \llgm{\cos \phi_R & \sin\phi_R \cr -\sin\phi_R & \cos\phi_R}\rrgm
   \llgm{\lambda^- \cr \tilde H^-_1}\rrgm, \cr 
   \llgm{\lambda^+_{1L} \cr \lambda^+_{2L}}\rrgm =&
   \llgm{1 & 0 \cr 0 & \epsilon_L}\rrgm
   \llgm{\cos \phi_L & \sin\phi_L \cr -\sin\phi_L & \cos\phi_L}\rrgm
   \llgm{\lambda^+ \cr \tilde H^+_2}\rrgm. 
}\eqno(2.10)
$$
Two orthogonal matrices in (2.10) diagonalizes the mass matrix,
$$
   {\cal M}_C\equiv 
   \llgm{ M_2 & \sqrt 2 M_W \cos\beta \cr \sqrt 2 M_W \sin\beta &\mu}\rrgm,
\eqno(2.11)
$$ 
as
$$
   \llgm{\cos \phi_L & \sin\phi_L \cr -\sin\phi_L & \cos\phi_L}\rrgm
  {\cal M}_C
   \llgm{\cos \phi_R & -\sin\phi_R \cr \sin\phi_R & \cos\phi_R}\rrgm 
  =\llgm{ m_{\tilde c_1} & 0 \cr 0 & m_{\tilde c_2} }\rrgm,
\eqno(2.12)
$$
where we set the ordering of the two charginos such that
$\vert m_{\tilde c_1}\vert < \vert m_{\tilde c_2}\vert$. 
The parameter $\mu$ and $M_2$ are  a Higgs-Higgs coupling constant  
and an SUSY breaking parameter as will be defined in (3.1) and (3.2). 
The mixing parameter
$\cos\beta$ and $\sin\beta$ are related to the ratio of the two vacuum
expectation values, $v_1$ and $v_2$,
$$
     \tan\beta = {{v_2}\over{v_1}},~~~~ 
     \cos\beta = {{v_1}\over{\sqrt{v_1^2+v_2^2}}},~~~~
     \sin\beta = {{v_2}\over{\sqrt{v_1^2+v_2^2}}}.
\eqno(2.13)
$$
The diagonal matrix with $\epsilon_L$ in (2.10) is to take care of the
possible negative eigenvalue for $m_{\tilde c_2}$.  
( We can always choose the mixing parameters $\phi_R$ and $\phi_L$ such 
that $m_{\tilde c_1}>0$).  Practically,  from (2.11) we find
$$
     \epsilon_L = {\rm sign}(M_2\mu -M_W^2  \sin2\beta).
\eqno(2.14)
$$
The physical masses of charginos are given by
$$
    m_{\tilde \chi_1^\pm} = m_{\tilde c_1},~~~~~~~~m_{\tilde \chi_2^\pm} = 
    \epsilon_Lm_{\tilde c_2},
\eqno(2.15)
$$
with $\tilde \chi^\pm_1$ lighter than $\tilde \chi^\pm_2$. 
The mixing angles are given by
$$
\eqalign{
   \tan\phi_L =& {{m^2_{\tilde\chi^\pm_1} -M_2^2-2M_W^2\cos^2\beta}
          \over{\sqrt 2M_W(M_2\sin\beta+\mu \cos\beta)}}
              ={{\sqrt 2M_W(M_2\sin\beta+\mu \cos\beta)}
          \over{m^2_{\tilde\chi^\pm_1} -\mu^2-2M_W^2\sin^2\beta}}, \cr
   \tan\phi_R =& {{m^2_{\tilde\chi^\pm_1} -M_2^2-2M_W^2\sin^2\beta}
          \over{\sqrt 2M_W(M_2\cos\beta+\mu \sin\beta)}}
              ={{\sqrt 2M_W(M_2\cos\beta+\mu \sin\beta)}
          \over{m^2_{\tilde\chi^\pm_1} -\mu^2-2M_W^2\cos^2\beta}}. \cr
}\eqno(2.16)
$$\par
    From the four neutral Weyl spinors, $\lambda$, $\lambda^0$, $\tilde H^0_1$
and $\tilde H_2^0$, four Majorana particles are constructed. They are 
denoted  by
$$
   {\rm neutralino}:~~~~~~~~~~
 \Psi(\tilde \chi^0_i) \equiv \llgm{\bar\lambda_i \cr \lambda_i}\rrgm,
 ~~~~~i=1\sim 4,
\eqno(2.17)
$$
where
$$
  \llgm{\lambda_1 \cr \lambda_2 \cr \lambda_3 \cr\lambda_4}\rrgm  
 = \llgm{\eta_1 & 0&0&0 \cr 0&\eta_2 & 0&0 \cr0&0&\eta_3 & 0 \cr0&0&0&\eta_4} 
    \rrgm \llgm {\cal O}_N  \rrgm 
  \llgm{\lambda \cr \lambda^0 \cr \tilde H_1^0 \cr\tilde H_2^0}\rrgm.
\eqno(2.18)
$$
The four-by-four matrix $O_N$ diagonalizes the 
symmetric mass matrix ${\cal M}_N$ of the neutral Weyl spinors 
$$
{\cal M}_N \equiv \llgm{M_1 & 0   & -M_Z \sin\theta_W \cos\beta
                             &  M_Z \sin\theta_W \sin\beta \cr
                     * & M_2 &  M_Z \cos\theta_W \cos\beta 
                             & -M_Z \cos\theta_W \sin\beta \cr
                     * & *   &  0              & -\mu \cr
                     * & *   &  *              & 0}\rrgm,  
\eqno(2.19)
$$
as
$$
     {\cal O}_N{\cal M}_N {\cal O}_N^T = 
          {\rm diag}(m_{\tilde n_1},~~ m_{\tilde n_2},~~m_{\tilde n_3},~~
            m_{\tilde n_4}),
\eqno(2.20)
$$
where the eigenvalues are arranged such that
$\vert m_{\tilde n_1}\vert <\vert m_{\tilde n_2}\vert <
 \vert m_{\tilde n_3}\vert <\vert m_{\tilde n_4}\vert$.
Here $M_1$ is another SUSY breaking parameter(see (3.2)).
In (2.18) a matrix with  $\eta_i$ is introduced in order to change the 
phase of the particle 
whose eigenvalue becomes negative by the diagonalization (2.20).  Namely,
$$
     \eta_i = \cases{1, &~~~~ $m_{\tilde n_i}>0$, \cr
                     i, &~~~~~$m_{\tilde n_i}<0$, \cr}
\eqno(2.21)
$$
and 
$$
     m_{\tilde\chi^0_i} = \eta^2_i m_{\tilde n_i},
\eqno(2.22)
$$
with $\tilde\chi_1^0$ being the lightest neutralino.\par
   Eight gluinos are colored Majorana particles with mass $M_3$ 
which comes from the SUSY breaking term in the lagrangian (3.2).
$$
   {\rm gluino}:~~~~~\Psi(\tilde g^\alpha) =
    \llgm{\overline{\tilde g^\alpha} \cr \tilde g^\alpha}\rrgm.
\eqno(2.23)
$$\par
     Next, we discuss the Higgs particles.  As we see in detail in 
section 4, 
the mass eigenstate of the charged Higgs, $H^\pm$, is given by 
$$
    \llgm{G^\pm \cr H^\pm}\rrgm =
    \llgm{\cos\beta & \sin\beta \cr -\sin\beta & \cos\beta}\rrgm
    \llgm{\phi_1^\pm \cr \phi_2^\pm}\rrgm,
\eqno(2.24)
$$
where $\phi_1^+=(\phi_1^-)^*$ and $\phi_2^-=(\phi_2^+)^*$ and   
$G^\pm$ is a massless Goldestone boson.\par
     From the four neutral Higgs, we construct two real Higgs scalars
with even CP, $H^0$, $h^0$, and one neutral Goldstone boson 
$G^0$  and one real Higgs $A^0$ which is CP odd,
$$
\eqalign{
  \llgm{H^0\cr h^0}\rrgm =& 
      \llgm{\cos\alpha & \sin\alpha \cr -\sin\alpha & \cos\alpha}\rrgm
      \llgm{\phi_1^0 \cr \phi_2^0}\rrgm, \cr
  \llgm{G^0\cr A^0}\rrgm =& 
      \llgm{\cos\beta & \sin\beta \cr -\sin\beta & \cos\beta}\rrgm
      \llgm{\chi_1^0 \cr \chi_2^0}\rrgm, \cr
}\eqno(2.25)
$$
where
$$
   \tan 2\alpha = \tan 2\beta {{M_A^2+M_Z^2}\over{M_A^2- M_Z^2}},~~
       -{\pi \over 2}< \alpha < 0.
\eqno(2.26)
$$
The masses of Higgs particles are given at tree level by
$$
\eqalign{
     M_A^2 =&~ m_1^2+m_2^2 = -m_{12}^2(\tan\beta + \cot\beta),\cr
     M_{H^0,h^0}^2 =&~ {1\over 2}[M_A^2+M_Z^2 \pm
       \sqrt{(M_A^2+M_Z^2)^2-4M_A^2M_Z^2\cos^22\beta} ],\cr
     M^2_{H^\pm}=&~ M_A^2 + M_W^2,
}\eqno(2.27)
$$
where $m_1^2$, $m_2^2$ and $m^2_{12}$ are defined in 
(4.9) by the parameters appearing in the MSSM
lagrangian discussed in detail in the next section. 
  
\vskip 2 cm

\noindent{\bf 3. Lagrangian}\par
    In this section, first  I show the lagrangian written by the superfields.
By $\theta$ integrations and then eliminating the 
auxiliary fields by the equation of motion, I decompose the MSSM
lagrangian in component fields. 
In Appendix C, the lagrangian is expressed in terms of physical
states which are mass eigenstates and are related to the component fields
as shown in section 2.  
The general form of the $\theta$ integration 
is given in Appendix B.\par

    The basic lagrangian consists of two part, the supersymmetric part and
softly breaking part.
$$
\eqalignno{
   {\cal L} =&~ \int d^2\theta {1\over 4} 
               [2 Tr({\bf W}{\bf W}) + WW + 2Tr({\bf W_s}{\bf W_s})] + h.c. 
               &(3.1a)\cr
      &+ \int d^2\theta d^2\bar\theta
          {\bf \Phi_\ell}^\dagger \exp[2(g{{\tau^a}\over 2}V^a
            + g^\prime {{Y_\ell}\over 2}V)] {\bf \Phi_\ell} & (3.1b)\cr
      &+ \int d^2\theta d^2\bar\theta \Phi_e^\dagger \exp(g^\prime Y_e V) 
            \Phi_e &(3.1c)\cr
      &+ \int d^2\theta d^2\bar\theta 
           {\bf \Phi_q}^\dagger \exp[2(g{{\tau^a}\over 2}V^a+ 
             g^\prime {{Y_q}\over 2}V
          + g_s{{\lambda^\alpha}\over 2} V_s^\alpha)]{\bf \Phi_q} &(3.1d)\cr
      &+ \int d^2\theta d^2\bar\theta \Phi_u ^\dagger \exp(g^\prime Y_u V
          -g_s\lambda^{\alpha *}V_s^\alpha) \Phi_u &(3.1e)\cr
      &+ \int d^2\theta d^2\bar\theta \Phi_d^\dagger \exp(g^\prime Y_d V
          -g_s\lambda^{\alpha *}V_s^\alpha)\Phi_d &(3.1f)\cr
      &+ \int d^2\theta d^2\bar\theta 
        {\bf \Phi_{H1}}^\dagger \exp[2(gT^aV^a+ g^\prime {{Y_{H1}}\over 2}V)] 
               {\bf \Phi_{H1}} &(3.1g)\cr
      &+ \int d^2\theta d^2\bar\theta 
          {\bf \Phi_{H2}}^\dagger\exp[2(gT^aV^a+ g^\prime {{Y_{H2}}\over 2}V)] 
               {\bf \Phi_{H2}} &(3.1h)\cr
      &+{{\sqrt 2 m_e}\over v_1} \int d^2\theta {\bf \Phi_{H_1}}
               {\bf \Phi_\ell}\Phi_e + h.c.&(3.1i)\cr
      &-{{\sqrt 2 m_u}\over v_2} \int d^2\theta {\bf \Phi_{H_2}}
               {\bf\Phi_q}\Phi_u + h.c. &(3.1j)\cr
      &+{{\sqrt 2 m_d}\over v_1} \int d^2\theta {\bf \Phi_{H_1}}
               {\bf\Phi_q}\Phi_d + h.c.&(3.1k)\cr
      &-\mu \int d^2\theta  {\bf \Phi_{H1}}{\bf \Phi_{H2}}+ h.c. & (3.1\ell)\cr
      &+ {\cal L}_{soft} & (3.1m)\cr
      &+ {\cal L}_{gf}+{\cal L}_{ghost}, & (3.1n)\cr  
}$$
where
$$
\eqalign{
     {\bf W}_\alpha =& -{1\over 4}\bar{\cal D}\bar{\cal D} e^{-{\bf V}} 
                      {\cal D}_\alpha e^{\bf V},~~{\rm with}~~~
      {\bf V} = \sum_1^3 {{\tau^a}\over 2}V^a, ~~~~~~
     W_\alpha = -{1\over 4} \bar{\cal D} \bar{\cal D}{\cal D}_\alpha V, \cr
    {\bf W}_{s\alpha} =& -{1\over 4}\bar{\cal D}\bar{\cal D} e^{-{\bf V_s}} 
                      {\cal D}_\alpha e^{\bf V_s},~~{\rm with}~~~
      {\bf V}_s = \sum_1^8 {{\lambda^\alpha}\over 2}V_s^\alpha,
}\eqno(3.1o)
$$
and $\lambda^\alpha$ in (3.1d), (3.1e) and (3.1f) stands for the 
$SU(3)_c$ Gell-Mann matrix.
   Note the minus and "$*$" signs in the exponents of (3.1e) and (3.1f).  
This is due to the fact that $\Phi_u$ and $\Phi_d$ are the left-handed 
chiral superfields for $\bar u$- and $\bar d$-quark, respectively. 
The lagrangian (3.1a) gives 
the kinetic part of the gauge bosons and gauginos, while (3.1b)-(3.1f) 
give the kinetic part of the matter(fermions and sfermions) fields and
their interaction lagrangians. (3.1i), (3.1j) and (3.1k) give the Yukawa
interaction of matter fields with Higgs and higgsino fields.
Note that in (3.1j) the overall sign is minus.  Higgs 
kinetic part and Higgs interactions with gauge bosons and gauginos are 
obtained from (3.1g) and (3.1h), which also produce the gauge boson masses. 
(3.1$\ell$) gives the higgsino off-diagonal mass.
From the $D$-terms, we obtain the four scalar vertices, which include
the  quartic Higgs self-interaction terms in the Higgs potential.
From the $F$-terms, we obtain another four scalar vertices, but they contain
always at least two  sfermions and they don't contribute to the Higgs 
potential.  Instead, $F$-terms contribute to the Higgs masses (quadratic
terms).\par
     The soft SUSY breaking part(3.1m) has the following form\footnote{*}
{One often includes the interaction (3.1$\ell$) in the soft breaking part.
This is because the coupling constant $\mu$ has a dimension of mass like 
the other soft breaking terms in (3.2). However, I don't classify it 
in the softly broken interaction
since it is still supersymmetry invariant.}
which contains as many as $6+ 8N_G$ parameters where $N_G=3$ is a 
number of fermion generation,
$$
\eqalign{
     {\cal L}_{soft} =& -{1\over 2}M_1\lambda\lambda 
                        -{1\over 2}M_2\lambda^a\lambda^a 
                        -{1\over 2}M_3\tilde g^\alpha\tilde g^\alpha + h.c.\cr
                     & -\tilde m_1^2{\bf H}_1^*{\bf H}_1 
                       -\tilde m_2^2{\bf H}_2^*{\bf H}_2
                       -(\tilde m_{12}^2{\bf H}_1{\bf H}_2 + h.c.)
                       - \sum _{\tilde f_i} \tilde m^2_{\tilde f_i}
                        \tilde f^*_i \tilde f_i \cr
      &- {{\sqrt 2 m_u}\over v_2}A_u {\bf H_2}{\bf A}(q_L)A(u_R)
      + {{\sqrt 2 m_d}\over v_1}A_d {\bf H_1}{\bf A}(q_L)A(d_R)+h.c.\cr
      &+ {{\sqrt 2 m_e}\over v_1}A_e {\bf H_1}{\bf A}(\ell_L)A(e_R)
       + h.c.
}\eqno(3.2)
$$
where
$$
\eqalign{
 {\bf H}^*_1{\bf H}_1 =&~ \vert H_1^0 \vert^2+ \vert H_1^-\vert ^2
            = {1\over 2}(\vert v_1+\phi_1^0 \vert^2 + \vert\chi_1^0 \vert^2)
              + \vert \phi_1^-\vert^2, \cr
 {\bf H}^*_1{\bf H}_2 =&~ H_1^{0*}H_2^+ + H_1^+H_2^0,\cr
 {\bf H}_1 {\bf H}_2 =&~ H_1^0 H_2^0 - H_1^- H_2^+.
}\eqno(3.2a)
$$
and the summation is taken over all left-handed and right-handed sfermions. 
Note that $SU(2)$ invariance requires
the common breaking parameters for the members of each left-handed doublet 
sfermions, e.g., $\tilde m_{\tilde u_L} = \tilde m_{\tilde d_L}$,
$\tilde m_{\tilde e_L} = \tilde m_{\tilde \nu_L}$, etc.\par
     The last two lagrangians in (3.1n) are for quantization.  They 
represent the gauge fixing terms and the ghost interactions.\par

\medskip

\noindent$\underline{3.1~ {\rm Matter~ gauge~interactions}}$\par
     We start with the interaction lagrangians which come from (3.1b)-(3.1f)
and from (3.1i)-(3.1k).  Using
$$
\eqalign{
          {g\over 2}\tau^aW_\mu^a + {{g^\prime}\over 2}Y B_\mu
      =&~ {g\over{\sqrt 2}}\llgm{0 & 1 \cr 0 & 0}\rrgm W_\mu^+ 
      + {g\over{\sqrt 2}}\llgm{0 & 0 \cr 1 & 0}\rrgm W_\mu^-  \cr
      &  + g_Z(T_3-s_W^2 Q)Z_\mu + eQA_\mu,      
}\eqno(3.3)
$$
with $g_Z={g\over{c_W}}$ and 
$s_W\equiv \sin\theta_W$, $c_W\equiv\cos\theta_W$, 
we obtain from the term\footnote{**}{Note that the sign convention of 
the gauge fields is fixed by the convention of the covariant derivatives
given by (A.16).}
linear to the gauge couplings $g$, $g^\prime$
and $g_s$,
$$
\eqalign{
   {\cal L}_{ffV} =& 
    - {g\over{\sqrt 2}}\sum_{(f_{\uparrow},f_{\downarrow})}
	       [\bar\Psi(f_{\uparrow})\gamma^\mu L 
               \Psi(f_{\downarrow}) W_\mu^+
      + \bar\Psi(f_{\downarrow})\gamma^\mu L \Psi(f_{\uparrow}) W_\mu^-]\cr
  & -g_Z\sum_f\bar\Psi(f)\gamma^\mu[(T_{3f}-s_W^2Q_f)L -s_W^2 Q_f R]
                     \Psi(f)Z_\mu \cr
  & -e\sum_f Q_f\bar\Psi(f)\gamma^\mu \Psi(f) A_\mu \cr
  & -g_s \sum_q\bar\Psi(q)\gamma^\mu {{\lambda^\alpha}\over 2}
                           \Psi(q) g_\mu^\alpha,
}\eqno(3.4)
$$
where $f_{\uparrow}$ and $f_{\downarrow}$ stand for the up and down
components of $SU(2)_L$ fermion doublets, and 
the summation $f$  is taken over the fermion species, $f=\nu, e, u$ 
and $d$, and the summation $q$ is taken over all the quark species.
We have used (A.9) and (A.13) in rewriting the interaction lagrangian 
in terms of the four-component spinors.  Note the property,
$$
  Q_{f_R}\equiv {{Y_R}\over 2}= -Q_{f_L}\equiv -[T_{3f_L}+{{Y_{f_L}}\over 2}],
  ~~~~~~ {\rm for~ each} ~f
\eqno(3.5)
$$
which comes from the fact that the quantum numbers of $f_R$ are those of
the corresponding lefthanded antiparticle.  The lagrangian (3.3) 
agrees with the fermion-gauge boson interaction of SM.  
In a similar way we obtain
$$
\eqalign{
   {\cal L}_{\tilde f \tilde f V} =&
   -i{g\over{\sqrt 2}} \sum_f
     (\tilde f^*_{\uparrow L}\partrl^\mu \tilde f_{\downarrow L} W^+_\mu
     +\tilde f^*_{\downarrow L}\partrl^\mu \tilde f_{\uparrow L} W^-_\mu)\cr
  & -ig_Z\sum_f [
          \tilde f_L^*\partrl^\mu(T_{3f}-s_W^2Q_f)\tilde f_L
         -\tilde f_R^*\partrl^\mu (s_W^2 Q_f) \tilde f_R ] Z_\mu\cr
  & -ie\sum_fQ_f[\tilde f_L^*\partrl^\mu \tilde f_L 
              + \tilde f_R^*\partrl^\mu \tilde f_R] A_\mu \cr
  & -ig_s\sum_q
    ( \tilde q_L^*{{\lambda^\alpha}\over 2}\partrl^\mu\tilde q_L
    +\tilde q_R^*{{\lambda^\alpha}\over 2}\partrl^\mu\tilde q_R)g_\mu^\alpha.
}\eqno(3.6)
$$
This confirms that $\tilde e_L$ and $\tilde e_R$ have indeed the same charge
as electron.\par
   The fermion-sfermion-gaugino interaction\footnote{*}
{Here, I used the terminology "gaugino", since the lagrangian is 
still expressed in terms of Weyl spinors
which are the partners of gauge bosons.  Using (2.9), (2.10) and (2.17), 
we have  to express (3.7) in terms of charginos and neutralinos  which are
mixtures of gauginos and higgsinos.} is extracted from the remaining part of 
trilinear coupling in  (3.1b)-(3.1f),
$$
\eqalign{
  {\cal L}_{\tilde f f \tilde V}=& - g\sum_f[
     (*,\lambda^-)L\Psi(f_\uparrow)\tilde f_{\downarrow L}^* + 
     (*,\lambda^+)L\Psi(f_\downarrow)\tilde f_{\uparrow L}^*] \cr
   & -\sqrt 2 \sum_f[\{
        g^\prime(Q_f-T_{3fL})(*,\lambda)
       +gT_{3fL}(*,\lambda^0)\}L\Psi(f)\tilde f_L^*]\cr
   & + \sqrt 2 g^\prime  \sum_f Q_f 
        (\bar\lambda,*)R\Psi(f)\tilde f_R^*\cr
   & -\sqrt 2 g_s \sum_q\lbrack
      \overline{\Psi(q)}R \Psi(\tilde g^\alpha)
        {{\lambda^\alpha}\over 2} \tilde q_L
     -\overline{\Psi(q)}L \Psi(\tilde g^\alpha) 
        {{\lambda^\alpha}\over 2}\tilde q_R  \rbrack \cr
   & + h.c..
}\eqno(3.7)
$$
Here, $"*"$ in the spinor means that the entry does not appear in 
the lagrangian 
and  is not necessary to be specified at this stage.
In the actual calculation, using (2.10), one obtains the lagrangian in 
terms of the mass eigenstates by replacing
$$
\eqalign{
 (*,\lambda^-) ~\to&~~  \cos\phi_R \bar\Psi(\tilde \chi_1^+)
                     -\sin\phi_R \bar\Psi(\tilde \chi_2^+),\cr
 (*,\lambda^+) ~\to&~~  \cos\phi_L \bar\Psi(\tilde \chi_1^-)
                -\epsilon_L \sin\phi_L \bar\Psi(\tilde \chi_2^-),\cr
 (*,\lambda)   ~\to&~~ ({\cal O}_N)_{j1}\eta^*_j\bar\Psi(\tilde\chi^0_j),\cr
 (\bar\lambda,*)   ~\to&~~ ({\cal O}_N)_{j1}\eta_j\bar\Psi(\tilde\chi^0_j),\cr
 (*,\lambda^0)   ~\to&~~ ({\cal O}_N)_{j2}\eta^*_j\bar\Psi(\tilde\chi^0_j).\cr
}\eqno(3.8)
$$
Therefore, depending on the convention of calling 
a positive chargino  as a particle or a negative chargino, 
either one of the two terms of the 
first line of (3.7) violates the fermion
number conservation, typical to the chargino interactions.\par
      From the gauge coupling squared terms in the matter lagrangian, 
(3.1b)-(3.1f), we obtain the $\tilde f \tilde f VV$ 
interaction terms;
$$
\eqalign{
   {\cal L}_{\tilde f\tilde f VV} =
    & ~~~{{g^2}\over 2}(\sum_f \tilde f^*_L\tilde f_ L) W^+_\mu W^{-\mu} \cr
    & -{{gg_Z}\over {\sqrt 2}}s_W^2\sum_f Y_{fL}
       (\tilde f^*_{\uparrow L}\tilde f^*_{\downarrow L}W^+_\mu 
      +\tilde f^*_{\downarrow L}\tilde f^*_{\uparrow L}W^-_\mu)Z^\mu \cr 
    & +{{ge}\over {\sqrt 2}}\sum_f Y_{fL}
       (\tilde f^*_{\uparrow L}\tilde f^*_{\downarrow L}W^+_\mu 
      +\tilde f^*_{\downarrow L}\tilde f^*_{\uparrow L}W^-_\mu)A^\mu \cr 
    & +g^2_ZF_{ZZ}Z_\mu Z^\mu
     +2eg_Z F_{ZA}Z_\mu A^\mu + e^2F_{AA}A_\mu A^\mu\cr
    & + g_s^2 \sum_q( 
   \tilde q_L^*{{\lambda^\alpha}\over 2} {{\lambda^\beta}\over 2}\tilde q_L
 + \tilde q_R^*{{\lambda^\alpha}\over 2} {{\lambda^\beta}\over 2}\tilde q_R)
           g_\mu^\alpha g^{\mu\beta}\cr
    &+{{g_s g}\over {\sqrt 2}}\{
     \tilde u_L^*\lambda^\alpha \tilde d_L W^{+\mu} 
    + \tilde d_L^*\lambda^\alpha \tilde u_L W^{-\mu}\}g_\mu^\alpha\cr
    &+g_s g_Z\sum_q\{(T_{3q}-s_W^2Q_q)\tilde q_L\lambda^\alpha\tilde q_L
           -s_W^2Q_q\tilde q_R\lambda^\alpha\tilde q_R\}Z^\mu g_\mu^\alpha \cr
    &+g_s e \sum_q Q_q(\tilde q_L\lambda^\alpha \tilde q_L
                  +\tilde q_R\lambda^\alpha \tilde q_R)A^\mu g_\mu^\alpha,
}\eqno(3.9)
$$
where
$$
\eqalign{
     F_{ZZ} =& \sum_f[~(T_{3f}-s_W^2Q_f)^2\tilde f^*_L\tilde f_L
              + s_W^4Q_f^2 \tilde f^*_R\tilde f_R~],\cr       
     F_{ZA} =& \sum_f [~Q_f(T_{3f}-s_W^2Q_f) \tilde f^*_L\tilde f_L
               -s_W^2Q_f^2\tilde f^*_R\tilde f_R~ ], \cr
     F_{AA} =& \sum_f Q_f^2(\tilde f^*_L\tilde f_L+\tilde f^*_R\tilde f_R).
}\eqno(3.9a)
$$\par
   From the Yukawa interaction of matter fields with Higgs fields, 
(3.1i), (3.1j) and (3.1k), we obtain the fermion mass terms
and the following $ffH$ interaction as well as 
$\tilde f f~ \tilde H$ interaction,
$$
\eqalign{
    {\cal L}_{f f H}=
     & -\sum_f m_f\bar\Psi(f)\Psi(f) \cr
     &+{{\sqrt 2 m_u}\over{v_2}}\bar\Psi(u)L\Psi(d)\phi_2^+ 
      -{{\sqrt 2 m_d}\over{v_1}}\bar\Psi(u)R\Psi(d)\phi_1^+ + h.c.\cr
     &-{{\sqrt 2 m_e}\over{v_1}}\bar\Psi(\nu)R\Psi(e)\phi_1^+ + h.c.\cr 
     &-{{m_u}\over{v_2}}[
         \bar\Psi(u)\Psi(u)\phi_2^0 -i\bar\Psi(u)\gamma_5\Psi(u)\chi_2^0]\cr
     &-{{m_d}\over{v_1}}[
         \bar\Psi(d)\Psi(d)\phi_1^0 +i\bar\Psi(d)\gamma_5\Psi(d)\chi_1^0]\cr
     &-{{m_e}\over{v_1}}[
         \bar\Psi(e)\Psi(e)\phi_1^0 +i\bar\Psi(e)\gamma_5\Psi(e)\chi_1^0],\cr
}\eqno(3.10a)
$$
$$
\eqalign{
    {\cal L}_{\tilde f f \tilde H}=
     &~{{\sqrt 2 m_u}\over{v_2}}\lbrack
       \bar\Psi(u)L(*,\tilde H_2^+)^t\tilde d_L 
      -\bar\Psi(u)L(*,\tilde H_2^0)^t\tilde u_L \cr
     &~~~~~~~~~~+(\bar\Psi(d)R(\overline{\tilde H_2^+},*)^t 
       -\bar\Psi(u)R(\overline{\tilde H_2^0},*)^t)\tilde u_R\rbrack + h.c.\cr
     &-{{\sqrt 2 m_d}\over{v_1}}\lbrack
        \bar\Psi(d)L(*,\tilde H_1^0)^t\tilde d_L
       -\bar\Psi(d)L(*,\tilde H_1^-)^t\tilde u_L  \cr
     &~~~~~~~~~~+(\bar\Psi(d)R(\overline{\tilde H_1^0},*)^t 
        -\bar\Psi(u)R(\overline{\tilde H_1^-},*)^t)\tilde d_R\rbrack + h.c.\cr
     &-{{\sqrt 2 m_e}\over{v_1}}\lbrack
        \bar\Psi(e)L(*,\tilde H_1^0)^t\tilde e_L
       -\bar\Psi(e)L(*,\tilde H_1^-)^t\tilde \nu_L  \cr
     &~~~~~~~~~~+(\bar\Psi(e)R(\overline{\tilde H_1^0},*)^t 
      -\bar\Psi(\nu)R(\overline{\tilde H_1^-},*)^t)\tilde e_R\rbrack + h.c.,\cr
}\eqno(3.10b)
$$
where
$$
\eqalign{
     (*,\tilde H_1^-)^t=& \sin\phi_R\Psi(\tilde\chi_1^-)
                         +\cos\phi_R\Psi(\tilde\chi_2^-),\cr
     (*,\tilde H_2^+)^t=& \sin\phi_L\Psi(\tilde\chi_1^+)
               +\epsilon_L\cos\phi_L\Psi(\tilde\chi_2^+),\cr
     (\overline{\tilde H_1^-},*)^t=& \sin\phi_R\Psi(\tilde\chi_1^+)
                       +\cos\phi_R\Psi(\tilde\chi_2^+),\cr
     (\overline{\tilde H_2^+},*)^t=& \sin\phi_L\Psi(\tilde\chi_1^-)
                       +\epsilon_L \cos\phi_L\Psi(\tilde\chi_2^-),\cr
     (*,\tilde H_1^0)^t=& ({\cal O}_N)_{j3}\eta^*_j\Psi(\tilde\chi_j^0),
     ~~~~~~~~
     (*,\tilde H_2^0)^t= ({\cal O}_N)_{j4}\eta^*_j\Psi(\tilde\chi_j^0),\cr
     (\overline{\tilde H_1^0},*)^t=& ({\cal O}_N)_{j3}\eta_j
                                       \Psi(\tilde\chi_j^0),~~~~~~~~
     (\overline{\tilde H_2^0},*)^t= ({\cal O}_N)_{j4}\eta_j
                                 \Psi(\tilde\chi_j^0).\cr
}\eqno(3.10c)
$$\par
\medskip

\noindent$\underline{3.2~ F\hbox{-terms~ and}~ D\hbox{-terms}}$\par
     The remaining interaction including matter fields come from the 
auxiliary fields $\vert F\vert ^2$  and $D^2$.
From (B.12) and (B.13), we find that the $F$-terms in the matter 
superfield provide supersymmetric sfermion mass terms,  $\tilde f\tilde f 
H(G)$ and $\tilde f\tilde f HH$ interactions;
$$         
    -\sum_{f_i}\vert F(f_i)\vert^2 = {\cal L}_{\tilde f~mass}
                     +{\cal L}_{\tilde f \tilde f H} 
                     +{\cal L}_{\tilde f \tilde f HH}. 
\eqno(3.11)
$$
The sfermion mass terms are given as
$$
    {\cal L}_{\tilde f~mass} = -\sum_f m_f^2(\tilde f_L^*\tilde f_L 
                                            +\tilde f_R^*\tilde f_R),
     ~~~~~~~~~~
\eqno(3.11a)
$$
where the sum is over the matter fields, $u_L$, $d_L$, $u_R$, $d_R$, 
$\nu_L$, $e_L$ and $e_R$.
$$
\eqalign{
    {\cal L}_{\tilde f \tilde f H(G)} =
    & -2{{m_u^2}\over {v_2}}
      (\tilde u_L^* \tilde u_L+ \tilde u_R\tilde u_R)\phi_2^0
      -2{{m_d^2}\over{v_1}} 
      (\tilde d_L^* \tilde d_L+ \tilde d_R\tilde d_R)\phi_1^0\cr
    & -2{{m_e^2}\over {v_1}}  
      (\tilde e_L^* \tilde e_L+ \tilde e^*_R\tilde e_R)\phi_1^0\cr
    & +\sqrt 2 g{{m_um_d}\over{M_W\sin2\beta}}
         (\tilde u^*_R \tilde d_R H^+ + \tilde d^*_R \tilde u_R H^-)\cr
    &+\sqrt 2\tilde u^*_L\tilde d_L
    ({{m_u^2}\over {v_2}}\phi_2^+ - {{m_d^2}\over {v_1}}\phi_1^+)
     -\sqrt 2{{m_e^2}\over {v_1}}\tilde \nu^*_L\tilde e_L\phi_1^+ +hc.,
}\eqno(3.11b)
$$
$$ 
\eqalign{
     {\cal L}_{\tilde f\tilde f HH}=
   &-{{m_u^2}\over{v^2_2}}\vert \chi_2^0\vert^2
        (\tilde u^*_R\tilde u_R + \tilde u^*_L\tilde u_L) 
    -{{m_d^2}\over{v^1_2}}\vert \chi_1^0\vert^2
        (\tilde d^*_R\tilde d_R + \tilde d^*_L\tilde d_L) \cr
   &-{{m_e^2}\over{v^1_2}}\vert \chi_1^0\vert^2
        (\tilde e^*_R\tilde e_R + \tilde e^*_L\tilde e_L) \cr
   & -\bigg\vert {{m_u}\over{v_2}}\phi_2^0\tilde u_R^*
          + \sqrt 2{{m_d}\over {v_1}}\phi_1^-\tilde d_R^*\bigg\vert^2  
     -\bigg\vert \sqrt 2{{m_u}\over {v_2}}\phi_2^+ \tilde u_R^*
          -{{m_d}\over{v_1}}\phi_1^0\tilde d_R^* \bigg\vert^2\cr
   & -{{m_u^2}\over{v_2^2}} \vert \phi_2^0\tilde u^*_L
          -\sqrt 2\phi_2^-\tilde d^*_L\vert^2  
     -{{m_d^2}\over {v_1^2}} \vert \phi_1^0\tilde d^*_L
          +  \sqrt 2\phi_1^+ \tilde u^*_L \vert^2\cr
   & -{{m_e^2}\over {v_1^2}}[2\vert\phi_1^-\vert^2 + (\phi_1^0)^2]  
                               \tilde e_R^*\tilde e_R
     -{{m_e^2}\over {v_1^2}} \vert \phi_1^0\tilde e^*_L
          +  \sqrt 2\phi_1^+ \tilde \nu^*_L \vert^2\cr
   &-i\sqrt 2({{m_u^2}\over{v_2^2}}\chi_2^0\phi_2^+           
             -{{m_d^2}\over{v_1^2}}\chi_1^0\phi_1^+)
             (\tilde u^*_L\tilde d_L)+h.c. \cr
   &-i\sqrt 2{{m_um_d}\over{v_1v_2}}
             ( \chi_2^0\phi_1^+ - \chi_1^0\phi_2^+)
             (\tilde u^*_R\tilde d_R)+h.c.\cr
}\eqno(3.11c)
$$\par
     From ${\bf F}(H_1)$ and ${\bf F}(H_2)$, one finds a part of the 
Higgs potential, $\tilde f_L\tilde f_R$ mixing terms and two 
kinds of interaction,
$$
    -\sum\vert F(H_i)\vert^2 = {\cal L}_{V(F)}
    +{\cal L}_{\tilde f_L\tilde f_R}+{\cal L}_{\tilde f\tilde f H(G)}
    +{\cal L}_{\tilde f\tilde f\tilde f\tilde f},
\eqno(3.12)
$$    
where the suffix $V(F)$ of the first lagrangian on the right-hand side
signifies that it is the Higgs potential contributed from the $F$  terms.
$$
 {\cal L}_{V(F)} = -\mu^2({\bf H}_1^*{\bf H}_1 + {\bf H}_2^*{\bf H}_2).
\eqno
(3.12a)
$$
$$
\eqalign{
  {\cal L}_{\tilde f_L\tilde f_R}=&~~~~~
       m_u~\mu~ \cot\beta( \tilde u^*_R\tilde u_L + \tilde u^*_L\tilde u_R)\cr 
    & +m_d~\mu~ \tan\beta( \tilde d^*_R\tilde d_L + \tilde d^*_L\tilde d_R)\cr 
    & +m_e~\mu~ \tan\beta( \tilde e^*_R\tilde e_L + \tilde e^*_L\tilde e_R),\cr
}\eqno(3.12b)
$$
$$
\eqalign{
    {\cal L}_{\tilde f \tilde fH(G)} = &~ {{\mu}\over{v_1}}\lbrack
             ( m_d\tilde d^*_R \tilde d_L
             + m_e\tilde e^*_R \tilde e_L)(\phi_2^0 - i\chi_2^0)\cr
     &~~~~~~~+\sqrt 2( m_d\tilde d^*_R \tilde u_L
               +m_e\tilde e^*_R \tilde \nu_L)\phi_2^-\rbrack\cr
     &+{{m_u\mu}\over{v_2}}\lbrack
             (\tilde u^*_R\tilde u_L)(\phi_1^0+i\chi_1^0)
            -\sqrt 2(\tilde u^*_R\tilde d_L)\phi_1^+ \rbrack\cr
     &+ h.c.,  
}\eqno(3.12c)
$$
$$
\eqalign{
   {\cal L}_{\tilde f\tilde f\tilde f\tilde f} =&
      - {2\over{v_1^2}}\vert m_d\tilde d^*_R \tilde d_L
                           + m_e\tilde e^*_R \tilde e_L\vert^2
      - {2\over{v_1^2}}\vert m_d\tilde d^*_R \tilde u_L
                           + m_e\tilde e^*_R \tilde \nu_L\vert^2\cr
    &-{{2m_u^2}\over{v_2^2}}(\vert \tilde u^*_R\tilde d_L\vert^2
                     +\vert \tilde u^*_R\tilde u_L\vert^2).
}\eqno(3.12d)
$$\par
     As three generations of fermions must be taken into account, we 
have to sum over the fermions in three generations before taking  square 
in each term of (3.12d).\par 
     From the $D$-terms (B.18), we obtain a part of the Higgs potential, 
sfermion masses, $\tilde f\tilde f H(G)$, $\tilde f\tilde f HH$ 
as well as quartic sfermion interactions,
$$
  D\hbox{-terms}=-{1\over 2}\sum\vert D\vert^2
           ={\cal L}_{V(D)}+{\cal L}_{\tilde f\tilde f}
  +{\cal L}_{\tilde f\tilde f H(G)}+{\cal L}_{\tilde f\tilde f HH} 
  + {\cal L}_{\tilde f\tilde f\tilde f\tilde f}.
\eqno(3.13)
$$
The contribution from the $D$-term to the Higgs potential, ${\cal L}_{V(D)}$
produces  the Higgs self-interaction terms, 
$$
\eqalign{
   {\cal L}_{V(D)}  =&- {1\over 8}(g^2 + g^{\prime 2}) 
            ({\bf H}^*_1{\bf H}_1 -{\bf H}^*_2{\bf H}_2)^2
       -{{g^2}\over 2}\vert{\bf H}^*_1{\bf H}_2\vert^2\cr
 = &~~ {\rm constant}  + {\rm mass~ terms} +{\cal L}_{3H} + {\cal L}_{4H},\cr
}\eqno(3.14)
$$
where ${\cal L}_{3H}$ and ${\cal L}_{4H}$  are three body and four body
Higgs and Goldstone boson self-interactions.
$$
\eqalign{
     {\cal L}_{3H} =
  &- {g\over{4c}}M_Z[\cos(\beta+\alpha)H^0-\sin(\beta+\alpha)h^0]X\cr
  &-gM_W[\cos(\beta-\alpha)H^0H^+H^-+ \sin(\beta-\alpha)h^0H^+H^-]\cr
  &-{g\over 2}M_W[\sin(\beta-\alpha)H^0-\cos(\beta-\alpha)h^0]
                 (H^+G^-+H^-G^+)  \cr
  &-i{g\over 2}M_W A^0(H^+G^--H^-G^+),\cr
}\eqno(3.14a)
$$
where
$$
\eqalign{
     X=& \cos2\alpha((H^0)^2-(h^0)^2) -2\sin2\alpha H^0h^0 
                   -\cos2\beta(2\vert H^+\vert^2+ (A^0)^2) \cr
       & -2\sin2\beta (G^0A^0 +H^+G^-+H^-G^+)
         +\cos2\beta(2\vert G^+\vert^2 + (G^0)^2),\cr
}\eqno(3.14b)
$$
$$
\eqalign{
      {\cal L}_{4H}=
  &-{{g^2}\over{32c^2}} X^2\cr
  &-{{g^2}\over 4}\lbrack
    \vert \cos(\beta-\alpha)(H^0H^+-h^0G^+) 
        + \sin(\beta-\alpha)(H^0G^++h^0H^+)\vert^2\cr
  &~~~~~~~~+\vert H^+G^0 - A^0G^+\vert^2 \rbrack\cr
  &-i{{g^2}\over 4}(H^+G^--H^-G^+)\cr
  &~~~~~~~[~\cos(\beta-\alpha)(A^0H^0 - h^0G^0)
           -\sin(\beta-\alpha)(A^0h^0 + H^0G^0)].\cr
}\eqno(3.14c)
$$
The mass terms and the constant terms are calculated in section 4 when
the full Higgs potential, resulted also from the SUSY breaking term 
and from the $F$ terms, is taken into account. 
$$
 {\cal L}_{\tilde f\tilde f} =-M_Z^2 \cos2\beta
 [~ (T_{3f}-s_W^2Q_f)\tilde f^*_L\tilde f_L + s_W^2Q_f\tilde f^*_R\tilde f_R)~].
\eqno(3.15)
$$
$$
\eqalign{
    {\cal L}_{\tilde f\tilde f H(G)} =
    & -g_Z M_Z(\cos(\beta+\alpha)H^0 - \sin(\beta+\alpha)h^0)\cr
    &~~~~~~~~~~~~
    [(T_{3f}-s_W^2Q_f)\tilde f^*_L\tilde f_L+s_W^2Q_f\tilde f^*_R\tilde f_R]\cr
&+{g\over{\sqrt 2}} M_W(\cos2\beta G^+-\sin2\beta H^+)\tilde f^*_{\uparrow L}
      \tilde f_{\downarrow L} + h.c.
}\eqno(3.16)
$$
$$
\eqalign{
  {\cal L}_{\tilde f\tilde fHH} =
    & -{{g_Z^2}\over 4} (\sum_f F_{LL}\tilde f^*_L\tilde f_L
                         + F_{RR}\sum_f s^2Q_f\tilde f^*_R\tilde f_R) \cr
    & -{{g^2}\over{2\sqrt 2}}F_{\uparrow\downarrow}^{(1)}   
       \sum_f \tilde f^*_{\uparrow L} \tilde f_{\downarrow L}
      +i{{g^2}\over{2\sqrt 2}}F_{\uparrow\downarrow}^{(2)}   
       \sum_f \tilde f^*_{\uparrow L} \tilde f_{\downarrow L}+ h.c.,\cr
}\eqno(3.17)      
$$
with
$$
\eqalign{
      F_{LL}=&~~~\{ \cos2\alpha((H^0)^2-(h^0)^2) -2\sin2\alpha H^0h^0 \cr
      &~~+ \cos2\beta((G^0)^2-(A^0)^2) -2\sin2\beta G^0A^0\}
                     (T_{3f}-s^2Q_f)\cr
      & -2\{ \cos2\beta (\vert G^+\vert^2-\vert H^+\vert^2 )
           -\sin2\beta(H^+G^-+H^-G^+)\}( T_{3f}+ s^2Q_{f^\prime}),\cr
      F_{RR}=&~~ \cos2\alpha((H^0)^2-(h^0)^2) -2\sin2\alpha H^0h^0 \cr
      & + \cos2\beta((G^0)^2-(A^0)^2) -2\sin2\beta G^0A^0 \cr
      & + 2\cos2\beta(\vert G^+\vert^2-\vert H^+\vert^2)
                +2\sin2\beta(H^+G^-+H^-G^+), \cr
      F_{\uparrow\downarrow}^{(1)} =
      & \cos(\beta+\alpha)(H^+h^0-H^0G^+)+\sin(\beta+\alpha)(H^+H^0+h^0G^+),\cr
      F_{\uparrow\downarrow}^{(2)} =
      & \cos2\beta(A^0H^+ -G^+G^0)+\sin2\beta(A^0G^++H^+G^0), \cr
}\eqno(3.17a)
$$
where $f^\prime$ is the $SU(2)_L$ partner of fermion $f$ in the doublet. 
Four-sfermion interactions are given by (B.18),
$$
\eqalign{
     {\cal L}_{\tilde f\tilde f\tilde f\tilde f} =
     &-{{g^2}\over 8}[4\vert \tilde u_L^* \tilde d_L
                              +\tilde \nu^* \tilde e_L \vert ^2
         + \vert \tilde u_L^*\tilde u_L -\tilde d_L^*\tilde d_L
         +\tilde \nu^*\tilde \nu -\tilde e_L^*\tilde e_L\vert^2] \cr
     &-{{g^{\prime 2}}\over 8}\lbrack
       -\tilde \nu^*_L\tilde\nu_L-\tilde e^*_L\tilde e_L
     +{1\over 3}(\tilde u^*_L\tilde u_L + \tilde d^*_L\tilde d_L)
     +2\tilde e^*_R\tilde e_R -{4\over 3}\tilde u^*_R\tilde u_R 
     +{2\over 3}\tilde d^*_R\tilde d_R \rbrack^2\cr
   & -{{g_s^2}\over 8}\{
   ~ {4\over 3}\sum_{q,i}\vert \tilde q_i^* \tilde q_i \vert^2 
   +4\sum_{q< q^\prime,i}  \vert \tilde q_i^* \tilde q^\prime_i \vert^2 
    -{4\over 3}\sum_{q< q^\prime,i}(\tilde q_i^*\tilde q_i)
                           (\tilde q^{\prime *}_i \tilde q^\prime_i) \}\cr
    &+{{g_s^2}\over 4}\{                           
   ~ 2 \sum_q\vert \tilde q_L^* \tilde q_R \vert^2 
   +4\sum_{q< q^\prime}  \vert \tilde q_L^* \tilde q^\prime_R \vert^2 
    -{2\over 3}\sum_{q,q^\prime}(\tilde q_L^*\tilde q_L)
                           (\tilde q^{\prime *}_R \tilde q^\prime_R) \},\cr
}\eqno(3.18)
$$
where for saving the space  the strong interaction part proportional 
to $g_s^2$ is not explicitly expanded.
Here again, as three generations of fermions are considered, following 
(B.18) one has to sum over the generations before taking squares at 
the $g^{\prime 2}$ term,
while in the $g^2$ term, according to (B.18) all the possible doublet 
combinations must be considered.\par

\medskip

\noindent$\underline{3.3~ {\rm Higgs~ gauge~interactions}}$\par 
    From (3.1g), (3.1h) as well as from the $D$-terms, we obtain  
the various Higgs interaction terms.  The $D$-terms provide the 
quartic part of the Higgs potentials shown in (B.18).
From (3.1g) and (3.1h), 
we obtain the gaugino-higgsino mixing terms and the following four 
interactions.
$$
\eqalign{
     {\cal L}_{HHV}=& -{g\over 2}i
           [(v_1\partial^\mu H_1^- - v_2\partial^\mu H_2^-)W^+_\mu
           -(v_1\partial^\mu H_1^+ - v_2\partial^\mu H_2^+)W^-_\mu]\cr
     &-{g\over{2c_W}}(v_1\partial^\mu\chi_1^0+v_2\partial^\mu\chi_2^0)Z_\mu\cr
     &+{g\over 2}i\lbrack
               \cos(\beta-\alpha)(H^0\partrl^\mu G^- + h^0\partrl^\mu H^-) \cr
     &~~~~~~ - \sin(\beta-\alpha)(H^0\partrl^\mu H^- - h^0\partrl^\mu G^-) \cr
     &~~~~~~ +iG^0\partrl^\mu G^- +iA^0\partrl^\mu H^- 
                  \rbrack W^+_\mu + h.c.\cr
     &+{g\over{2c_W}}i\lbrack
         (1-2s_W^2)(G^+\partrl^\mu G^-+H^+\partrl^\mu H^-)\cr
     &~~~~~~-i\cos(\beta-\alpha)(G^0\partrl^\mu H^0 + A^0\partrl^\mu h^0) \cr
     &~~~~~~-i\sin(\beta-\alpha)(G^0\partrl^\mu h^0 - A^0\partrl^\mu H^0) 
          \rbrack Z_\mu \cr
     &+ie[G^+\partrl^\mu G^- + H^+\partrl^\mu H^-] A_\mu.
}\eqno(3.19)
$$     
Here, the first two lines can be removed by properly choosing the gauge
fixing terms.
$$
\eqalign{
   {\cal L}_{\tilde H\tilde H V} =&+{g\over{\sqrt 2}}
       [\tilde H_1^-\sigma^\mu \bar{\tilde H_1^0}
       +\tilde H_2^0\sigma^\mu \bar{\tilde H_2^+}]W^+_\mu +h.c.\cr
    &+{g\over{2c_W}}\lbrack
         \tilde H_1^0\sigma^\mu \bar{\tilde H_1^0} 
        -\tilde H_2^0\sigma^\mu \bar{\tilde H_2^0} \cr
    &~~~~~~+(1-2s_W^2)(\tilde H_2^+\sigma^\mu \bar{\tilde H_2^+}
             -\tilde H_1^-\sigma^\mu \bar{\tilde H_1^-})\rbrack Z_\mu\cr
    &-e[ \tilde H_1^-\sigma^\mu\bar{\tilde H_1^-}
       - \tilde H_2^+\sigma^\mu\bar{\tilde H_2^+}]A_\mu.
}\eqno(3.20)
$$
In (3.20) and in the following (3.22b), in the case of neutralino-neutralino 
interactions, $\tilde \chi_i^0\tilde \chi_j^0 V$, 
$\tilde \chi_i^0\tilde \chi_j^0 H$,  $\tilde \chi_i^0\tilde \chi_j^0 G$,
one has to impose the Majorana condition for neutralinos in order to
find the Feynman rule from the lagrangian.  Namely, using
$$
     \overline{\Psi(\tilde \chi^0_i)}\Gamma\Psi(\tilde \chi^0_j)
    =\overline{\Psi(\tilde \chi^0_j)}\Gamma^c\Psi(\tilde \chi^0_i),
\eqno(3.21)
$$
where
$$
     \Gamma^c = C\Gamma C^{-1},
\eqno(3.21a)
$$
with
$$
     C(1,\gamma_5, \gamma_\mu, \gamma_\mu\gamma_5, \sigma_{\mu\nu})C^{-1}
    = (1,\gamma_5, -\gamma_\mu, \gamma_\mu\gamma_5, -\sigma_{\mu\nu}),
\eqno(3.21b)
$$
the lagrangian must be put in the form,
$$
\eqalign{
    {\cal L} \sim \sum_{i,j} 
       \overline{\Psi(\tilde \chi^0_i)}\Gamma(i,j)\Psi(\tilde \chi^0_j)
    =& \sum_i \overline{\Psi(\tilde \chi^0_i)}\Gamma(i,i)
                         \Psi(\tilde \chi^0_i)\cr
    +& \sum_{i<j} 
       \overline{\Psi(\tilde \chi^0_i)}[\Gamma(i,j)+ \Gamma(j,i)^c]
       \Psi(\tilde \chi^0_j).
}\eqno(3.21c)
$$
$$
\eqalign{
     {\cal L}_{\tilde H \tilde V}= 
     & -\sqrt 2 M_W[ \cos\beta \varphi(\tilde H^-_1)\varphi(\lambda^+) 
                    +\sin\beta \varphi(\tilde H^+_2)\varphi(\lambda^-)]+h.c.\cr
     &-M_Z[\cos\beta \varphi(\tilde H^0_1)-\sin\beta\varphi(\tilde H^0_2)]
          [c_W\varphi(\lambda^0)-s_W\varphi(\lambda)]+h.c.\cr
}\eqno(3.22a)
$$
$$
\eqalign{
     {\cal L}_{H\tilde H \tilde V}= 
     &~[               g           \varphi(\tilde H^0_1)\varphi(\lambda^-) 
        -{g\over {\sqrt 2}}        \varphi(\lambda^0)\varphi(\tilde H^-_1) 
        -{{g^\prime}\over{\sqrt 2}}\varphi(\lambda)\varphi(\tilde H_1^-)] 
                                \phi^+_1 \cr
     &-[               g           \varphi(\lambda^+)\varphi(\tilde H^0_2)
        +{g\over {\sqrt 2}}        \varphi(\tilde H^+_2)\varphi(\lambda^0)
        +{{g^\prime}\over{\sqrt 2}}\varphi(\tilde H^+_2)\varphi(\lambda)]                                        \phi^-_2 \cr
     &-[  {g\over {\sqrt 2}}       \varphi(\lambda^+)\varphi(\tilde H^-_1) 
         +{g \over 2}              \varphi(\tilde H_1^0)\varphi(\lambda^0)
         -{{g^\prime}\over 2}      \varphi(\tilde H_1^0)\varphi(\lambda)] 
            (\phi^0_1 + i\chi_1^0)\cr
     &-[   {g\over {\sqrt 2}}      \varphi(\tilde H^+_2)\varphi(\lambda^-)
          -{g\over 2}              \varphi(\tilde H_2^0)\varphi(\lambda^0)
          +{{g^\prime}\over 2}     \varphi(\tilde H_2^0)\varphi(\lambda)] 
            (\phi^0_2 - i\chi_2^0)\cr
     & + h.c.\cr       
}\eqno(3.22b)
$$     
Using (2.9), (2.10), (2.14) and (2.15), we can easily rewrite 
the above lagrangian in terms of the mass eigenstates:  the rule is 
simply to replace each product of two Weyl spinors  according to the 
following rule,
$$
    \varphi(a)\varphi(b) \to \bar\Psi(a)L\Psi(b),
\eqno(3.22c)
$$
where
$$
\eqalign{
 \bar\Psi(a) =& \cases{({\cal O}_N)_{ia}\eta^*_i\bar\Psi(\tilde\chi_i^0),
                  ~~~~~~~~~~~~~~~~~~~~~~~$a= neutral$ \cr
                           ({\cal O}_{CR})_{ia}
                           (\delta_{i1}+\epsilon_L\delta_{i2})
                           \bar\Psi(\tilde\chi_i^-),~~~~~~~~$a=positive$\cr}\cr
    \Psi(b) =& \cases{({\cal O}_N)_{ib}\eta^*_i\Psi(\tilde\chi_i^0),
                  ~~~~~~~~~~~~~~~~~~~~~~~$b= neutral$ \cr
               ({\cal O}_{CL})_{ib}\Psi(\tilde\chi_i^-),~
                  ~~~~~~~~~~~~~~~~~~~~~~~$b=negative$\cr}
}\eqno(3.22d)
$$
with  ${\cal O}_{CL}$ and ${\cal O}_{CR}$ being two orthogonal matrices 
appearing in (2.10).

From the (gauge~ coupling)$^2$ terms, we have
$$
     {\cal L}_{Vmass}+ {\cal L}_{H(G)VV}+{\cal L}_{HHVV},
\eqno(3.23)
$$
where
$$
     {\cal L}_{Vmass}= {{g^2}\over 4}(v_1^2+v_2^2)W^+_\mu W^{-\mu}
                      +{{g_Z^2}\over 8}(v_1^2+v_2^2)Z_\mu Z^\mu,
\eqno(3.23a)
$$
$$
\eqalign{
       {\cal L}_{H(G)VV}=
       &~~ M_W G^-(eA^\mu-g_Zs_W^2Z^\mu)W^+_\mu + h.c.\cr
       &+gM_W(\cos(\beta-\alpha)H^0+ \sin(\beta-\alpha)h^0)W_\mu^+ W^{-\mu} \cr
       &+{{g_Z}\over 2}M_Z(\cos(\beta-\alpha)H^0+ \sin(\beta-\alpha)h^0)
                   Z_\mu Z^\mu, \cr
}\eqno(3.23b)
$$
$$
\eqalign{
      {\cal L}_{HHVV}=
      &~ {{g^2}\over 4} W_\mu^+ W^{-\mu}            
       [(H^0)^2 + (h^0)^2 + (A^0)^2+(G^0)^2+2\vert G^+\vert^2 + 
        2\vert H^+\vert^2] \cr
   &+ {{g_Z^2}\over 8}Z_\mu Z^\mu \lbrack (H^0)^2+(h^0)^2+(A^0)^2+(G^0)^2 \cr
   &~~~~~~~~~~    +2(1-2s_W^2)^2 (\vert G^+\vert^2 + \vert H^+\vert^2)\rbrack\cr
      &+(c_W^2-s_W^2)eg_Z  Z_\mu A^\mu(\vert G^+\vert^2 + \vert H^+\vert^2)\cr
      &+e^2A_\mu A^\mu(\vert G^+\vert^2 + \vert H^+\vert^2)\cr
      &+{{ge}\over 2} W^+_\mu (A^\mu-{{s_W}\over{c_W}}Z^\mu)F_W + h.c.,
}\eqno(3.23c)
$$
where
$$
\eqalign{
     F_W =&~~ \cos(\beta-\alpha)(H^0G^- + h^0H^-) 
          + \sin(\beta-\alpha)(h^0G^--H^0H^-)\cr
          &+i(G^0G^-+A^0H^-).
}\eqno(3.23d)
$$\par
\medskip

\noindent$\underline{3.4~{\rm SUSY~ breaking~ interactions}}$\par
     The SUSY breaking lagrangian (3.1m) or (3.2) 
contributes to the mass terms as well as the Yukawa interaction
terms which give the $\tilde f\tilde f H(G)$ interactions.  From the
third and fourth lines of (3.2) one obtains,
$$
\eqalign{
  {\cal L}_{\tilde f\tilde f H(G)} =
  &~  m_uA_u\tilde u^*_R\tilde u_L + m_dA_d\tilde d^*_R\tilde d_L 
   + m_eA_e\tilde e^*_R\tilde e_L +h.c.\cr
  &+{{m_u}\over{v_2}}A_u(-\sqrt 2 \phi^+_2\tilde u^*_R\tilde d_L
     + (\phi^0_2+i\chi_2^0)\tilde u^*_R\tilde u_L)+h.c.\cr
  &+{{m_d}\over{v_1}}A_d(\sqrt 2 \phi^-_1\tilde d^*_R\tilde u_L
     + (\phi^0_1-i\chi_1^0)\tilde d^*_R\tilde d_L)+h.c.\cr
  &+{{m_e}\over{v_1}}A_e(\sqrt 2 \phi^-_1\tilde e^*_R\tilde \nu_L
     + (\phi^0_1-i\chi_1^0)\tilde e^*_R\tilde e_L)+h.c.\cr
}\eqno(3.24)
$$\par
\medskip

\noindent$\underline{3.5~ {\rm Kinetic ~term~ (3.1a)}}$\par
     From the lagrangian (3.1a) we obtain 
the (nonlinear) gauge boson kinetic terms which includes gauge boson 
self-interaction, gaugino kinetic terms and ${\cal L}_{\tilde V
\tilde V V}$ interaction.
$$
\eqalign{
  (3.1a) =&-{1\over 4}W_{\mu\nu}^aW^{a\mu\nu}
           -{1\over 4}B_{\mu\nu} B^{\mu\nu}
           -{1\over 4}g_{\mu\nu}^\alpha g^{\alpha\mu\nu} \cr
          &+{\cal L}_{\tilde V~kin}+ {\cal L}_{\tilde V \tilde V V} 
           + D\hbox{-terms}.
}\eqno(3.25)
$$
The gauge boson kinetic part, the first line of (3.25), is the same as the SM.
The gaugino kinetic part is given as
$$
\eqalign{
   {\cal L}_{\tilde V~kin}=
       &~i\lambda^a\sigma^\mu\partial_\mu\overline{\lambda^a}
       +i\lambda\sigma^\mu\partial_\mu\overline{\lambda}
       +i\tilde g^\alpha \sigma^\mu\partial_\mu\overline{\tilde g}^\alpha\cr
      =&~i\lambda^+\sigma^\mu\partial_\mu\overline{\lambda^+}
       +i\lambda^-\sigma^\mu\partial_\mu\overline{\lambda^-}
       +i\lambda^0\sigma^\mu\partial_\mu\overline{\lambda^0}
       +i\lambda  \sigma^\mu\partial_\mu\overline\lambda
       +i\tilde g^\alpha \sigma^\mu\partial_\mu\overline{\tilde g}^\alpha.\cr
}\eqno(3.25a)
$$
Note that
$$
  \overline{\lambda^\pm} = {{\bar\lambda^1\pm i\bar\lambda^2}\over {\sqrt 2}}
                         = (\bar\lambda)^\mp.
\eqno(3.25b)
$$
$$
\eqalign{
    {\cal L}_{\tilde V\tilde V V}=
          &~~g(\lambda^-\sigma^\mu\overline{\lambda^0} 
             - \lambda^0\sigma^\mu\overline{\lambda^+})W^+_\mu + h.c.\cr
          &+cg(\lambda^+\sigma^\mu\overline{\lambda^+}
             - \lambda^-\sigma^\mu\overline{\lambda^-})Z_\mu\cr
          &+e (\lambda^+\sigma^\mu\overline{\lambda^+}
              -\lambda^-\sigma^\mu\overline{\lambda^-})A_\mu \cr
         & +{{g_s}\over 2}if^{\alpha\beta\gamma}\overline{\Psi(\tilde g^\alpha)}
            \gamma^\mu \Psi(\tilde g^\beta)g_\mu^\gamma.      
}\eqno(3.25c)
$$
    Combining (3.25a) with the kinetic part of the fermions, sfermions
Higgs and higgsino fields, 
$$
\eqalign{
   {\cal L}_{kinet} =
   &~~+ i\sum_f\bar\Psi(f)\slash\partial\Psi(f)\cr 
   &+ i\sum_i\bar\Psi(\tilde\chi_i^+)\slash \partial\Psi(\tilde\chi_i^+)
    + {i\over 2}\sum_i\bar\Psi(\tilde\chi_i^0)\slash \partial
                          \Psi(\tilde\chi_i^0)
    + {i\over 2}\sum_\alpha\bar\Psi(\tilde g^\alpha)\slash \partial
                          \Psi(\tilde g^\alpha)\cr
   &+ \sum_{f, i}(\partial^\mu\tilde f^*_i \partial_\mu\tilde f_i) \cr  
   &+ \sum_H \partial^\mu H^* \partial_\mu H 
    + \sum_G \partial^\mu G^* \partial_\mu G.\cr
}\eqno(3.26)
$$
Note that the factor ${1\over 2}$ is properly reproduced for neutralino
and gluino kinetic terms.\par

\medskip

\noindent$\underline{3.6~{\rm Gauge~fixing~terms~and~ghost~interacitons}}$\par
     Upon using (2.24)and (2.25) and partial integration, the first two 
lines of (3.19) reduces to
$$
     -iM_WG^-\partial^\mu W^+_\mu + iM_WG^+\partial^\mu W^-_\mu
                                  + M_ZG^0\partial^\mu Z_\mu.
\eqno(3.27)
$$
Therefore, the gauge fixing term can be chosen such that this part disappears
from the lagrangian:
$$
     {\cal L}_{gf} =
      -{1\over {\xi_W}}F^+ F^- -{1\over {2\xi_Z}}\vert F^Z \vert^2 
      -{1\over {2\xi_\gamma}}\vert F^\gamma \vert^2  
      -{1\over {2\xi_g}}\sum_\alpha\vert F^{g\alpha} \vert^2,  
\eqno(3.28)
$$
where
$$
\eqalign{
     F^\pm=&~ \partial^\mu W^\pm_\mu \pm iM_W\xi_WG^\pm, \cr 
     F^Z  =&~ \partial^\mu Z_\mu +M_Z\xi_ZG^0,\cr 
     F^\gamma=&~ \partial^\mu A_\mu,\cr 
     F^{g\alpha}=&~\partial^\mu g_\mu^\alpha.\cr 
}\eqno(3.29)
$$
This is exactly the same form as the SM gauge fixing terms, ensuring 
that the gauge boson propagators and the Goldstone boson propagators have 
the same form as the standard model.\par
\medskip

     Ghost interaction comes from the $SU(2)$ and $U(1)$ variation of the 
gauge fixing functions $F^\pm$ etc.  First we note that the matter and 
gauge fields transform as\footnote{*}
{The non-Abelian gauge transformation is defined in my convention as
$$
    {\bf W}_\mu \equiv \sum {{\tau^a}\over 2} W^a_\mu \to
     U(x) {\bf W}_\mu U(x)^{-1} -{i\over g} U(x)\partial_\mu U(x)^{-1},
$$
with
$$
     U(x) \equiv \exp[-i \sum {{\tau^a}\over 2} u(x)^a]. 
$$}
$$
\eqalign{
   SU(2)_L: & \cases{\delta W_\mu^i = \epsilon^{ijk}u^jW^k_\mu
                     +{1\over g}\partial_\mu u^i, \cr
              \delta B_\mu =0,\cr       
              \delta({\bf H}_i-<{\bf H}_i>_0) = -i\left\lgroup
              {{\vec \tau}\over 2}\cdot \vec u\right\rgroup {\bf H}_i,\cr} \cr
    U(1): &  \cases{\delta W_\mu^i =0,\cr
             \delta B_\mu = {1\over{g^\prime}}\partial_\mu\alpha,\cr
             \delta({\bf H}_i-<{\bf H}_i>_0) = -iY_i{\alpha\over 2}
             {\bf H}_i, \cr} \cr
   SU(3)_c: & ~~~~\delta g_\mu^\alpha = 
                  f^{\alpha\beta\gamma}u^\beta g_\mu^\gamma
              +{1\over {g_s}}\partial^\mu u^\alpha_\mu,          
}\eqno(3.30)
$$
where $u^i$ and $\alpha$ are the gauge transformation parameters. 
Or, equivalently,
$$
   SU(2)_L:  \cases{\delta W_\mu^\pm = \pm i(W_\mu^3u^\pm-W_\mu^\pm u^3)
                     +{1\over g}\partial_\mu u^\pm, \cr
              \delta Z_\mu =c_W[-iW_\mu^- u^+ +iW_\mu^+ u^-
                     +{1\over g}\partial_\mu u^3 ],\cr
              \delta A_\mu =s_W[-iW_\mu^- u^+ +iW_\mu^+ u^-
                     +{1\over g}\partial_\mu u^3 ],\cr}
\eqno(3.31)
$$
where
$$
u^\pm = {{u^1\mp i u^2}\over{\sqrt 2}}.
\eqno(3.32)
$$
The transformation of the gauge  fixing functions
$F^\pm$ etc is expressed as
$$
 \delta\llgm{F^+\cr F^- \cr F^Z\cr F^\gamma}\rrgm 
     = \llgm{ \tilde {\cal M}_1 & \tilde {\cal M}_2 \cr
              \tilde {\cal M}_3 & \tilde {\cal M}_4 }\rrgm
       \llgm{ -u^+/g \cr -u^-/g \cr -u^3/g \cr -\alpha/g^\prime}\rrgm,
       ~~~~~~~~~~
\eqno(3.33)
$$
$$
     \delta F^{g\alpha} =[ -\partial^\mu\partial_\mu\delta_{\alpha\beta}
                        -g_sf^{\alpha\beta\gamma}\partial^\mu g_\mu^\gamma]
                           (-u^\beta/g_s).
$$
Changing the base from $u^3$, $\alpha$ to $u^z$ and $u^\gamma$ by
$$
    \llgm{ u^z/g_Z \cr u^\gamma/e}\rrgm = 
    \llgm{\cos\theta_W\ & -\sin\theta_W \cr \sin\theta_W & \cos\theta_W}\rrgm
    \llgm{u^3/g \cr \alpha/g^\prime}\rrgm,
\eqno(3.34)
$$
so that the kinetic and mass terms become diagonal, and then separating 
the kinetic term and mass term 
one finds the final expression for the interaction lagrangian,
$$
\eqalign{
    {\cal L}_{ghost} =& ~~ \partial_\mu\bar\omega_+\partial_\mu\omega_+
                      -\xi_W M_W^2 \bar\omega_+\omega_+
                   + \partial^\mu\bar\omega_-\partial_\mu\omega_-
                      -\xi_W M_W^2 \bar\omega_-\omega_- \cr
            & + \partial^\mu\bar\omega_z\partial_\mu\omega_z
                      -\xi_Z M_Z^2 \bar\omega_z\omega_z
          ~~ + \partial^\mu\bar\omega_\gamma\partial_\mu\omega_\gamma
          ~~ + \partial^\mu\bar\omega^\alpha\partial_\mu \omega^\alpha\cr
                 & + \left\lgroup 
     \bar\omega_+, \bar\omega_-, \bar\omega_z, \bar\omega_\gamma\right\rgroup
      \llgm{ {\cal M}_{int~1} & {\cal M}_{int~2}\cr
                       {\cal M}_{int~3} & {\cal M}_{int~4}\cr}\rrgm 
                       \llgm{\omega_+ \cr \omega_- \cr 
                             \omega_z \cr \omega_\gamma}\rrgm \cr
            & +g_sf^{\alpha\beta\gamma} 
              \partial^\mu\bar\omega^\alpha \omega^\beta g_\mu^\gamma,
}\eqno(3.35)
$$
with
$$
\eqalign{
      \llgm{ {\cal M}_{int~1} & {\cal M}_{int~2}\cr
                       {\cal M}_{int~3} & {\cal M}_{int~4}\cr}\rrgm 
           =&  \llgm{ \tilde{\cal M}_1 & \tilde {\cal M}_2\cr
                       \tilde{\cal M}_3 & \tilde {\cal M}_4\cr}\rrgm
                \llgm{ 1 & 0 & 0 & 0 \cr
                       0 & 1 & 0 & 0 \cr
                       0 & 0 & c_W & s_W \cr
                       0 & 0 & -s_W & c_W \cr}\rrgm \cr
            &~~ - ({\rm kinetic~ terms})-({\rm mass~ terms}).
}\eqno(3.36)
$$              
Here $\omega$ stands for the ghost particles which are scalar but behave
as if they were fermion.  The explicit form of the matrices 
${\cal M}_{int~1}$ etc is given as follows,
$$
\eqalignno{
   {\cal M}_{int~1}=& -ig (c_W \partial^\mu Z_\mu+s_W\partial^\mu A_\mu)
                     \llgm{1 & 0 \cr 0 & -1\cr}\rrgm \cr
         &-{g\over 2}M_W\xi_W 
             [\cos(\beta-\alpha)H^0+ \sin(\beta-\alpha)h^0]
             \llgm{1 & 0 \cr 0 & 1\cr}\rrgm \cr
         & -i {g\over 2}M_W\xi_W G^0\llgm{1 & 0 \cr 0 & -1 \cr}\rrgm, 
                                                          & (3.37a)\cr
   {\cal M}_{int~2}=&~ig\llgm
            {c_W\partial^\mu W^+_\mu & s_W\partial^\mu W^+_\mu \cr
            -c_W\partial^\mu W^-_\mu & -s_W\partial^\mu W^-_\mu \cr}\rrgm\cr
         & -M_W\xi_W\llgm{{{g_Z}\over 2}(c_W^2-s_W^2)G^+ & eG^+\cr   
                  {{g_Z}\over 2} (c_W^2-s_W^2)G^- & eG^-\cr}\rrgm, &(3.37b)\cr
   {\cal M}_{int~3}=&~ ig\llgm
            {c_W\partial^\mu W^-_\mu & -c_W\partial^\mu W^+_\mu \cr
             s_W\partial^\mu W^-_\mu & -s_W\partial^\mu W^+_\mu \cr}\rrgm\cr
         &+ {g\over 2}M_Z\xi_Z\llgm{G^- & G^+\cr 0 & 0\cr}\rrgm, &(3.37c)\cr 
   {\cal M}_{int~4}=& - {{g_Z}\over 2}M_Z\xi_Z 
             [\cos(\beta-\alpha)H^0+\sin(\beta-\alpha)h^0 ]
             \llgm{1 & 0 \cr 0 & 0\cr}\rrgm. &(3.37d)\cr
}$$

They provide three interactions ${\cal L}_{\bar\omega\omega V}$, 
${\cal L}_{\bar\omega\omega H}$ and ${\cal L}_{\bar\omega\omega G}$
where $G$ stands for the Goldstone bosons.
$$
\eqalign{
   {\cal L}_{\bar\omega\omega V}=
   &~igc_W[\partial^\mu\bar\omega_+\omega_+ -\partial^\mu\bar\omega_-\omega_-]
     Z_\mu \cr
   &+ie[\partial^\mu\bar\omega_+\omega_+ -\partial^\mu\bar\omega_-\omega_-]
     A_\mu \cr
   &+ig[c_W(\partial^\mu\bar\omega_z\omega_- -\partial^\mu\bar\omega_+\omega_z)
      +s_W(\partial^\mu\bar\omega_\gamma\omega_- 
          -\partial^\mu\bar\omega_+\omega_\gamma)]W^+_\mu\cr
   &+ig[c_W(\partial^\mu\bar\omega_-\omega_z -\partial^\mu\bar\omega_z\omega_+)
       +s_W(\partial^\mu\bar\omega_-\omega_\gamma 
          -\partial^\mu\bar\omega_\gamma\omega_+)]W^-_\mu\cr
   &+ g_sf^{\alpha\beta\gamma}
     \partial\bar\omega^\alpha \omega^\beta g_\mu^\gamma. \cr       
}\eqno(3.38)
$$
$$
\eqalign{
  {\cal L}_{\bar\omega\omega H}=
   &-{1\over 2}[gM_W\xi_W(\bar\omega_+\omega_+ + \bar\omega_-\omega_-)
                + g_ZM_Z\xi_Z\bar\omega_z\omega_z]\cr
   &~~  [\cos(\beta-\alpha)H^0+\sin(\beta-\alpha)h^0].\cr
}\eqno(3.39)
$$
$$
\eqalign{
    {\cal L}_{\bar\omega\omega G} =   
   &  -i{g\over 2}M_W\xi_W [\bar\omega_+\omega_+-\bar\omega_-\omega_-]G^0 \cr
   &  -(c_W^2-s_W^2){{g_Z}\over 2}M_W\xi_W[\bar\omega_+\omega_z G^+
                                          +\bar\omega_-\omega_z G^-]\cr
   &-eM_W\xi_W
       [\bar\omega_+\omega_\gamma G^+ + \bar\omega_-\omega_\gamma G^-]\cr
   &   + {g\over 2} M_Z\xi_Z[\bar\omega_z\omega_+G^- 
                           + \bar\omega_z\omega_- G^+].\cr
}\eqno(3.40)
$$\par

\vskip 2 truecm

\noindent{\bf 4. Mass~ matrix}\par
   In this section, I will list the mass matrices of various particles,
which are diagonalized by the orthogonal matrices as discussed in section 2.
\par
   (1)$\underline{\rm Fermion}$  Their  masses come from the interaction 
(3.1i), (3.1j), (3.1k) as explicitly shown in (3.10a).
\par
   (2)$\underline{\rm Gauge~ bosons}$  Their masses come from (3.1g) and (3,1h) 
as shown in (3.23a).  The mass term becomes
$$
\eqalign{
     {\cal L}_m =& {1\over 4} g^2(v_1^2+v_2^2) W_\mu^+W^{-\mu} \cr
                 &+ {1\over 8}(g^2+g^{\prime 2})(v_1^2+v_2^2)
              \left\lgroup W^3_\mu, B_\mu \rrgm
              \llgm {g^2 & -gg^\prime \cr -gg^\prime & g^{\prime 2}\cr}\rrgm
                  \llgm{W^{3\mu} \cr B^\mu}\rrgm,
}\eqno(4.1)
$$
from which, we obtain the mass eigenvalues (2.5) and the mixing angle (2.4b).
Photon and gluons  remain massless.\par
   (3)$\underline{\rm Sfermion}$  The diagonal mass matrix elements come from 
the matter $F$-terms (3.11a) and the $D$-terms given by (3.15) while the 
off-diagonal elements come from the Higgs $F$-terms (3.12b)
as well as from the soft breaking term (3.24).
$$
    \llgm{m^2_{\tilde f_L} & m^2_{\tilde f_{LR}} \cr 
          m^{2*}_{\tilde f_{LR}} & m^2_{\tilde f_R} }\rrgm,
\eqno(4.2)
$$
with $$
\eqalign{
    m^2_{\tilde f_L} =&~ \tilde m^2_{\tilde f_L}+ m^2_f 
                        +M_Z^2\cos 2\beta(T_{3f}-Q_f s^2_W),\cr
    m^2_{\tilde f_R} =&~ \tilde m^2_{\tilde f_R}+ m^2_f 
                        +M_Z^2\cos2\beta Q_fs^2_W,\cr
    m^2_{\tilde f_{LR}}=& \cases{-m_u(\mu \cot\beta+A_u^*), & $f=u$\cr
                                 -m_f(\mu \tan\beta+A_f^*), & $f=d,e$\cr}
}\eqno(4.3)
$$
    (4)$\underline{\rm Charginos}$  They are mixture of gaugino $\lambda^\pm$ 
and higgsino $\tilde H_1^-$ and $\tilde H_2^+$.  Higgsino mass comes from 
(3.1$\ell$), which reads as
$$
\eqalign{
    (3.1\ell)=& -\mu[ \varphi(\tilde H_1^-)\varphi(\tilde H_2^+)     
                    +\bar\varphi(\tilde H_1^-)\bar\varphi(\tilde H_2^+)
                    -\bar\Psi(\tilde H_1^0)\Psi(\tilde H_2^0)]\cr
              & + F(H_i)\hbox{-terms},
}\eqno(4.4)
$$
while gauginos acquire  masses only from the SUSY breaking interaction 
(3.1m).  The gaugino higgsino mixing terms come from (3.22a). 
Collecting them all, we find, (see (2.11))
$$
   {\cal L}_m = -\left\lgroup \varphi(\lambda^+), \varphi(\tilde  H_2^+)\rrgm
                \llgm{M_2 & \sqrt 2 M_W\cos\beta\cr
                      \sqrt 2 M_W \sin\beta & \mu}\rrgm
                \llgm{ \varphi(\lambda^-) \cr \varphi(\tilde H_1^-)}\rrgm
                + h.c.
\eqno(4.5)
$$
The mass eigenvalues are
$$
     m^2_{\tilde \chi^\pm_{1,2}}= {1\over 2}(M_2^2+\mu^2+2M_W^2\mp\sqrt C),
\eqno(4.6)
$$
where
$$
     C = (M_2^2+\mu^2+2M_W^2)^2 -4(M_2\mu -M_W^2\sin 2\beta)^2.
\eqno(4.6a)
$$
\par                                            
    (5)$\underline{\rm Neutralinos}$  They are mixture of neutral gauginos 
$\lambda$, $\lambda^0$ 
and neutral higgsinos $\tilde H_1^0$ and $\tilde H_2^0$.  As in the 
case of charginos, gaugino acquires masses from SUSY breaking lagrangian
while the higgsino mass comes from (4.4).  The gaugino-higgsino mixing
term come from (3.22a).  Assembling them together, we find (see (2.19))
$$
\eqalign{
   {\cal L}_m =& -{1\over 2}
   \llgm{\lambda \cr\lambda^0\cr \tilde H_1^0\cr \tilde H_2^0}\rrgm^t 
     \llgm{M_1 & 0   & -M_Z~s_W\cos\beta & ~M_Z~s_W\sin\beta \cr
           *   & M_2 & ~M_Z~c_W\cos\beta & -M_Z~c_W\sin\beta \cr
           *   & *   & 0                & -\mu               \cr
           *   & *   & *                & 0                 \cr}\rrgm
   \llgm{ \lambda \cr\lambda^0 \cr \tilde H_1^0 \cr \tilde H_2^0 }\rrgm \cr
       & + h.c.
}\eqno(4.7)
$$\par
   (6)$\underline{\rm Gluinos}$  Being a SUSY partner of gluons, gluinos are 
massless in the SUSY limit.  Acquiring  the contribution only from the
SUSY breaking term (3.2),  their mass is $M_3$.\par
   (7)$\underline{\rm Higgs}$  Their  masses come from the quadratic part 
of the Higgs potential, 
which  arises from the $F(H_i)$-terms,(3.12a),  $D$-terms (3.14) and 
the SUSY breaking term (3.1m).  It has the following form,
$$
\eqalign{
     V =&~ m_1^2{\bf H}^*_1{\bf H}_1+ m_2^2{\bf H}^*_2{\bf H}_2 + 
           (m_{12}^2 {\bf H}_1 {\bf H}_2 +h.c.)\cr
        &+ {1\over 8}(g^2 + g^{\prime 2}) 
            ({\bf H}^*_1{\bf H}_1 -{\bf H}^*_2{\bf H}_2)^2
       +{{g^2}\over 2}\vert{\bf H}^*_1{\bf H}_2\vert^2,
}\eqno(4.8)
$$
where
$$
\eqalign{
     m_1^2 =&~ \tilde m_1^2 + \vert\mu\vert^2, \cr 
     m_2^2 =&~ \tilde m_2^2 + \vert\mu\vert^2, \cr 
     m_{12}^2 =&~ \tilde m_{12}^2.
}\eqno(4.9)
$$
and the product of two doublets is defined  in (3.2a).  
Since the minimum of the above potential is realized by 
$$
    {\bf H}_1 = \llgm{{{v_1}\over{\sqrt 2}} \cr 0}\rrgm,~~~~~      
    {\bf H}_2 = \llgm{0 \cr {{v_1}\over{\sqrt 2}}}\rrgm,  
\eqno(4.10)
$$
$v_1$, $v_2$ and $m_1^2$ etc are related at tree level as follows.
$$
\eqalign{
     {{v_1v_2}\over{v_1^2+v_2^2}}
         =&-{{m_{12}^2}\over{m_1^2+m_2^2}} (= {1\over 2} \sin 2\beta), \cr
     v_1^2+v_2^2 =& -{8\over{g^2+g^{\prime 2}}}
            {{m_1^2\cos^2\beta -m_2^2\sin^2\beta}\over{\cos2\beta}}.
}\eqno(4.11)
$$
Upon using (4.11), the Higgs potential (4.8) can be rewritten as
$$
\eqalign{
     V =&~ m_1^2{\bf H}^*_1{\bf H}_1+ m_2^2{\bf H}^*_2{\bf H}_2  
          -(m_1^2+m_2^2)\cos\beta \sin\beta( {\bf H}_1 {\bf H}_2 +h.c.)\cr
        & + {{g^2+g^{\prime 2}}\over 8} 
            ({\bf H}^*_1{\bf H}_1 -{\bf H}^*_2{\bf H}_2)^2
       +{{g^2}\over 2}\vert{\bf H}^*_1{\bf H}_2\vert^2.
}\eqno(4.12)
$$
The straightforward calculation yields the following mass matrix
for neutral Higgs,
$$
\eqalign{
     {\cal L}_m =& 
            -{1\over 2}(m_1^2+m_2^2)\left\lgroup \chi_1^0, \chi_2^0
             \right\rgroup
             \llgm{\sin^2\beta & -\sin\beta \cos\beta \cr
                   -\sin\beta \cos\beta & \cos^2\beta }\rrgm
            \llgm{\chi_1^0 \cr \chi_2^0}\rrgm\cr
            &-{1\over 2}\llgm{\phi_1^0 \cr\phi_2^0}\rrgm^t
            \llgm{ M_A^2\sin^2\beta+M_Z^2\cos^2\beta &
                 -(M_A^2+M_Z^2)\sin\beta\cos\beta \cr
                 -(M_A^2+M_Z^2)\sin\beta\cos\beta &
                   M_A^2 \cos^2\beta+M_Z^2\sin^2\beta}\rrgm
            \llgm{\phi_1^0 \cr \phi_2^0}\rrgm,
}\eqno(4.13)
$$
from which one obtains the masses of three neutral Higgses (see (2.27)),
$$
\eqalign{
     M_A^2 =&~ m_1^2+m_2^2 = -m_{12}^2(\tan\beta + \cot\beta),\cr
     M_{H^0,h^0}^2 =&~ {1\over 2}[M_A^2+M_Z^2 \pm
       \sqrt{(M_A^2+M_Z^2)^2-4M_A^2M_Z^2\cos^22\beta} ],\cr
     M^2_{H^\pm}=&~ M_A^2 + M_W^2.
}\eqno(4.14)
$$
The corresponding mixing angles are given in
(2.25) with (2.26).   For charged Higgs, we have 
$$
     {\cal L}_m = -(M_A^2+M_W^2)\left\lgroup \phi^+_1~ \phi_2^+\right\rgroup
     \llgm{ \sin^2\beta & -\sin\beta \cos\beta \cr
           -\sin\beta \cos\beta & \cos^2\beta }\rrgm
     \llgm{ \phi_1^- \cr \phi_2^-}\rrgm,
\eqno(4.15)
$$
with the eigenvalue $M^2_{H^\pm}= M_A^2+M_W^2$  and the mixing matrix given by
(2.24).  The final form of the Higgs potential looks as follows,
$$
   -V = {{g^2}\over{32c^2}}(v_1^2-v_2^2)^2 + (4.13)+(4.15)+(3.14a)+(3.14c).
\eqno(4.16)
$$\par

\vskip 2truecm

\noindent{\bf 5. Choice of input parameters}\par
    MSSM has four couplings, $g$, $g^\prime$, $g_s$ and $\mu$,  and 
$6+8N_G$ free independent parameters for SUSY breaking, all together 
$10 + 8N_G$  parameters to be fixed.
Among them, 8 parameters for each generation appear only in the
sfermion mass matrix and their mixing. Therefore, it is more convenient 
to choose 5 physical masses and 3 mixing angles of sfermions in each
generation as input parameters. From (2.8) one finds
$$
\eqalign{
    m^2_{\tilde f_L} =&~ \cos^2\theta_f m^2_{\tilde f_1}
                       +\sin^2\theta_f m^2_{\tilde f_2}, \cr
    m^2_{\tilde f_R} =&~ \sin^2\theta_f m^2_{\tilde f_1}
                       +\cos^2\theta_f m^2_{\tilde f_2}, \cr
    m^2_{\tilde f_{LR}} =&~ \cos\theta_f\sin\theta_f 
                         ( m^2_{\tilde f_1}-m^2_{\tilde f_2}).\cr
}\eqno(5.1)
$$
Using (4.2), (4.3) and (5.1), one can easily express the original parameters 
appearing  in the lagrangian,  $\tilde m^2_{\tilde f_L}$, 
$\tilde m^2_{\tilde f_R}$ and $A_f$, in terms of the physical 
input parameters (the physical masses and the mixing angles of sfermions).  
If we  neglect a fermion mass 
($m_f=0$), there is no mixing between $\tilde f_L$ and $\tilde f_R$. 
Consequently, we need one less parameter; 
we can discard the mixing angle $\theta_f$, or equivalently $A_f$ 
in terms of the parameter appearing in the lagrangian.  \par
     In case we neglect the masses of the fermions belonging to the 
first and second  generations, the input parameters in sfermion sector are
the following 20 quantities( 5+5+8=18 independent quantities):
$$
\matrix{
     m^2_{\tilde u_L}=m^2_{\tilde d_L}, & m^2_{\tilde\nu e_L}=m^2_{\tilde e_L}, &
     m^2_{\tilde u_R},& m^2_{\tilde d_R}, & m^2_{\tilde e_R},\cr
     m^2_{\tilde c_L}=m^2_{\tilde s_L},  & 
     m^2_{\tilde\nu \mu_L}=m^2_{\tilde \mu_L}, &
     m^2_{\tilde c_R},& m^2_{\tilde s_R}, & m^2_{\tilde \mu_R},\cr
}$$
$$
\matrix{
     m^2_{\tilde b_1},& m^2_{\tilde b_2}, & \theta_b,  &
     m^2_{\tilde t_1},& m^2_{\tilde t_2}, & \theta_t, \cr 
     m^2_{\tilde \tau_1},& m^2_{\tilde \tau_2}, & \theta_\tau,  &
     m^2_{\tilde \nu\tau_L}, & & \cr
}\eqno(5.2)
$$
with the constraints coming from the $SU(2)_L$ invariance of ${\cal L}_{soft}$,
$\tilde m^2_{{\tilde f}_{L\uparrow}}=\tilde m^2_{{\tilde f}_{L\downarrow}}$,
$$
\eqalign{
   \cos^2\theta_t m^2_{\tilde t_1} + \sin^2\theta_t m^2_{\tilde t_2} 
   -m_t^2=&~ 
   \cos^2\theta_b m^2_{\tilde b_1} + \sin^2\theta_b m^2_{\tilde b_2} 
   -m_b^2+M_W^2 \cos 2\beta,\cr
   m^2_{\tilde \nu\tau}
   =&~ \cos^2\theta_\tau m^2_{\tilde \tau_1} 
     +\sin^2\theta_\tau m^2_{\tilde \tau_2} -m_\tau^2+ M_W^2 \cos2\beta.\cr
}\eqno(5.3)
$$\par
     Concerning the remaining 10 parameters,
$$
  g,~ g^\prime,~ g_s,~ \mu,~ M_1,~ M_2,~ M_3,~ \tilde m_1^2,~ \tilde m_2^2,~
   \tilde m_{12}^2,
\eqno(5.4)
$$
it is customary to use, in place of
$\tilde m_1^2$,  $\tilde m_2^2$ and  $\tilde m_{12}^2$,
the pseudoscalar Higgs mass $M_A^2$ as given 
by (4.14) and two vacuum expectation values $v_1$ and $v_2$  which are
defined by (4.11).  This is because $\tilde m_1^2$,  $\tilde m_2^2$ and  
$\tilde m_{12}^2$ appear only in the Higgs potential (4.8) through (4.9) and
as we see from (4.12), (4.13) and (4.14), Higgs masses, mixing angles and
Higgs potential are more naturally expressed in terms of $M_A^2$
and two vacuum expectation values.  For reference sake, relations 
between the original three parameters and the $M_A^2$, $v_1$ and $v_2$ are 
shown here, 
$$
\eqalign{
     \tilde m^2_1 =&~ M_A^2 \sin^2\beta -{{g^2+g^{\prime 2}}\over 8}
                          (v_1^2-v_2^2)-\mu^2,\cr
     \tilde m^2_2 =&~ M_A^2 \cos^2\beta +{{g^2+g^{\prime 2}}\over 8}
                          (v_1^2-v_2^2)-\mu^2,\cr
     \tilde {m_{12}^2}=&~ -{1\over 2} M_A^2 \sin2\beta.
}\eqno(5.5)
$$
   Therefore, as far as the Higgs sector is concerned, MSSM contains 
only one more parameter than SM, for which we can take $\tan\beta$.\par
     In place of $g$, $g^\prime$ and $v_1^2+v_2^2$, more physical 
parameters can be used, namely, $e$, $M_W$ and $M_Z$, which are related
to the former parameters according to (2.5) and (A.15). 
Consequently, the input parameters  we use are the following 10 quantities, 
$$
    e, ~~g_s,~~M_W, ~~M_Z, ~~\tan\beta, ~~M_A, ~~\mu, ~~M_1, ~~M_2,~~M_3,
\eqno(5.6)
$$
plus 20 sfermion parameters (5.2) under two constraints (5.3).

\vskip 2 truecm

\noindent{\bf Acknowledgment}\par
     I would like to thank Prof. Y. Shimizu for useful discussions
and for careful reading
of the manuscript,  and Dr. T. Ishikawa for pointing out mistypings in 
the equations of the manuscript.  

\vfill\eject

\noindent{\bf Appendix A.  Notations, Conventions and Mathematical formulae}
\par
     In this note the Greek letters, $\Phi$, $\varphi$, $\chi$ and $\Psi$
are used in the following meaning,
$$
\eqalign{
  \Phi: & ~~~~~ \hbox{left-handed~ chiral~ superfield} \cr
  \varphi,~\chi:& ~~~~~ {\rm two~ component~ Weyl~ spinors} \cr
  \Psi:& ~~~~~{\rm four~ component~ spinors}\cr
       & ~~~~~\llgm{\bar\varphi^{\dot a} \cr \chi_b}\rrgm~~~~ 
                 {\rm Dirac~ fermion}\cr
       & ~~~~~~\llgm{\bar\varphi \cr \varphi}\rrgm~~~~
                 {\rm Majorana~ fermion}.\cr
}\eqno(A.1)
$$
We use the convention of gamma matrices,
$$
     \gamma^\mu = \llgm{ 0 & \bar\sigma^\mu \cr \sigma^\mu & 0}\rrgm,
\eqno(A.2)
$$
where
$$
     \sigma^\mu = ( 1, \vec \sigma),~~~~~
 \bar\sigma^\mu = ( 1, -\vec \sigma).
\eqno(A.3)
$$
Note the following relation:
$$
     \chi\sigma^\mu\bar\varphi = -\bar\varphi\bar\sigma^\mu \chi.
\eqno(A.4)
$$
The $\gamma_5$ is then given by
$$
     \gamma_5 = i\gamma^0\gamma^1\gamma^2\gamma^3 =
             \llgm{1 & 0 \cr 0 & -1}\rrgm,
\eqno(A.5)
$$
so that the right-handed and left-handed projections are given as
follows;
$$
     R = \llgm{1 & 0 \cr 0 & 0}\rrgm,~~~~~
     L = \llgm{0 & 0 \cr 0 & 1}\rrgm,~~~~~
     \gamma^\mu R = \llgm{0 & 0 \cr \sigma^\mu  & 0}\rrgm,~~~~~
     \gamma^\mu L = \llgm{0 & \bar\sigma^\mu \cr 0 &0}\rrgm.
\eqno(A.6)
$$
In this convention of  the $\gamma$ matrices, the charge conjugation
matrix $C$, which must fulfill the following conditions (see, e.g. 
Appendix A of ref.[8].),
$$
     C^T = -C,~~~~C^\dagger = C^{-1} ({\rm unitarity}),~~~~ 
     C\gamma_\mu^T C^{-1} = -\gamma_\mu,
\eqno(A.7)
$$
is expressed as
$$
    C=- i\gamma^2\gamma^0 = 
    - \llgm{i(\bar\sigma^2\sigma^0)^{\dot\alpha}~_{\dot\beta} & 0 \cr
             0 & i(\sigma^2\bar\sigma^0)_\alpha~^\beta \cr}\rrgm
                         = \llgm{ 0 & 1 & 0 & 0 \cr
                                 -1 &  0 & 0 & 0 \cr
                                  0 &  0 & 0 & -1 \cr
                                  0 &  0 & 1 & 0 \cr}\rrgm.
\eqno(A.8)
$$
Therefore, for a given spinor (with upper indices for anti-Weyl spinors
and down indices for Weyl spinors),
$$
    \Psi = \llgm{ \bar\varphi \cr \chi}\rrgm
         = \llgm{ \bar\varphi^{\dot 1} \cr \bar\varphi^{\dot 2} \cr 
                  \chi_1 \cr \chi_2 }\rrgm,
\eqno(A.9)
$$
one finds
$$
\eqalign{
     \bar\Psi =& \llgm{\bar\chi, \varphi}\rrgm
              =\llgm{\bar\chi_{\dot 1}, \bar\chi_{\dot 2}, 
                     \varphi^1, \varphi^2}\rrgm,   \cr
     \Psi^c   =& C(\bar\Psi)^T  
      =C \llgm{ \bar\chi_{\dot 1} \cr \bar\chi_{\dot 2} \cr
                \varphi^1\cr\varphi^2}\rrgm
      =C \llgm{- \bar\chi^{\dot 2} \cr \bar\chi^{\dot 1} \cr \varphi_2\cr
                   - \varphi_1}\rrgm
      =\llgm{\bar \chi \cr \varphi}\rrgm, \cr
     \overline{\Psi^c} =& -\Psi^T C^{-1} = (\bar\varphi,\chi),
}\eqno(A.10)
$$
where the following properties 
$$
   \bar\chi^{\dot\alpha} =
           (\chi^\alpha)^*=\epsilon^{\dot\alpha\dot\beta}\bar\chi_{\dot\beta},
           ~~~~~
   \varphi_\alpha =
           (\bar\varphi_{\dot\alpha})^*=\epsilon_{\alpha\beta}
           \varphi^\beta,
\eqno(A.11)
$$
with
$$
     \epsilon^{\alpha\beta} = \epsilon^{\dot\alpha\dot\beta}=
     \llgm{ 0 & 1 \cr -1 & 0 }\rrgm,~~~~~
     \epsilon_{\alpha\beta} = \epsilon_{\dot\alpha\dot\beta} = 
     \llgm{ 0 & -1 \cr 1 & 0 }\rrgm,~~~~~
     \epsilon_{\alpha\beta}\epsilon^{\beta\gamma} =
                              \delta_\alpha^\gamma,
\eqno(A.12)
$$
are used.\par
     The bilinear of the four-component spinors are written in terms 
of two-component Weyl spinors as,
$$
\eqalign{
   \bar\Psi_2 \Psi_1 =& \bar\chi_2\bar\varphi_1 + \varphi_2\chi_1, \cr
   \bar\Psi_2 \gamma^\mu \Psi_1 = & \bar\chi_2 \bar\sigma^\mu\chi_1
                               + \varphi_2\sigma^\mu\bar\varphi_1 
                             =  \bar\chi_2 \bar\sigma^\mu\chi_1
                               - \bar\varphi_1 \bar\sigma^\mu\varphi_2,\cr 
   \bar\Psi_2 \gamma_5 \Psi_1 = & \bar\chi_2 \bar\varphi_1-\varphi_2\chi_1,\cr
   \bar \Psi_2 R \Psi_1 =& \bar\chi_2\bar\varphi_1,\cr
   \bar \Psi_2 L \Psi_1 =& \varphi_2\chi_1,\cr
   \bar\Psi_2 \gamma_\mu R \Psi_1 = & \varphi_2\sigma^\mu\bar\varphi_1
                                  = -\bar\varphi_1\bar\sigma^\mu\varphi_2, \cr
   \bar\Psi_2 \gamma_\mu L \Psi_1 = & \bar\chi_2 \bar\sigma^\mu\chi_1.\cr
}\eqno(A.13)
$$
In particular,
$$
\eqalign{
     \bar\Psi \Psi =& \bar\varphi\bar\chi + \varphi\chi,~~~~~~~~~~~~~~~~~~~~
                                         ({\rm mass~ term}) \cr
     \bar\Psi\slash\partial \Psi =& \chi\sigma^\mu\partial_\mu\bar\chi                                             +\varphi\sigma^\mu\partial_\mu\bar\varphi,
     ~~~~~~({\rm kinetic~ term})\cr
}\eqno(A.14)
$$\par
     The gauge mixing angle $\theta_W$ and gauge couplings are related as 
follows, which is the same as the SM,
$$
\eqalign{
    c_W\equiv \cos\theta_W =& {g\over{\sqrt{g^2+g^{\prime 2}}}},~~~~
    s_W\equiv \sin\theta_W = {g^\prime\over{\sqrt{g^2+g^{\prime 2}}} },\cr
     e =& g~s_W =g^\prime c_W = {{g g^\prime}\over
                   {\sqrt{g^2+g^{\prime 2}}}}.
}\eqno(A.15)
$$
   I use the following convention for the covariant derivative
$$
    {\cal D}_\mu = \partial_\mu +ig {{\tau^a}\over 2} W^a_\mu 
                 + ig^\prime{Y\over 2}B_\mu+ ig_s{{\lambda^\alpha}\over 2} 
                 g^\alpha_\mu,
\eqno(A.16)
$$
where $\lambda^\alpha$ stands for the $SU(3)$ Gell-Mann matrix.
Note that the signs of the coupling constants or gauge fields are
opposite to the Kyoto convention [9] adopted in the SM part of {\tt GRACE}.  

\vfill\eject

\noindent{\bf Appendix B. $\theta$ integrals, $F$-terms and $D$-terms}\par
    In this Appendix, I show the general result of the $\theta$ 
integrals of each superfield interaction given in (3.1).
First we note that the left-handed chiral superfield is given as,
$$
\eqalign{
       y =&~ x -i\theta\sigma\bar\theta, \cr
       z =&~ x +i\theta\sigma\bar\theta, \cr
     \Phi =&~ A(y) + \sqrt 2 \theta\varphi(y) + \theta\theta F(y) \cr
          =&~ A(x) + \sqrt 2\theta\varphi(x) + \theta\theta F(x) 
             - i\theta\sigma^\mu\bar\theta\partial_\mu A(x) \cr
           & + {i\over{\sqrt 2}}\theta\theta \partial_\mu 
               \varphi(x) \sigma^\mu \bar \theta
             -{1\over 4}\theta\theta\bar\theta\bar\theta
             \partial_\mu\partial^\mu A(x),\cr
 \Phi^\dagger =&~ A^*(z) + \sqrt 2 \bar\theta\bar\varphi(z) + 
               \bar\theta\bar\theta F^*(z) \cr
          =&~ A^*(x) + \sqrt 2\bar\theta\bar\varphi(x) 
              + \bar\theta\bar\theta F^*(x) 
             + i\theta\sigma^\mu\bar\theta\partial_\mu A^*(x) \cr
           & - {i\over{\sqrt 2}}\bar\theta\bar\theta \theta 
               \sigma^\mu \partial_\mu \bar\varphi(x)
             -{1\over 4}\theta\theta\bar\theta\bar\theta
             \partial_\mu\partial^\mu A^*(x),\cr
}\eqno(B.1)
$$
and the gauge superfield is decomposed in the Wess-Zumino gauge as,
$$
V(x,\theta,\bar\theta) = \theta\sigma^\mu\bar\theta V_\mu 
               + \theta\theta\bar\theta\bar\lambda  
               + \bar\theta\bar\theta\theta\lambda 
               +{1\over 2} \theta\theta\bar\theta\bar\theta D.
\eqno(B.2)
$$
Here $F(x)$ and $D(x)$ are spurious fields, called auxiliary fields, and
they are eliminated upon using the Euler's equation of motion.\par
    The kinetic part of the gauge superfields  becomes
$$
\eqalign{
    (3.1a)=
     &~{1\over 2} D^aD^a + i\lambda^a\sigma^\mu {\cal D}_\mu \bar\lambda^a
      -{1\over 4}  V_{\mu\nu}^a V^{a\mu\nu}\cr
    +&{1\over 2} D^2 + i\lambda\sigma^\mu \partial_\mu \bar\lambda
      -{1\over 4}  V_{\mu\nu} V^{\mu\nu}\cr
    +&{1\over 2} D^\alpha D^\alpha + i\tilde g^\alpha\sigma^\mu{\cal D}_\mu 
                                     \bar{\tilde g^\alpha}
      -{1\over 4} g^\alpha_{\mu\nu} g^{\alpha\mu\nu},\cr
}\eqno(B.3)
$$
where
$$
\eqalign{
     {\cal D}_\mu \lambda^a =&~ \partial_\mu\lambda^a
              -g\epsilon^{abc}V_\mu^b\lambda^c,\cr
      V_{\mu\nu}^a =&~ \partial_\mu V_\nu^a -\partial_\nu V_\mu^a 
              -g\epsilon^{abc}V_\mu^b V_\nu^c,\cr
     {\cal D}_\mu\tilde g^\alpha =&~ \partial_\mu\tilde g^\alpha 
     -g_sf^{\alpha\beta\gamma}g_\mu^\beta \tilde g^\gamma,\cr
     g^\alpha_{\mu\nu} = &~ \partial_\mu g_\nu^\alpha 
                         - \partial_\nu g_\mu^\alpha 
                         -g_sf^{\alpha\beta\gamma}g_\mu^\beta g_\nu^\gamma,\cr
}\eqno(B.4)
$$
with $\epsilon^{123}=1$.\par
    For the doublets,
$$
\eqalign{
     \int d^2\theta d^2\bar\theta
     {\bf \Phi}^\dagger& \exp[2(g{{\tau^a}\over 2}V^a +{{g^\prime}\over 2}Y V
        + g_s{{\lambda^\alpha}\over 2}V_s^\alpha)] {\bf\Phi} \cr
     =&~ \vert {\bf F} \vert^2 + \vert \partial_\mu {\bf A}\vert ^2 
        + i{\bf \varphi} \sigma^\mu \partial _\mu \bar{\bf \varphi} \cr
      & +{g \over 2}\lbrack
        -(\bar{\bf\varphi}\bar\sigma^\mu\tau^a{\bf\varphi} 
          +i{\bf A}^*\tau^a\partial^\mu {\bf A}       )W^a_\mu 
     -\sqrt 2 (  \bar{\bf\varphi}\tau^a{\bf A}\bar\lambda^a 
                  + {\bf A}^*\tau^a{\bf\varphi}\lambda^a)\cr
     & ~~~~~~+{\bf A}^*\tau^a{\bf A} D^a \rbrack \cr
     &+ {{g^\prime}\over 2}Y\lbrack
        -(   \bar{\bf\varphi}\bar\sigma^\mu{\bf\varphi} 
            +i{\bf A}^*\partial^\mu{\bf A}         )B_\mu 
       -\sqrt 2  ( \bar{\bf\varphi}{\bf A}\bar\lambda 
                    +{\bf A}^*{\bf\varphi}\lambda) \cr
     &~~~~~~ + {\bf A}^*{\bf A} D\rbrack \cr
     &+{{g_s}\over 2}\lbrack-(\bar\varphi\bar\sigma^\mu\lambda^\alpha\varphi
                + i{\bf A}^*\lambda^\alpha\partial^\mu {\bf A})g_\mu^\alpha
     -\sqrt 2(\bar\varphi\lambda^\alpha {\bf A}\overline{\tilde g^\alpha}
              + {\bf A}^*\lambda^\alpha\varphi\tilde g^\alpha)\cr
     &~~~~~~ + {\bf A}^*\lambda^\alpha{\bf A} D^\alpha\rbrack\cr
     &+{1\over 4} [(g^2W_\mu^aW^{a\mu} + Y^2g^{\prime 2}B^\mu B_\mu)
                   {\bf A}^*{\bf A} 
                  +2Ygg^\prime W^a_\mu B^\mu({\bf A}^*\tau^a{\bf A})]\cr
     &+{1\over 4}\lbrack g_s^2g_\mu^\alpha g^{\mu\beta}
        {\bf A}^*(\lambda^\alpha\lambda^\beta){\bf A}
      + 2Yg_sg^\prime g_\mu^\alpha B^\mu({\bf A}^*\lambda^\alpha{\bf A})\cr
     &~~~~~+ 2 g_sg g_\mu^\alpha W^{a\mu}{\bf A}^*(\tau^a\lambda^\alpha){\bf A}
     \rbrack.
}\eqno(B.5)
$$
$$
\eqalign{
      \int d^2\theta d^2\bar\theta\Phi^\dagger& \exp[g^\prime Y V
             + \lambda^\alpha V_s^\alpha]\Phi \cr
     =&~ \vert F \vert^2 + \vert \partial_\mu A \vert ^2 
        + i{\bf \varphi} \sigma^\mu \partial _\mu \bar{\bf \varphi} \cr
     &+ {{g^\prime}\over 2}Y\lbrack
        -(   \bar\varphi\bar\sigma^\mu\varphi 
            +i A^*\partial^\mu A         )B_\mu 
       -\sqrt 2  ( \bar\varphi A\bar\lambda 
                    + A^*\varphi\lambda) +  A^* A D\rbrack \cr
     &+{{g_s}\over 2}\lbrack-(\bar\varphi\bar\sigma^\mu\lambda^\alpha\varphi
                + i A^*\lambda^\alpha\partial^\mu A)g_\mu^\alpha
     -\sqrt 2(\bar\varphi\lambda^\alpha A\overline{\tilde g^\alpha}
              + A^*\lambda^\alpha\varphi\tilde g^\alpha)\cr
     &~~~~~~ +  A^*\lambda^\alpha A D^\alpha\rbrack\cr
     &+{1\over 4} [ Y^2g^{\prime 2}B^\mu B_\mu (A^* A)] \cr
     &+{1\over 4}\lbrack g_s^2g_\mu^\alpha g^{\mu\beta}
         A^*(\lambda^\alpha\lambda^\beta) A
      + 2Yg_sg^\prime g_\mu^\alpha B^\mu( A^*\lambda^\alpha A)\rbrack.
}\eqno(B.6)
$$
Note that in (B.5), (B.6) as well as in (B.14), (B.15) below,
$\lambda^\alpha$ with a Greek index $\alpha$ is the $SU(3)$
Gell-Mann matrix, while $\lambda$ and $\lambda^a$ with a Roman 
index $a$ are the component fields of the gauge superfields.
In (B.5) and (B.6) color indices are not explicitly shown.  For example,
in (B.5)
$$
     {\bf A}^*(\tau^a\lambda^\alpha){\bf A} \equiv
     \sum_{i,j=1,2}\sum_{\rho,\sigma=1,3} A^*_{i,\rho}(\tau^a)_{ij}
                  (\lambda^\alpha)_{\rho\sigma} A_{j,\sigma}.
$$
In deriving the above expression, we  have used the following formula,
$$
  (\bar\theta\bar\varphi)(\theta\sigma^\mu\bar\theta)(\theta\chi) = 
   -{1\over 4}\theta\theta\bar\theta\bar\theta(\bar\varphi\bar\sigma^\mu\chi).
\eqno(B.7)
$$
The product of two chiral superfields becomes
$$
  \int d^2\theta {\bf\Phi}_1(y) {\bf\Phi}_2(y)  =
   {\bf F}_1(y){\bf A}_2(y) + {\bf A}_1(y){\bf F}_2(y)  - 
   {\bf \varphi}_1(y){\bf\varphi}_2(y),
\eqno(B.8)
$$
while the product of three chiral superfields becomes
$$
\eqalign{
\int d^2\theta {\bf\Phi}_1(y) {\bf\Phi}_2(y) \phi_3(y) =&
  ~{\bf F}_1(y){\bf A}_2(y) A_3(y) + {\bf A}_1(y){\bf F}_2(y) A_3(y) + 
   {\bf A}_1(y){\bf A}_2(y)F_3(y)\cr
  & - {\bf A}_1(y){\bf\varphi}_2(y)\varphi_3(y) 
    - {\bf\varphi}_1(y){\bf A}_2(y)\varphi_3(y) 
    - {\bf\varphi}_1(y){\bf\varphi}_2(y)A_3(y).
}\eqno(B.9)
$$\par
     The auxiliary fields $F_i$ have a simple structure.  They appear
in the lagrangian always in the combination
$$
     {\cal L}\sim \vert F_i \vert^2 + F_i f_i(...) + F_i^*f_i^*(...),
\eqno(B.10)
$$
where $f_i$ is a function of scalar fields and is obtained from the
superpotential $W$ by differentiation,
$$
      f_i(...) = {{\partial W}\over {\partial A_i}}.
\eqno(B.11)
$$
The superpotential, $W$, is  obtained from the Higgs-matter Yukawa 
coupling part of the lagrangian (from (3,1i) to (3.1k)) and the Higgs 
self-interaction (3.1$\ell$) by replacing the superfields,$\Phi_i$, by the 
corresponding scalar fields, $A_i$.  Explicit expression of $F_i$ 
is given as follows,
$$
\eqalign{
   F(u_L)^*=&-m_u(1+{{\phi_2^0+i\chi_2^0}\over {v_2}})\tilde u^*_R
             -{{\sqrt 2 m_d}\over {v_1}}\phi_1^-\tilde d^*_R, \cr
   F(d_L)^*=&~{{\sqrt 2 m_u}\over {v_2}}\phi_2^+\tilde u^*_R
             -m_d(1+{{\phi_1^0-i\chi_1^0}\over {v_1}})\tilde d^*_R, \cr
   F(u_R)^*=&{{\sqrt 2 m_u}\over {v_2}}\phi^+_2\tilde d_L
            - m_u(1+{{\phi_2^0+i\chi_2^0}\over {v_2}})\tilde u_L,\cr
   F(d_R)^*=& -m_d(1+{{\phi_1^0-i\chi_1^0}\over {v_1}})\tilde d_L, 
            -{{\sqrt 2 m_d}\over {v_1}}\phi_1^-\tilde u_L, \cr
   F(\nu_L)^*=&-{{\sqrt 2 m_e}\over {v_1}}\phi_1^-\tilde e^*_R,\cr        
   F(e_L)^*=&-m_e(1+{{\phi_1^0-i\chi_1^0}\over {v_1}})\tilde e^*_R, \cr
   F(e_R)^*=& -m_e(1+{{\phi_1^0-i\chi_1^0}\over {v_1}})\tilde e_L 
            -{{\sqrt 2 m_e}\over {v_1}}\phi_1^-\tilde \nu_L, ~~~~~~\cr
}\eqno(B.12a)
$$
$$
\eqalign{
     F(H_1^0)^*=& -{{\sqrt 2 m_d}\over {v_1}}\tilde d^*_R\tilde d_L
                -{{\sqrt 2 m_e}\over {v_1}}\tilde e^*_R\tilde e_L
                +\mu {{v_2+\phi_2^0+i\chi_2^0}\over {\sqrt 2}}, \cr
     F(H_1^-)^*=& +{{\sqrt 2 m_d}\over {v_1}}\tilde d^*_R\tilde u_L
                +{{\sqrt 2 m_e}\over {v_1}}\tilde e^*_R\tilde \nu_L
                -\mu \phi_2^+, \cr
     F(H_2^+)^*=& ~{{\sqrt 2 m_u}\over {v_2}}\tilde u^*_R\tilde d_L
                ~~~~~~~~~~~~~~~~~~~~~+\mu \phi_1^-, \cr
     F(H_2^0)^*=& -{{\sqrt 2 m_u}\over {v_2}}\tilde u^*_R\tilde u_L
                ~~~~~~~~~~~~~~~~~~
                +\mu {{v_1+\phi_1^0-i\chi_1^0}\over {\sqrt 2}}. \cr
}\eqno(B.12b)
$$
Upon eliminating $F_i$, the part of the lagrangian that contains 
$F_i$ is given by
$$
  (B.10) = -\sum_i \vert F_i\vert^2 
         = -\bigg | {{\partial W}\over {\partial A_i}}\bigg |^2.
\eqno(B.13)
$$
From the $F$-terms of the matter fields, the masses of the sfermion as well as
$\tilde f \tilde f H$  and $\tilde f \tilde f H H$ interactions are
created, while from the $F$-terms of the Higgs fields, one obtains,
above all,  charged Higgs mass, neutral Higgs mass as well as $\tilde f_L 
\tilde f_R$ mixing terms.
\par
     The auxiliary fields $D$ contained in gauge superfields  appear in the
lagrangian as
$$
\eqalign{
     &{1\over 2} D^aD^a + {1\over 2} D^2 + {1\over 2} D^\alpha D^\alpha\cr
    +&{g\over 2}[\sum_{\rm doublet}{\bf A}^*\tau^a {\bf A}] D^a
    + {{g^\prime} \over 2} [\sum_{\rm doublet} Y {\bf A}^*{\bf A} 
                           +\sum_{\rm singlet} Y A^*A]D \cr
    +&{{g_s}\over 2}[\sum_{\rm doublet}\tilde q_L^*\lambda^\alpha \tilde q_L
         -\sum_{\rm singlet}\tilde q_R^*\lambda^\alpha\tilde q_R]D^\alpha, 
}\eqno(B.14)
$$
where $Y$ is the hypercharge of the multiplet.  From the equation of 
motion for $D^a$ and $D$, one finds,
$$
\eqalign{
     D =& - {{g^\prime} \over 2} [\sum_{\rm doublet} Y {\bf A}^*{\bf A} 
                           +\sum_{\rm singlet} Y A^*A],\cr
     D^a=&- {g\over 2}[\sum_{\rm doublet}{\bf A}^*\tau^a {\bf A}],\cr
     D^\alpha=&-{{g_s}\over 2}
             [\sum_{\rm doublet}\tilde q_L^*\lambda^\alpha \tilde q_L
             -\sum_{\rm singlet}\tilde q_R^*\lambda^\alpha\tilde q_R], 
}\eqno(B.15)
$$
and (B.14) becomes simply
$$
  (B.14) = -{1\over 2}D^a D^a -{1\over 2} D^2 -{1\over 2} D^\alpha D^\alpha.
\eqno(B.16)
$$
Using the property of generators of $SU(n)$,
$$
   \sum_{\alpha=1}^{n^2-1} \lambda^\alpha_{ab}\lambda^\alpha_{cd}
    = 2\delta_{ad}\delta_{bc}-{2\over n}\delta_{ab}\delta_{cd},
\eqno(B.17)
$$
where  $\lambda^\alpha = 2\times({\rm generator})$, namely 
$\lambda^\alpha=\tau^\alpha$ for $SU(2)$ and $\lambda^\alpha$ is the 
Gell-Mann matrix for $SU(3)$, one finds
$$
\eqalign{
     (B.14)=& -{{g^2}\over 8} \sum_a
            \vert \sum_{\rm doublet}{\bf A}^*\tau^a{\bf A}\vert^2 \cr
             &-{{g^{\prime 2}}\over 8}
             [\sum_{\rm doublet} Y {\bf A}^*{\bf A} +\sum_{\rm singlet} Y A^*A]^2\cr
   & -{{g_s^2}\over 8}\{
   ~ {4\over 3}\sum_{q,i}\vert \tilde q_i^* \tilde q_i \vert^2 
   +4\sum_{q < q^\prime,i}  \vert \tilde q_i^* \tilde q^\prime_i \vert^2 
    -{4\over 3}\sum_{q < q^\prime,i}(\tilde q_i^*\tilde q_i)
                           (\tilde q^{\prime *}_i \tilde q^\prime_i) \}\cr
    &+{{g_s^2}\over 4}\{                           
   ~ 2 \sum_q\vert \tilde q_L^* \tilde q_R \vert^2 
   +4\sum_{q < q^\prime}  \vert \tilde q_L^* \tilde q^\prime_R \vert^2 
    -{2\over 3}\sum_{q, q^\prime}(\tilde q_L^*\tilde q_L)
                           (\tilde q^{\prime *}_R \tilde q^\prime_R) \},\cr
}\eqno(B.18)
$$
where in the third and fourth lines the sums are taken over 
$q,q^\prime=u,d,\cdots$ and $i=R,L$.  Extracting from the first two lines
the terms which are quartic in Higgs fields ${\bf A} = {\bf H}_1$ 
and ${\bf A}={\bf H}_2$, one obtains 
the SUSY part of the Higgs potential given in (4.8).

\vfill\eject

\noindent{\bf Appendix C. Explicit expression of the MSSM lagrangian in 
terms of mass eigenstates}\par

   The lagrangian derived in section 3 is not explicitly expressed on
the basis of the mass eigenstates; they are still written in terms  of two
component Weyl spinors (in the case of charginos, neutralinos and gluinos)
or in terms of unmixed states (in the case of sfermions).  
In this Appendix, the MSSM lagrangian is given explicitly in the
mass eigenstates, so that we can easily read out the Feynman rule of MSSM
from the lagrangian. For sfermions (except for (C.2)-(C.5)), in order to 
save the space I use the base $\tilde f_L$ and $\tilde f_R$ 
instead of $\tilde f_1$ and $\tilde f_2$.\par
    Since the lagrangian consists of a huge number of terms, for the 
reference sake  I first list in Table 2  the type of interactions and the 
corresponding equation numbers.  Also listed in Table 2 are the equation
numbers  where the interactions are explained.\par
     
\medskip
   
\topinsert
{\offinterlineskip \tabskip=0pt
\halign{ \strut
  \vrule#& \quad # \quad &
  \vrule#& \quad # \quad &
  \vrule#& \quad # \quad &
  \vrule#& \quad # \quad &
  \vrule#  \cr
\noalign{\hrule}
  &              && type         && equation && explanation &\cr
\noalign{\hrule}
  & Gauge~coupl. && $Vff$                   &&(C.1)         && (3.4) &\cr
  & (3-point)~~~~~~~&& $V\tilde f \tilde f$    &&(C.2)-(C.5)   && (3.6) &\cr 
  &              && $V\tilde\chi\tilde\chi$ &&(C.6)-(C.10) &&(3.20)(3.25c)&\cr
  &              && $VHH$                   &&(C.11)        && (3.19) &\cr
  &              && $VGH$                   &&(C.12)        && (3.19) &\cr
  &              && $VGG$                   &&(C.13)        && (3.19) &\cr  
\noalign{\hrule}
  & Higgs~coupl. && $Hff$                   &&(C.14)(C.15)  && (3.10a) &\cr
  & (3-point)~~~~~~~&& $HVV$                   &&(C.16)        && (3.23b) &\cr
  
  &              && $H\tilde f\tilde f$     &&(C.17)-(C.20) 
                                            && (3.11b)(3.12c)(3.16)(3.24) &\cr
  &              && $H\tilde\chi\tilde\chi$ &&(C.21)-(C.23) && (3.22b) &\cr
\noalign{\hrule}
  & Goldstone    && $Gff$                   &&(C.24)        && (3.10a) &\cr
  & (3-point)~~~~~~~&& $GVV$                   &&(C.25)        && (3.23b) &\cr
  &              && $G\tilde f\tilde f$     &&(C.26)        
                                            && (3.11b)(3.12c)(3.16)(3.24)&\cr
  &              && $G\tilde\chi\tilde\chi$ &&(C.27)-(C.29)  && (3.22b) &\cr
\noalign{\hrule}
  & other~3-point && $\tilde f f\tilde\chi$  &&(C.30)-(C.32) &&(3.7)(3.10b)&\cr
\noalign{\hrule}
  & 4-point~coupling&&$VV\tilde f\tilde f$  &&(C.33)-(C.37)  &&(3.9)   &\cr
  &              && $HHVV$                 &&(C.38)         &&(3.23c) &\cr
  &              && $HGVV$                 &&(C.39)         &&(3.23c) &\cr
  &              && $GGVV$                 &&(C.40)         &&(3.23c) &\cr
  &              && $\tilde f\tilde f HH$  &&(C.41)-(C.46)  &&(3.11c)(3.17)&\cr
  &              && $\tilde f\tilde f GH$  &&(C.47)-(C.50)  &&(3.11c)(3.17)&\cr
  &              && $\tilde f\tilde f GG$  &&(C.51)-(C.53)  &&(3.11c)(3.17)&\cr
  &              && $\tilde f\tilde f\tilde f\tilde f$&&(C.54)-(C.58)
                                           && (3.12d)(3.18)&\cr
\noalign{\hrule}
  & Gauge~  && $VVV$   && (C.59)      && 1st~line~of~(3.25) &\cr
  & selfinteraction && $VVVV$  && (C.60)      && 1st~line~of~(3.25) &\cr
\noalign{\hrule}
  & Higgs~  && $HHH$   && (C.61)-(C.62)     &&  (3.14a)     &\cr
  & selfinteraction  && $HHHH$  && (C.63)-(C.65)     &&  (3.14c)     & \cr
\noalign{\hrule}
  & Goldstone-Higgs   && $HHG$   && (C.66)       &&  (3.14a)    &\cr
  & interaction   && $HGG$   && (C.67)       &&  (3.14a) &\cr
  &              && $HHHG$  && (C.68)       &&  (3.14c) &\cr
  &              && $HHGG$  && (C.69)       &&  (3.14c) &\cr
  &              && $HGGG$  && (C.70)       &&  (3.14c) &\cr
  &              && $GGGG$  && (C.71)       &&  (3.14c) &\cr
\noalign{\hrule}
  & ghost        && $\bar\omega\omega V$ && (C.72)   && (3.38) &\cr
  &              && $\bar\omega\omega H$ && (C.73)   && (3.39) &\cr
  &              && $\bar\omega\omega G$ && (C.74)   && (3.40) &\cr
\noalign{\hrule}   }}
\medskip
\noindent{Table 2. Type of interactions in MSSM.}\par
\endinsert
\medskip

\noindent{\bf Gauge-boson-Fermion-Fermion}\par
$$
\eqalign{
  {\cal L} =& -{g\over{\sqrt 2}}\sum_{(f_\uparrow, f_\downarrow)}
    [\bar f_\uparrow \gamma^\mu {{1-\gamma_5}\over 2} f_\downarrow W_\mu^+
   + \bar f_\downarrow \gamma^\mu {{1-\gamma_5}\over 2} f_\uparrow W_\mu^-]\cr
      & -g_Z\sum_f \bar f \gamma^\mu[(T_{3f}-s_W^2Q_f)
    {{1-\gamma_5}\over 2}-s_W^2Q_f {{1+\gamma_5}\over 2}]f Z_\mu \cr
      & -e\sum_f Q_f \bar f \gamma^\mu f A_\mu\cr
      & - g_s \sum_q \bar q \gamma^\mu {{\lambda^\alpha}\over 2} q 
                g_\mu^\alpha.
}\eqno(C.1)
$$
\noindent{\bf W-Sfermion-Sfermion}\par
$$
 \eqalign{
   {\cal L} =& -i{g\over{\sqrt 2}}\sum_{(f_\uparrow,f_\downarrow)}\lbrack
      c_{f_\uparrow}c_{f_\downarrow}
         (\tilde f^*_{\uparrow 1}\partrl^\mu \tilde f_{\downarrow 1})
     -c_{f_\uparrow}s_{f_\downarrow}
         (\tilde f^*_{\uparrow 1}\partrl^\mu \tilde f_{\downarrow 2}) \cr
   & ~~~~~~~~-s_{f_\uparrow}c_{f_\downarrow}
         (\tilde f^*_{\uparrow 2}\partrl^\mu \tilde f_{\downarrow 1})
     +s_{f_\uparrow}s_{f_\downarrow}
         (\tilde f^*_{\uparrow 2}\partrl^\mu \tilde f_{\downarrow 2}) \rbrack
           W_\mu^+  + h.c., \cr
}\eqno(C.2)
$$
where $f_\uparrow$ and $f_\downarrow$ stand for the up- and 
down-components of the fermion doublet, and 
$c_{f_\uparrow}\equiv \cos \theta_{f_\uparrow}$ and 
$s_{f_\uparrow}\equiv \sin\theta_{f_\uparrow}$ etc.\par
\noindent{\bf Z-Sfermion-Sfermion}\par
$$
\eqalign{
     {\cal L} =& -ig_Z\sum_f\lbrace
        [T_{3f}c_f^2 -s_W^2 Q_f ]
         (\tilde f^*_1\partrl^\mu \tilde f_1)\cr
      &~~~~~~~~~~~~~ - T_{3f}c_f s_f
         [(\tilde f^*_1\partrl^\mu \tilde f_2)
         +(\tilde f^*_2\partrl^\mu \tilde f_1)]\cr
      &~~~~~~~~~~~~~ +  [T_{3f}s_f^2 -s_W^2 Q_f ]
         (\tilde f^*_2\partrl^\mu \tilde f_2)
         \rbrace Z_\mu.\cr
}\eqno(C.3)$$
\noindent{\bf $\gamma$-Sfermion-Sfermion}\par
$$
    {\cal L} = -ie\sum_f Q_f[ (\tilde f^*_1\partrl^\mu \tilde f_1)
               +(\tilde f^*_2\partrl^\mu \tilde f_2)] A_\mu.
\eqno(C.4)
$$
\noindent{\bf Gluon-Sfermion-Sfermion}\par
$$
    {\cal L} = -ig_s\sum_q [ (\tilde q^*_1{{\lambda^\alpha}\over 2}
                             \partrl^\mu \tilde q_1)
               +(\tilde q^*_2 {{\lambda^\alpha}\over 2}
                             \partrl^\mu \tilde q_2)] g_\mu^\alpha.
\eqno(C.5)
$$
\noindent{\bf W-Chargino-Neutralino}\par
$$
\eqalign{
    {\cal L} = +g\sum_{i,j}\lbrack
    \bar{\tilde\chi}_i^+ \gamma^\mu
    (\ell_{ij}^{+0} {{1-\gamma_5}\over 2}+r_{ij}^{+0}{{1+\gamma_5}\over 2})
     \tilde\chi_j^0 W_\mu^+   
   + \bar{\tilde\chi}_i^0 \gamma^\mu
    (\ell_{ij}^{0+} {{1-\gamma_5}\over 2}+r_{ij}^{0+}{{1+\gamma_5}\over 2})
     \tilde\chi_j^+ W_\mu^-\rbrack, 
}\eqno(C.6)$$
where
$$
\eqalign{
   \ell_{i1}^{0+} =&~\eta_i[({\cal O}_N)_{i2} \cos\phi_L 
                    -{1\over{\sqrt 2}}({\cal O}_N)_{i4} \sin\phi_L], \cr
   \ell_{i2}^{0+} =&~\eta_i\epsilon_L[-({\cal O}_N)_{i2} \sin\phi_L 
                    -{1\over{\sqrt 2}}({\cal O}_N)_{i4} \cos\phi_L], \cr
   r_{i1}^{0+} =&~\eta^*_i[({\cal O}_N)_{i2} \cos\phi_R 
                    +{1\over{\sqrt 2}}({\cal O}_N)_{i3} \sin\phi_R], \cr
   r_{i2}^{0+} =&~\eta^*_i[-({\cal O}_N)_{i2} \sin\phi_R 
                    +{1\over{\sqrt 2}}({\cal O}_N)_{i3} \cos\phi_R], \cr
}\eqno(C.6a)
$$
and $\ell_{ij}^{+0} =(\ell_{ji}^{0+})^*$ ~, $r_{ij}^{+0} =(r_{ji}^{0+})^*$.
\par
\noindent{\bf Z-Chargino-Chargino}\par
$$
    {\cal L} = +g_Z\sum_{i,j} \bar{\tilde\chi}^+_i \gamma^\mu
               (v_{ij}^+ + a_{ij}^+\gamma_5) \tilde\chi_j^+ Z_\mu,
\eqno(C.7)
$$
where
$$
\eqalign{
  v_{11}^+ =&~ {1\over 4}(\sin^2\phi_R + \sin^2\phi_L) -c_W^2,\cr
  a_{11}^+ =&~ {1\over 4}(\sin^2\phi_R - \sin^2\phi_L),\cr
  v_{22}^+ =&~ {1\over 4}(\cos^2\phi_R + \cos^2\phi_L) -c_W^2,\cr
  a_{22}^+ =&~ {1\over 4}(\cos^2\phi_R - \cos^2\phi_L),\cr
  v_{12}^+ =&~ v_{21}^+ = {1\over 4}(\cos\phi_R\sin\phi_R 
                 + \epsilon_L \cos\phi_L \sin\phi_L),\cr
  a_{12}^+ =&~ a_{21}^+ = {1\over 4}(\cos\phi_R\sin\phi_R 
                 - \epsilon_L \cos\phi_L \sin\phi_L).
}\eqno(C.7a)
$$
\noindent{\bf $\gamma$-Chargino-Chargino}\par
$$
   {\cal L} = -e(\bar{\tilde\chi}^+_1\gamma^\mu \tilde\chi^+_1
               + \bar{\tilde\chi}^+_2\gamma^\mu \tilde\chi^+_2)A_\mu.
\eqno(C.8)
$$
\noindent{\bf Z-Neutralino-Neutralino}\par
$$
   {\cal L} = {{g_Z}\over 2} \sum_i a_{ii}^0
        \overline{\tilde \chi^0_i} \gamma^\mu\gamma_5\tilde\chi^0_i Z_\mu
        -g_Z\sum_{i<j}\overline{\tilde\chi^0_i}\gamma^\mu
                       (v_{ij}^0-a_{ij}^0\gamma_5)\tilde\chi^0_j Z_\mu,
\eqno(C.9)
$$
where
$$
\eqalign{
    v_{ij}^0 =&~ {i\over 2} {\cal I}m(\eta_i\eta_j^*)
             [({\cal O}_N)_{i3}({\cal O}_N)_{j3}     
             -({\cal O}_N)_{i4}({\cal O}_N)_{j4}],\cr     
    a_{ij}^0 =&~ {1\over 2} {\cal R}e(\eta_i\eta_j^*)
             [({\cal O}_N)_{i3}({\cal O}_N)_{j3}     
             -({\cal O}_N)_{i4}({\cal O}_N)_{j4}].\cr
}\eqno(C.9a)
$$                  
\noindent{\bf Gluon-Gluino-Gluino}\par
$$
    {\cal L} = {{g_s}\over 2} i f^{\alpha\beta\gamma}
       \overline{\tilde g^\alpha}\gamma^\mu \tilde g^\beta g^\gamma_\mu.         \eqno(C.10)
$$
\noindent{\bf Gauge-boson-Higgs-Higgs}\par
$$
\eqalign{
   {\cal L} =& +{g\over 2}i[ \cos(\alpha-\beta)(h^0\partrl^\mu H^-) 
               +\sin(\alpha-\beta)(H^0\partrl^\mu H^-)]W^+_\mu + h.c. \cr
             & -{g\over 2}(A^0\partrl^\mu H^-)W_\mu^+  +h.c.\cr
             & -{{g_Z}\over 2}[ \cos(\alpha-\beta)(h^0\partrl^\mu A^0)
                     +\sin(\alpha-\beta)(H^0\partrl^\mu A^0)]Z_\mu \cr
             &+ig_Z({1\over 2}-s_W^2)(H^+\partrl^\mu H^-)Z_\mu
              +ie (H^+\partrl^\mu H^-)A_\mu.\cr        
}\eqno(C.11)
$$
\noindent{\bf Gauge-boson-Higgs-Goldstone}\par
$$
\eqalign{
   {\cal L}=&-{g\over 2}i[\sin(\alpha-\beta)(h^0\partrl^\mu G^-)
                 -\cos(\alpha-\beta)(H^0\partrl^\mu G^-)]W_\mu^+ +h.c.\cr
            &+{{g_Z}\over 2}[\sin(\alpha-\beta)(h^0\partrl^\mu G^0)
                 -\cos(\alpha-\beta)(H^0\partrl^\mu G^0)]Z_\mu.\cr
}\eqno(C.12)
$$
\noindent{\bf Gauge-boson-Goldstone-Goldstone}\par
$$
\eqalign{
    {\cal L}=& -{g\over 2}[G^0\partrl^\mu G^-)W^+_\mu 
                         +(G^0\partrl^\mu G^+)W^-_\mu] \cr
             &+ig_Z({1\over 2}-s_W^2)(G^+\partrl^\mu G^-)Z_\mu
              +ie(G^+\partrl^\mu G^-)A_\mu.\cr 
}\eqno(C.13)
$$
\noindent{\bf H$^0$(h$^0$, A$^0$)-Fermion-Fermion}\par
$$
\eqalign{
    {\cal L}=&
     +{{gm_e\sin\alpha}\over{2M_W\cos\beta}} \bar e e h^0 
     -{{gm_e\cos\alpha}\over{2M_W\cos\beta}} \bar e e H^0 
     +i{{gm_e}\over{2M_W}}\tan\beta \bar e \gamma_5 e A^0 \cr
   & -{{gm_u\cos\alpha}\over{2M_W\sin\beta}} \bar u u h^0 
     -{{gm_u\sin\alpha}\over{2M_W\sin\beta}} \bar u u H^0 
     +i{{gm_u}\over{2M_W}}\cot\beta \bar u \gamma_5 u A^0 \cr
   & +{{gm_d\sin\alpha}\over{2M_W\cos\beta}} \bar d d h^0 
     -{{gm_d\cos\alpha}\over{2M_W\cos\beta}} \bar d d H^0 
     +i{{gm_d}\over{2M_W}}\tan\beta \bar d \gamma_5 d A^0. \cr
}\eqno(C.14)
$$
\noindent{\bf H$^\pm$-Fermion-Fermion}\par
$$
  {\cal L}= {g\over{\sqrt 2 M_W}}[m_e\tan\beta(\bar\nu{{1+\gamma_5}\over 2}e)
           + m_u\cot\beta(\bar u{{1-\gamma_5}\over 2} d) 
           + m_d\tan\beta(\bar u {{1+\gamma_5}\over 2} d)]H^+
                 +h.c.
\eqno(C.15)
$$
\noindent{\bf Higgs-Gauge-boson-Gauge-boson}\par
$$
\eqalign{
   {\cal L} =&
    +gM_W[\cos(\alpha-\beta)H^0-\sin(\alpha-\beta)h^0]W_\mu^+W^{-\mu} \cr
    & +{{g_Z}\over 2}M_Z[\cos(\alpha-\beta)H^0-\sin(\alpha-\beta)h^0]
        Z_\mu Z^\mu.
}\eqno(C.16)
$$
In the following four equations (C.17), (C.18), (C.19) and (C.20),
in order to avoid the complication of the 
equation I use $\tilde f_L$ and $\tilde f_R$ instead of the mass eigenstates
$\tilde f_1$ and $\tilde f_2$.  Using (2.6), one can easily find the
proper interactions of the Higgs-sfermion-sfermion.\par
\noindent{\bf H$^0$-Sfermion-Sfermion}\par
$$
\eqalign{
   {\cal L} =
   &-{{g_Z}\over 2} M_Z \cos(\alpha+\beta) \tilde\nu^* \tilde \nu H^0\cr
   &-({{gm_e^2\cos\alpha}\over{M_W\cos\beta}} + g_ZM_Z\cos(\alpha+\beta)
       (-{1\over 2}+s_W^2))\tilde e^*_L\tilde e_L H^0\cr
   &-({{gm_e^2\cos\alpha}\over{M_W\cos\beta}} - g_ZM_Z\cos(\alpha+\beta)
      s_W^2)\tilde e^*_R\tilde e_R H^0\cr
   & +{{gm_e}\over{2M_W\cos\beta}}(A_e\cos\alpha + \mu\sin\alpha)
      (\tilde e^*_L\tilde e_R H^0 + \tilde e^*_R\tilde e_L H^0)
}$$  
$$
\eqalign{
~~~~&-({{gm_u^2\sin\alpha}\over{M_W\sin\beta}} + g_ZM_Z\cos(\alpha+\beta)
       ({1\over 2}-{2\over3}s_W^2))\tilde u^*_L\tilde u_L H^0\cr
   &-({{gm_u^2\sin\alpha}\over{M_W\sin\beta}} +{2\over 3}g_ZM_Z
           \cos(\alpha+\beta) s_W^2)\tilde u^*_R\tilde u_R H^0\cr
   & +{{gm_u}\over{2M_W\sin\beta}}(A_u\sin\alpha + \mu\cos\alpha)
      (\tilde u^*_L\tilde u_R H^0 + \tilde u^*_R\tilde u_L H^0)
}$$
$$
\eqalign{
~~~~&-({{gm_d^2\cos\alpha}\over{M_W\cos\beta}} + g_ZM_Z\cos(\alpha+\beta)
       (-{1\over 2}+{1\over3}s_W^2))\tilde d^*_L\tilde d_L H^0\cr
   &-({{gm_d^2\cos\alpha}\over{M_W\cos\beta}} -{1\over 3}g_ZM_Z
           \cos(\alpha+\beta)s_W^2)\tilde d^*_R\tilde d_R H^0\cr
   & +{{gm_d}\over{2M_W\cos\beta}}(A_d\cos\alpha + \mu\sin\alpha)
      (\tilde d^*_L\tilde d_R H^0 + \tilde d^*_R\tilde d_L H^0).
}\eqno(C.17)
$$
\noindent{\bf h$^0$-Sfermion-Sfermion}\par
$$
\eqalign{
   {\cal L} =
   &+{{g_Z}\over 2} M_Z \sin(\alpha+\beta) \tilde\nu^* \tilde \nu h^0\cr
   &+({{gm_e^2\sin\alpha}\over{M_W\cos\beta}} + g_ZM_Z\sin(\alpha+\beta)
       (-{1\over 2}+s_W^2))\tilde e^*_L\tilde e_L h^0\cr
   &+({{gm_e^2\sin\alpha}\over{M_W\cos\beta}} - g_ZM_Z\sin(\alpha+\beta)
      s_W^2)\tilde e^*_R\tilde e_R h^0\cr
   & -{{gm_e}\over{2M_W\cos\beta}}(A_e\sin\alpha - \mu\cos\alpha)
      (\tilde e^*_L\tilde e_R h^0 + \tilde e^*_R\tilde e_L h^0)
}$$  
$$
\eqalign{
~~~~&-({{gm_u^2\cos\alpha}\over{M_W\sin\beta}} - g_ZM_Z\sin(\alpha+\beta)
       ({1\over 2}-{2\over3}s_W^2))\tilde u^*_L\tilde u_L h^0\cr
   &-({{gm_u^2\cos\alpha}\over{M_W\sin\beta}} -{2\over 3}g_ZM_Z
           \sin(\alpha+\beta)s_W^2)\tilde u^*_R\tilde u_R h^0\cr
   & +{{gm_u}\over{2M_W\sin\beta}}(A_u\cos\alpha - \mu\sin\alpha)
      (\tilde u^*_L\tilde u_R h^0 + \tilde u^*_R\tilde u_L h^0)
}$$
$$
\eqalign{
~~~~&+({{gm_d^2\sin\alpha}\over{M_W\cos\beta}} + g_ZM_Z\sin(\alpha+\beta)
       (-{1\over 2}+{1\over3}s_W^2))\tilde d^*_L\tilde d_L h^0\cr
   &+({{gm_d^2\sin\alpha}\over{M_W\cos\beta}} -{1\over 3}g_ZM_Z
           \sin(\alpha+\beta)s_W^2)\tilde d^*_R\tilde d_R h^0\cr
   & -{{gm_d}\over{2M_W\cos\beta}}(A_d\sin\alpha - \mu\cos\alpha)
      (\tilde d^*_L\tilde d_R h^0 + \tilde d^*_R\tilde d_L h^0).
}\eqno(C.18)
$$
Note that (C.18) is obtained from (C.17) by the change
$$
     \sin\alpha \to \cos\alpha,~~~~~~~ \cos\alpha \to -\sin\alpha,~~~~~
     \sin(\alpha+\beta) \to \cos(\alpha+\beta), ~~~~~ 
     \cos(\alpha+\beta) \to -\sin(\alpha+\beta).
\eqno(C.18a)
$$     
\noindent{\bf A$^0$-Sfermion-Sfermion}\par
$$
\eqalign{
   {\cal L} =& -i{{gm_e}\over {2M_W}}(A_e\tan\beta-\mu)
                 (\tilde e_L^*\tilde e_R A^0 -\tilde e_R^*\tilde e_L A^0)\cr  
             & -i{{gm_u}\over {2M_W}}(A_u\cot\beta-\mu)
                 (\tilde u_L^*\tilde u_R A^0 -\tilde u_R^*\tilde u_L A^0)\cr  
             & -i{{gm_d}\over {2M_W}}(A_d\tan\beta-\mu)
                 (\tilde d_L^*\tilde d_R A^0 -\tilde d_R^*\tilde d_L A^0).\cr  
}\eqno(C.19)
$$
\noindent{\bf H$^\pm$-Sfermion-Sfermion}\par
$$
\eqalign{
   {\cal L} =& -{g\over {\sqrt 2M_W}}[
            (M_W^2\sin 2\beta-m_e^2\tan\beta)\tilde\nu^* \tilde e_L
           + m_e(A_e\tan\beta-\mu)\tilde\nu^* \tilde e_R ]H^+ \cr
           &-{g\over {\sqrt 2M_W}}[M_W^2\sin 2\beta-m_d^2\tan\beta
             -m_u^2\cot\beta](\tilde u^*_L \tilde d_L H^+) \cr
           &+{{\sqrt 2 g m_u m_d}\over{M_W\sin 2\beta}}
             (\tilde u_R^*\tilde d_R H^+)  \cr
           &-{g\over {\sqrt 2M_W}}[  
              m_u(A_u\cot\beta-\mu)\tilde u^*_R\tilde d_L 
             +m_d(A_d\tan\beta-\mu)\tilde u^*_L\tilde d_R] H^+\cr
           & + h.c.
}\eqno(C.20)
$$
\noindent{\bf Higgs-Chargino-Chargino}\par
$$
\eqalign{
   {\cal L}=-{g\over{\sqrt 2}}&\lbrace 
      s_{11}^{CCh}\bar{\tilde\chi}^+_1 \tilde\chi^+_1 h^0
    + s_{22}^{CCh}\bar{\tilde\chi}^+_2 \tilde\chi^+_2 h^0 \cr
   +&\bar{\tilde\chi}^+_1(\ell_{12}^{CCh}~{{1-\gamma_5}\over 2}
                    +r_{12}^{CCh}~{{1+\gamma_5}\over 2})\tilde\chi_2^+h^0
    + \bar{\tilde\chi}^+_2(\ell_{21}^{CCh}~{{1-\gamma_5}\over 2}
                +r_{21}^{CCh}~{{1+\gamma_5}\over 2})\tilde\chi_1^+h^0 \cr
   +& s_{11}^{CCH}\bar{\tilde\chi}^+_1 \tilde\chi^+_1 H^0
    + s_{22}^{CCH}\bar{\tilde\chi}^+_2 \tilde\chi^+_2 H^0 \cr
   +&\bar{\tilde\chi}^+_1(\ell_{12}^{CCH}~{{1-\gamma_5}\over 2}
                    +r_{12}^{CCH}~{{1+\gamma_5}\over 2})\tilde\chi_2^+H^0
    + \bar{\tilde\chi}^+_2(\ell_{21}^{CCH}~{{1-\gamma_5}\over 2}
                    +r_{21}^{CCH}~{{1+\gamma_5}\over 2})\tilde\chi_1^+H^0 
                                  \rbrace \cr
    -{{ig}\over{\sqrt 2}}&\lbrace 
      p_{11}^{CCA}\bar{\tilde\chi}^+_1 \gamma_5\tilde\chi^+_1 A^0
    + p_{22}^{CCA}\bar{\tilde\chi}^+_2 \gamma_5\tilde\chi^+_2 A^0 \cr
   +&\bar{\tilde\chi}^+_1(\ell_{12}^{CCA}~{{1-\gamma_5}\over 2}
                    +r_{12}^{CCA}~{{1+\gamma_5}\over 2})\tilde\chi_2^+A^0
    + \bar{\tilde\chi}^+_2(\ell_{21}^{CCA}~{{1-\gamma_5}\over 2}
                    +r_{21}^{CCA}~{{1+\gamma_5}\over 2})\tilde\chi_1^+A^0 
                               \rbrace, 
}\eqno(C.21)
$$
where
$$
\eqalign{
    s_{11}^{CCh}=&-\sin\alpha \cos\phi_L \sin\phi_R 
                  +\cos\alpha \sin\phi_L \cos\phi_R,\cr 
    s_{22}^{CCh}=&~\epsilon_L(\sin\alpha \sin\phi_L \cos\phi_R 
                  -\cos\alpha \cos\phi_L \sin\phi_R ),\cr 
 \ell_{12}^{CCh}=&~\epsilon_L(\sin\alpha \sin\phi_L \sin\phi_R
                 + \cos\alpha \cos\phi_L \cos\phi_R), \cr
    r_{12}^{CCh}=&-\sin\alpha \cos\phi_L \cos\phi_R 
                  -\cos\alpha \sin\phi_L \sin\phi_R, \cr
 \ell_{21}^{CCh}=&~ r_{12}^{CCh}, \cr
   r_{21}^{CCh} =&~ \ell_{12}^{CCh}, \cr
    s_{11}^{CCH}=&~ \cos\alpha \cos\phi_L \sin\phi_R
                  +\sin\alpha \sin\phi_L \cos\phi_R,\cr 
    s_{22}^{CCH}=&~\epsilon_L(-\cos\alpha \sin\phi_L \cos\phi_R 
                  -\sin\alpha \cos\phi_L \sin\phi_R ),\cr 
 \ell_{12}^{CCH}=&~\epsilon_L(-\cos\alpha \sin\phi_L \sin\phi_R
                  + \sin\alpha \cos\phi_L \cos\phi_R), \cr
   r_{12}^{CCH} =&~\cos\alpha \cos\phi_L \cos\phi_R 
                  - \sin\alpha \sin\phi_L \sin\phi_R, \cr
 \ell_{21}^{CCH}=&~ r_{12}^{CCH}, \cr
   r_{21}^{CCH} =&~ \ell_{12}^{CCH}, \cr
    p_{11}^{CCA}=&~\sin\beta \cos\phi_L \sin\phi_R
                 +\cos\beta \sin\phi_L \cos\phi_R,\cr 
    p_{22}^{CCA}=&~\epsilon_L(-\sin\beta \sin\phi_L \cos\phi_R 
                  -\cos\beta \cos\phi_L \sin\phi_R),\cr 
 \ell_{12}^{CCA}=&~\epsilon_L(\sin\beta \sin\phi_L \sin\phi_R
                 - \cos\beta \cos\phi_L \cos\phi_R), \cr
   r_{12}^{CCA} =&~\sin\beta \cos\phi_L \cos\phi_R 
                  - \cos\beta \sin\phi_L \sin\phi_R, \cr
 \ell_{21}^{CCA}=& -r_{12}^{CCA}, \cr
   r_{21}^{CCA} =& -\ell_{12}^{CCA}. 
}\eqno(C.21a)
$$
\noindent{\bf Higgs-Chargino-Neutralino}\par
$$
\eqalign{
     {\cal L}= &-g\sum_{i,j}\bar{\tilde\chi}_i^0
  (\ell_{ij}^{CNH}~{{1-\gamma_5}\over 2}+r_{ij}^{CNH}~{{1+\gamma_5}\over 2})
     \tilde\chi_j^+ H^- \cr
               &- g\sum_{i,j}\bar{\tilde\chi}_i^+
     ((r_{ji}^{CNH})^*{{1-\gamma_5}\over 2}+
      (\ell_{ji}^{CNH})^*{{1+\gamma_5}\over 2})  \tilde\chi_j^0 H^+,\cr
}\eqno(C.22)
$$
where
$$
\eqalign{
 \ell_{i1}^{CNH}=&~ \cos\beta[\cos\phi_L({\cal O}_N)_{i4}
                   +{1\over {\sqrt 2}}\sin\phi_L(({\cal O}_N)_{i2}
                   +{{s_W}\over{c_W}}({\cal O}_N)_{i1} )]\eta_i^*, \cr
    r_{i1}^{CNH}=&~ \sin\beta[\cos\phi_R({\cal O}_N)_{i3}
                   -{1\over {\sqrt 2}}\sin\phi_R(({\cal O}_N)_{i2}
                   +{{s_W}\over{c_W}}({\cal O}_N)_{i1} )]\eta_i, \cr
 \ell_{i2}^{CNH}=&~ \cos\beta\epsilon_L[-\sin\phi_L({\cal O}_N)_{i4}
                   +{1\over {\sqrt 2}}\cos\phi_L(({\cal O}_N)_{i2}
                   +{{s_W}\over{c_W}}({\cal O}_N)_{i1} )]\eta_i^*, \cr 
    r_{i2}^{CNH}=&~ \sin\beta[-\sin\phi_R({\cal O}_N)_{i3}
                    -{1\over {\sqrt 2}}\cos\phi_R(({\cal O}_N)_{i2}
                   +{{s_W}\over{c_W}}({\cal O}_N)_{i1} )]\eta_i.
}\eqno(C.22a)
$$
\noindent{\bf Higgs-Neutralino-Neutralino}\par
$$
\eqalign{
    {\cal L} =& -{g\over 4}\sum_is_{ii}^{NNH}
               \overline{\tilde\chi^0_i}\tilde\chi^0_i H^0 
                -{g\over 2}\sum_{i<j}\overline{\tilde\chi^0_i}
                (s_{ij}^{NNH}+ip_{ij}^{NNH}\gamma_5)\tilde\chi^0_j H^0 \cr
              & -{g\over 4}\sum_is_{ii}^{NNh}
               \overline{\tilde\chi^0_i}\tilde\chi^0_i h^0 
                -{g\over 2}\sum_{i<j}\overline{\tilde\chi^0_i}
                (s_{ij}^{NNh}+ip_{ij}^{NNh}\gamma_5)\tilde\chi^0_j h^0 \cr
              & -i{g\over 4}\sum_ip_{ii}^{NNA}
               \overline{\tilde\chi^0_i}\gamma_5\tilde\chi^0_i A^0 
                -{g\over 2}\sum_{i<j}\overline{\tilde\chi^0_i}
                (-s_{ij}^{NNA}+ip_{ij}^{NNA}\gamma_5)\tilde\chi^0_j A^0, \cr
}\eqno(C.23)
$$
where
$$
\eqalign{
     s_{ij}^{NNH} =&~ {\cal R}e(\eta_i\eta_j)\lbrace
        [({\cal O}_N)_{i2}-({\cal O}_N)_{i1}\tan\theta_W]    
        [\cos\alpha({\cal O}_N)_{j3}-\sin\alpha({\cal O}_N)_{j4}]\cr
                   & ~~~~~~~~+(i\leftrightarrow j)\rbrace,\cr    
     p_{ij}^{NNH} =&~ {\cal I}m(\eta_i\eta_j)\lbrace
        [({\cal O}_N)_{i2}-({\cal O}_N)_{i1}\tan\theta_W]    
        [\cos\alpha({\cal O}_N)_{j3}-\sin\alpha({\cal O}_N)_{j4}]\cr
                   & ~~~~~~~~+(i\leftrightarrow j)\rbrace,\cr    
     s_{ij}^{NNh} =&~ {\cal R}e(\eta_i\eta_j)\lbrace
        [({\cal O}_N)_{i2}-({\cal O}_N)_{i1}\tan\theta_W]    
        [-\sin\alpha({\cal O}_N)_{j3}-\cos\alpha({\cal O}_N)_{j4}]\cr
                   & ~~~~~~~~+(i\leftrightarrow j)\rbrace,\cr    
     p_{ij}^{NNh} =&~ {\cal I}m(\eta_i\eta_j)\lbrace
        [({\cal O}_N)_{i2}-({\cal O}_N)_{i1}\tan\theta_W]    
        [-\sin\alpha({\cal O}_N)_{j3}-\cos\alpha({\cal O}_N)_{j4}]\cr
                   & ~~~~~~~~+(i\leftrightarrow j)\rbrace,\cr    
     s_{ij}^{NNA} =&~ {\cal I}m(\eta_i\eta_j)\lbrace
        [({\cal O}_N)_{i2}-({\cal O}_N)_{i1}\tan\theta_W]    
        [\sin\beta({\cal O}_N)_{j3}-\cos\beta({\cal O}_N)_{j4}]\cr
                   & ~~~~~~~~+(i\leftrightarrow j)\rbrace,\cr    
     p_{ij}^{NNA} =&~ {\cal R}e(\eta_i\eta_j)\lbrace
        [({\cal O}_N)_{i2}-({\cal O}_N)_{i1}\tan\theta_W]    
        [\sin\beta({\cal O}_N)_{j3}-\cos\beta({\cal O}_N)_{j4}]\cr
                   & ~~~~~~~~+(i\leftrightarrow j)\rbrace.\cr    
}\eqno(C.23a)
$$
Note the minus sign in front of $s_{ij}^{NNA}$.  As in (C.16) and (C.17),
$s_{ij}^{NNh}$ and $p_{ij}^{NNh}$
are obtained from $s_{ij}^{NNH}$ and $p_{ij}^{NNH}$ by
$$
     \sin\alpha \to \cos\alpha,~~~~~~~ \cos\alpha \to -\sin\alpha,~~~~~
\eqno(C.23b)
$$
while $s_{ij}^{NNA}$ and $p_{ij}^{NNA}$
are obtained from $p_{ij}^{NNH}$ and $s_{ij}^{NNH}$ by
$$
     \cos\alpha \to \sin\beta,~~~~~\sin\alpha \to \cos\beta.
\eqno(C.23c)
$$
\noindent{\bf Goldstone-Fermion-Fermion}\par
$$
\eqalign{
  {\cal L}=& -i{g\over{2M_W}}
              [ m_e\bar e\gamma_5 eG^0 - m_u\bar u\gamma_5 u G^0
             + m_d\bar d\gamma_5 dG^0 ]\cr
           & -{g\over{\sqrt 2M_W}} 
              [m_e\bar\nu {{1+\gamma_5}\over 2} e G^+ 
             - m_u\bar u  {{1-\gamma_5}\over 2} d G^+ 
             + m_d\bar u  {{1+\gamma_5}\over 2} d G^+] + h.c.
}\eqno(C.24)
$$
\noindent{\bf Goldstone-Gauge-boson-Gauge-boson}\par
$$
   {\cal L} = -g_ZM_W s_W^2(G^-W_\mu^+Z^\mu) 
              +eM_W(G^-W_\mu^+A^\mu) + h.c. 
\eqno(C.25)
$$
\noindent{\bf Goldstone-Sfermion-Sfermion}\par
$$
\eqalign{
   {\cal L}=&+ {g\over{\sqrt 2M_W}}[  
                (M_W^2\cos 2\beta-m_e^2)\tilde\nu^*\tilde e_L G^+
               +(M_W^2\cos 2\beta+m_u^2-m_d^2)\tilde u_L^*\tilde d_L G^+]\cr 
            &+ {g\over{\sqrt 2M_W}}[
                m_e(A_e+\mu\tan\beta)\tilde \nu^*\tilde e_R G^+
               -m_u(A_u+\mu\cot\beta)\tilde u^*_R\tilde d_L G^+ \cr
            &~~~~~~~~~~~~ +m_d(A_d+\mu\tan\beta)\tilde u^*_L\tilde d_R G^+]\cr
            &+ i{g\over{2M_W}}[
                m_e(A_e+\mu\tan\beta)\tilde e^*_L\tilde e_R G^0
               -m_u(A_u+\mu\cot\beta)\tilde u^*_L\tilde u_R G^0\cr
            &~~~~~~~~~~~~   
               +m_d(A_d+\mu\tan\beta)\tilde d^*_L\tilde d_R G^0]\cr
            &  + h.c.
}\eqno(C.26)
$$
\noindent{\bf G$^0$-Chargino-Chargino}\par
$$
\eqalign{
     {\cal L}={{ig}\over{\sqrt 2}}&\lbrace 
     p_{11}^{CCG}\bar{\tilde\chi}^+_1\gamma_5 \tilde\chi_1^+ G^0
   + p_{22}^{CCG}\bar{\tilde\chi}^+_2\gamma_5 \tilde\chi_2^+ G^0 \cr
   +& \bar{\tilde\chi}_1^+(\ell_{12}^{CCG}{{1-\gamma_5}\over 2}
                  +r_{12}^{CCG}{{1+\gamma_5}\over 2})\tilde\chi^+_2G^0  
   +  \bar{\tilde\chi}_2^+(\ell_{21}^{CCG}{{1-\gamma_5}\over 2}
                  +r_{21}^{CCG}{{1+\gamma_5}\over 2})\tilde\chi^+_1G^0  
                     \rbrace, \cr
}\eqno(C.27)
$$
where
$$
\eqalign{
     p_{11}^{CCG} =&~ \cos\beta \cos\phi_L \sin\phi_R 
                    -\sin\beta \sin\phi_L \cos\phi_R, \cr
     p_{22}^{CCG} =&~ \epsilon_L(-\cos\beta \sin\phi_L \cos\phi_R 
                                +\sin\beta \cos\phi_L \sin\phi_R ), \cr
     \ell_{12}^{CCG}=&~\epsilon_L(\cos\beta \sin\phi_L \sin\phi_R 
                        +\sin\beta \cos\phi_L \cos\phi_R), \cr
        r_{12}^{CCG}=&~\cos\beta \cos\phi_L \cos\phi_R 
                        +\sin\beta \sin\phi_L \sin\phi_R, \cr
     \ell_{21}^{CCG}=& -r_{12}^{CCG},\cr
        r_{21}^{CCG}=& -\ell_{12}^{CCG}.\cr
}\eqno(C.27a)
$$
\noindent{\bf G$^0$-Neutralino-Neutralino}\par
$$
    {\cal L} = ~ i{g\over 4}\sum_ip_{ii}^{NNG}
               \overline{\tilde\chi^0_i}\gamma_5\tilde\chi^0_i G^0 
                +{g\over 2}\sum_{i<j}\overline{\tilde\chi^0_i}
                (-s_{ij}^{NNG}+ip_{ij}^{NNG}\gamma_5)\tilde\chi^0_j G^0, 
\eqno(C.28)
$$
where
$$
\eqalign{
     s_{ij}^{NNG} =&~ {\cal I}m(\eta_i\eta_j)\lbrace
        [({\cal O}_N)_{i2}-{{s_W}\over{c_W}}({\cal O}_N)_{i1}]    
        [\cos\beta({\cal O}_N)_{j3}+\sin\beta({\cal O}_N)_{j4}]\cr
                   & ~~~~~~~~+(i\leftrightarrow j)\rbrace,\cr    
     p_{ij}^{NNG} =&~ {\cal R}e(\eta_i\eta_j)\lbrace
        [({\cal O}_N)_{i2}-{{s_W}\over {c_W}}({\cal O}_N)_{i1}]    
        [\cos\beta({\cal O}_N)_{j3}+\sin\beta({\cal O}_N)_{j4}]\cr
                   & ~~~~~~~~+(i\leftrightarrow j)\rbrace.\cr    
}\eqno(C.28a)
$$
\noindent{\bf G$^\pm$-Chargino-Neutralino}\par
$$
\eqalign{
   {\cal L}=+&g\sum_{i,j}\bar{\tilde\chi}^0_i
               (\ell_{ij}^{CNG}~{{1-\gamma_5}\over 2}
             +r_{ij}^{CNG}~{{1+\gamma_5}\over 2})\tilde\chi^+_j G^- \cr
            +& g\sum_{i,j}\bar{\tilde\chi}^+_i
               ((r_{ji}^{CNG})^*{{1-\gamma_5}\over 2}
     +  (\ell_{ji}^{CNG})^*{{1+\gamma_5}\over 2})\tilde\chi^0_j G^+,
}\eqno(C.29)
$$
where
$$
\eqalign{
  \ell_{i1}^{CNG} =&~ \sin\beta[- \cos\phi_L({\cal O}_N)_{i4}
       -{1\over{\sqrt 2}}\sin\phi_L(({\cal O}_N)_{i2}
        +{{s_W}\over{c_W}}({\cal O}_N)_{i1} )]\eta^*_i, \cr
     r_{i1}^{CNG} =&~ \cos\beta[ \cos\phi_R({\cal O}_N)_{i3}
       -{1\over{\sqrt 2}}\sin\phi_R(({\cal O}_N)_{i2}
        +{{s_W}\over{c_W}}({\cal O}_N)_{i1} )]\eta_i, \cr
  \ell_{i2}^{CNG} =&~ \sin\beta\epsilon_L[\sin\phi_L({\cal O}_N)_{i4}
       -{1\over{\sqrt 2}}\cos\phi_L(({\cal O}_N)_{i2}
        +{{s_W}\over{c_W}}({\cal O}_N)_{i1} )]\eta^*_i. \cr
     r_{i2}^{CNG} =&~ \cos\beta[-\sin\phi_R({\cal O}_N)_{i3}
       -{1\over{\sqrt 2}}\cos\phi_R(({\cal O}_N)_{i2}
        +{{s_W}\over{c_W}}({\cal O}_N)_{i1} )]\eta_i. \cr
}\eqno(C.29a)
$$
\noindent{\bf Chargino-Fermion-Sfermion}\par
$$
\eqalign{
   {\cal L}=
    &+\overline{\tilde\chi_1^-}(-g\cos\phi_L{{1-\gamma_5}\over 2}
     + g{{m_e\sin\phi_R}\over{\sqrt 2 M_W\cos\beta}}{{1+\gamma_5}\over 2})
        e\tilde\nu^* \cr
    &+\overline{\tilde\chi_2^-}(g\epsilon_L\sin\phi_L{{1-\gamma_5}\over 2}
     + g{{m_e\cos\phi_R}\over{\sqrt 2 M_W\cos\beta}}{{1+\gamma_5}\over 2})
        e\tilde\nu^* \cr
    &-g\cos\phi_R\overline{\tilde\chi^+_1}{{1-\gamma_5}\over 2} \nu\tilde e^*_L 
     +g\sin\phi_R\overline{\tilde\chi^+_2}{{1-\gamma_5}\over 2} \nu\tilde e^*_L      \cr
   &+{{gm_e\sin\phi_R}\over{\sqrt 2 M_W \cos\beta}}
       \overline{\tilde\chi^+_1}{{1-\gamma_5}\over 2} \nu\tilde e^*_R
     +{{gm_e\cos\phi_R}\over{\sqrt 2 M_W \cos\beta}}
       \overline{\tilde\chi^+_2}{{1-\gamma_5}\over 2} \nu\tilde e^*_R \cr
}$$
$$
\eqalign{
~~~~ &+\overline{\tilde\chi_1^+}(-g\cos\phi_R{{1-\gamma_5}\over 2}
     + g{{m_u\sin\phi_L}\over{\sqrt 2 M_W\sin\beta}}{{1+\gamma_5}\over 2})
        u\tilde d^*_L \cr
    &+\overline{\tilde\chi_2^+}(g\sin\phi_R{{1-\gamma_5}\over 2}
     + g{{m_u\epsilon_L\cos\phi_L}\over{\sqrt 2 M_W\sin\beta}}
       {{1+\gamma_5}\over 2})  u\tilde d^*_L \cr
    &+{{gm_d\sin\phi_R}\over{\sqrt 2 M_W\cos\beta}}
       \overline{\tilde\chi_1^+}{{1-\gamma_5}\over 2}u \tilde d^*_R
     +g{{m_d\cos\phi_R}\over{\sqrt 2 M_W\cos\beta}}
       \overline{\tilde\chi_2^+}{{1-\gamma_5}\over 2}u \tilde d^*_R \cr
    &+\overline{\tilde\chi_1^-}(-g\cos\phi_L{{1-\gamma_5}\over 2}
     + g{{m_d\sin\phi_R}\over{\sqrt 2 M_W\cos\beta}}{{1+\gamma_5}\over 2})
        d\tilde u^*_L \cr
    &+\overline{\tilde\chi_2^-}(g\epsilon_L\sin\phi_L{{1-\gamma_5}\over 2}
     + g{{m_d\cos\phi_R}\over{\sqrt 2 M_W\cos\beta}}{{1+\gamma_5}\over 2})
        d\tilde u^*_L \cr
    &+{{gm_u\sin\phi_L}\over{\sqrt 2 M_W\sin\beta}}
       \overline{\tilde\chi_1^-}{{1-\gamma_5}\over 2}d \tilde u^*_R
     +g{{m_u\epsilon_L\cos\phi_L}\over{\sqrt 2 M_W\sin\beta}}
       \overline{\tilde\chi_2^-}{{1-\gamma_5}\over 2}d \tilde u^*_R \cr
    & + h.c.   
}\eqno(C.30)
$$
\noindent{\bf Neutralino-Fermion-Sfermion}\par
$$
\eqalign{
   {\cal L} =& -{g\over{\sqrt 2}}\sum_i\lbrace
        \overline{\tilde\chi^0_i}(\ell_i^{NfL} {{1-\gamma_5}\over 2}
                +r_i^{NfL} {{1+\gamma_5}\over 2}) f \tilde f^*_L
       + \overline{\tilde\chi^0_i}(\ell_i^{NfR} {{1-\gamma_5}\over 2}
                +r_i^{NfR} {{1+\gamma_5}\over 2}) f \tilde f^*_R \rbrace\cr
             & + h.c.,
}\eqno(C.31)
$$
where
$$
\eqalign{
   \ell_i^{N\nu L} =&~~ \eta^*_i[({\cal O}_N)_{i2}
                              -{{s_W}\over{c_W}}({\cal O}_N)_{i1}], \cr
      \ell_i^{NeL} =& -\eta^*_i[({\cal O}_N)_{i2}
                              +{{s_W}\over{c_W}}({\cal O}_N)_{i1}], \cr
      \ell_i^{NuL} =&~~ \eta^*_i[({\cal O}_N)_{i2}
                              +{{s_W}\over{3c_W}}({\cal O}_N)_{i1}], \cr
      \ell_i^{NdL} =& -\eta^*_i[({\cal O}_N)_{i2}
                              -{{s_W}\over{3c_W}}({\cal O}_N)_{i1}], \cr
      r_i^{N\nu L} =&~ 0, \cr
         r_i^{NeL} =&~ \eta_i{{m_e}\over{M_W\cos\beta}} ({\cal O}_N)_{i3},\cr
         r_i^{NuL} =&~ \eta_i{{m_u}\over{M_W\sin\beta}} ({\cal O}_N)_{i4},\cr
         r_i^{NdL} =&~ \eta_i{{m_d}\over{M_W\cos\beta}} ({\cal O}_N)_{i3},\cr
}$$
$$
\eqalign{
   \ell_i^{N\nu R} =&~ 0,\cr
      \ell_i^{NeR} =&~ \eta^*_i{{m_e}\over{M_W\cos\beta}}({\cal O}_N)_{i3},
      ~~~~~~~~~~~~~~ \cr
      \ell_i^{NuR} =&~ \eta^*_i{{m_u}\over{M_W\sin\beta}}({\cal O}_N)_{i4},\cr
      \ell_i^{NdR} =&~ \eta^*_i{{m_d}\over{M_W\cos\beta}}({\cal O}_N)_{i3},\cr
      r_i^{N\nu R} =&~ 0, \cr
         r_i^{NeR} =&~ 2\eta_i{{s_W}\over{c_W}}({\cal O}_N)_{i1},\cr
         r_i^{NuR} =& -{4\over 3}\eta_i{{s_W}\over{c_W}}
                      ({\cal O}_N)_{i1},\cr
         r_i^{NdR} =& ~{2\over 3}\eta_i{{s_W}\over{c_W}}
                      ({\cal O}_N)_{i1}.\cr
}\eqno(C.31a)
$$                
\noindent{\bf Gluino-Quark-Squark}\par
$$
   {\cal L}= -\sqrt 2 g_s \sum_q[
            \bar q{{1+\gamma_5}\over 2} \tilde g^\alpha 
            {{\lambda^\alpha}\over 2}\tilde q_L
           -\bar q{{1-\gamma_5}\over 2} \tilde g^\alpha 
            {{\lambda^\alpha}\over 2}\tilde q_R] + h.c.
\eqno(C.32)
$$
\noindent{\bf W-W-Sfermion-Sfermion}\par
$$
  {\cal L}= {{g^2}\over 2}(\tilde u_L^*\tilde u_L +\tilde d^*_L\tilde d_L 
        + \tilde \nu^*\tilde \nu +\tilde e^*_L\tilde e_L) W_\mu^+ W^{-\mu}.
\eqno(C.33)
$$
\noindent{\bf W-Z($\gamma$)-Sfermion-Sfermion}\par
$$
\eqalign{
     {\cal L}=& -{{gg_Z}\over {\sqrt 2}}\sin^2\theta_W \sum_f Y_{f_L}
           (\tilde f_{\uparrow L}^*\tilde f_{\downarrow L}W^+_\mu
           +\tilde f_{\downarrow L}^*\tilde f_{\uparrow L}W^-_\mu)Z^\mu,\cr
              & +{{ge}\over {\sqrt 2}} \sum_f Y_{f_L}
           (\tilde f_{\uparrow L}^*\tilde f_{\downarrow L}W^+_\mu
           +\tilde f_{\downarrow L}^*\tilde f_{\uparrow L}W^-_\mu)A^\mu.\cr
}\eqno(C.34)
$$
The hypercharge  $Y_{f_L}$ of the left-handed fermion doublets is defined as
$Q_f = T_{3f} + {{Y_{f_L}}\over 2}$ and its value is  given in Table 1.\par
\noindent{\bf (Z,$\gamma$)-(Z,$\gamma$)-Sfermion-Sfermion}\par
$$
\eqalign{
 {\cal L}=&+ g_Z^2\sum_f [~(T_{3f}-s_W^2Q_f)^2\tilde f^*_L\tilde f_L
              + s_W^4Q_f^2 \tilde f^*_R\tilde f_R~]Z_\mu Z^\mu\cr
              &+2eg_Z\sum_f [~Q_f(T_{3f}-s_W^2Q_f) 
              \tilde f^*_L\tilde f_L
               -s_W^2Q_f^2\tilde f^*_R\tilde f_R~ ]Z_\mu A^\mu \cr
              &+e^2\sum_f Q_f^2(\tilde f^*_L\tilde f_L+\tilde f^*_R\tilde f_R)
               A_\mu A^\mu.
}\eqno(C.35)
$$
\noindent{\bf Gluon-Gluon-Squark-Squark}\par
$$
    {\cal L} = + g_s^2 \sum_q( 
   \tilde q_L^*{{\lambda^\alpha}\over 2} {{\lambda^\beta}\over 2}\tilde q_L
 + \tilde q_R^*{{\lambda^\alpha}\over 2} {{\lambda^\beta}\over 2}\tilde q_R)
           g_\mu^\alpha g^{\mu\beta}.
\eqno(C.36)
$$
\noindent{\bf Gluon-(W,Z,$\gamma$)-Squark-Squark}\par
$$
\eqalign{
    {\cal L}=&+{{g_s g}\over {\sqrt 2}}\{
     \tilde u_L^*\lambda^\alpha \tilde d_L W^{+\mu} 
    + \tilde d_L^*\lambda^\alpha \tilde u_L W^{-\mu}\}g_\mu^\alpha\cr
    &+g_s g_z\sum_q\{(T_{3q}-s_W^2Q_q)
         \tilde q_L^*\lambda^\alpha\tilde q_L   -s_W^2Q_q
         \tilde q_R^*\lambda^\alpha\tilde q_R\}Z^\mu g_\mu^\alpha \cr
    &+g_s e \sum_q Q_q(\tilde q_L^*\lambda^\alpha \tilde q_L
                  +\tilde q_R^*\lambda^\alpha \tilde q_R)A^\mu g_\mu^\alpha.
}\eqno(C.37)
$$
\noindent{\bf Higgs-Higgs-Gauge-boson-Gauge-boson}\par
$$
\eqalign{
      {\cal L}=
      &+ {{g^2}\over 4} W_\mu^+ W^{-\mu}            
            [(H^0)^2 + (h^0)^2 + (A^0)^2+ 2H^+H^-] \cr
      &+ {{g_Z^2}\over 8}Z_\mu Z^\mu 
            [ (H^0)^2+(h^0)^2+(A^0)^2+2(1-2s_W^2)^2 H^+H^-]\cr
      &+eg_Z(1-2s_W^2) Z_\mu A^\mu H^+H^- +e^2A_\mu A^\mu H^+H^-\cr
      &-{{gg_Z}\over 2}s_W^2 W^+_\mu Z^\mu
         [\cos(\alpha-\beta)h^0H^- + \sin(\alpha-\beta)H^0H^-
          +iA^0H^-] + h.c.\cr
      &+{{ge}\over 2} W^+_\mu A^\mu
         [\cos(\alpha-\beta)h^0H^- + \sin(\alpha-\beta)H^0H^-
          +iA^0H^-] +h.c.\cr
}\eqno(C.38)
$$
\noindent{\bf Higgs-Goldstone-Gauge-boson-Gauge-boson}\par
$$
   {\cal L} = {{ge}\over 2} W^+_\mu 
              (A^\mu-{{s_W}\over{c_W}} Z^\mu). 
              [ \cos(\alpha-\beta)H^0G^- - \sin(\alpha-\beta)h^0G^-] + h.c.
\eqno(C.39)
$$
\noindent{\bf Goldstone-Goldstone-Gauge-boson-Gauge-boson}\par
$$
\eqalign{
      {\cal L}=
      &+{{g^2}\over 4} W_\mu^+ W^{-\mu}[(G^0)^2+2G^+G^-]\cr
      &+ {{g_z^2}\over 8}Z_\mu Z^\mu [(G^0)^2 
             +2(1-2s_W^2)^2  G^+G^-] \cr
      &+(c_W^2-s_W^2)eg_Z  Z_\mu A^\mu G^+G^- 
       +e^2A_\mu A^\mu G^+G^-\cr
      &+i{{ge}\over 2} W^+_\mu 
        ( A^\mu-{{s_W}\over{c_W}} Z^\mu)G^0G^- + h.c.
}\eqno(C.40)
$$
\noindent{\bf H$^0$-H$^0$-Sfermion-Sfermion}\par
$$
\eqalign{
    {\cal L}=& -{1\over 4}g_Z^2(T_{3\nu})\cos 2\alpha 
                 \tilde \nu^*\tilde \nu (H^0)^2\cr
             & - [{{g^2m_e^2\cos^2\alpha}\over{4M_W^2 \cos^2\beta}}
                  +{1\over 4}g_Z^2(T_{3e}-s_W^2Q_e )\cos 2\alpha]
                 \tilde e_L^*\tilde e_L (H^0)^2\cr 
             & - [{{g^2m_e^2\cos^2\alpha}\over{4M_W^2 \cos^2\beta}}
                  +{1\over 4}g_Z^2s_W^2Q_e\cos 2\alpha]
                 \tilde e_R^*\tilde e_R (H^0)^2\cr 
}$$
$$
\eqalign{
   ~~~~~~ & - [{{g^2m_u^2\sin^2\alpha}\over{4M_W^2 \sin^2\beta}}
           +{1\over 4}g_Z^2(T_{3u}-s_W^2Q_u)\cos 2\alpha]
            \tilde u_L^*\tilde u_L (H^0)^2\cr 
        & - [{{g^2m_u^2\sin^2\alpha}\over{4M_W^2 \sin^2\beta}}
             +{1\over 4}g_Z^2s_W^2Q_u\cos 2\alpha]
            \tilde u_R^*\tilde u_R (H^0)^2\cr 
        & - [{{g^2m_d^2\cos^2\alpha}\over{4M_W^2 \cos^2\beta}}
         +{1\over 4}g_Z^2(T_{3d}-s_W^2Q_d)\cos 2\alpha]
                 \tilde d_L^*\tilde d_L (H^0)^2\cr 
        & - [{{g^2m_d^2\cos^2\alpha}\over{4M_W^2 \cos^2\beta}}
                  +{1\over4 }g_Z^2s_W^2Q_d\cos 2\alpha]
                 \tilde d_R^*\tilde d_R (H^0)^2.\cr 
}\eqno(C.41)
$$
\noindent{\bf h$^0$-h$^0$-Sfermion-Sfermion}\par
$$
\eqalign{
    {\cal L}=& +{1\over 4}g_Z^2(T_{3\nu})\cos 2\alpha 
                 \tilde \nu^*\tilde \nu (h^0)^2\cr
             & - [{{g^2m_e^2\sin^2\alpha}\over{4M_W^2 \cos^2\beta}}
                  -{1\over 4}g_Z^2(T_{3e}-s_W^2Q_e )\cos 2\alpha]
                 \tilde e_L^*\tilde e_L (h^0)^2\cr 
             & - [{{g^2m_e^2\sin^2\alpha}\over{4M_W^2 \cos^2\beta}}
                  -{1\over 4}g_Z^2s_W^2Q_e\cos 2\alpha]
                 \tilde e_R^*\tilde e_R (h^0)^2\cr 
}$$
$$
\eqalign{
   ~~~~~~ & - [{{g^2m_u^2\cos^2\alpha}\over{4M_W^2 \sin^2\beta}}
           -{1\over 4}g_Z^2(T_{3u}-s_W^2Q_u)\cos 2\alpha]
            \tilde u_L^*\tilde u_L (h^0)^2\cr 
        & - [{{g^2m_u^2\cos^2\alpha}\over{4M_W^2 \sin^2\beta}}
             -{1\over 4}g_Z^2s_W^2Q_u\cos 2\alpha]
            \tilde u_R^*\tilde u_R (h^0)^2\cr 
        & - [{{g^2m_d^2\sin^2\alpha}\over{4M_W^2 \cos^2\beta}}
         -{1\over 4}g_Z^2(T_{3d}-s_W^2Q_d)\cos 2\alpha]
                 \tilde d_L^*\tilde d_L (h^0)^2\cr 
        & - [{{g^2m_d^2\sin^2\alpha}\over{4M_W^2 \cos^2\beta}}
                  -{1\over 4}g_Z^2s_W^2Q_d\cos 2\alpha]
                 \tilde d_R^*\tilde d_R (h^0)^2.\cr 
}\eqno(C.42)
$$
The interaction (C.42) is obtained from (C.41) by the interchange
$$
     \sin\alpha \to -\cos\alpha, ~~~~~ \cos\alpha \to \sin\alpha,~~~~~
     \cos 2\alpha \to -\cos 2\alpha.
\eqno(C.42a)
$$
\noindent{\bf H$^0$-h$^0$-Sfermion-Sfermion}\par
$$
\eqalign{
    {\cal L}=& +{1\over 2}g_Z^2(T_{3\nu})\sin 2\alpha 
                 \tilde \nu^*\tilde \nu H^0h^0\cr
             & + [{{g^2m_e^2\sin 2\alpha}\over{4M_W^2 \cos^2\beta}}
                  +{1\over 2}g_Z^2(T_{3e}-s_W^2Q_e )\sin 2\alpha]
                 \tilde e_L^*\tilde e_L H^0h^0\cr 
             & + [{{g^2m_e^2\sin 2\alpha}\over{4M_W^2 \cos^2\beta}}
                  +{1\over 2}g_Z^2s_W^2Q_e\sin 2\alpha]
                 \tilde e_R^*\tilde e_R H^0h^0\cr 
}$$
$$
\eqalign{
   ~~~~~~ & - [{{g^2m_u^2\sin 2\alpha}\over{4M_W^2 \sin^2\beta}}
           -{1\over 2}g_Z^2(T_{3u}-s_W^2Q_u)\sin 2\alpha]
            \tilde u_L^*\tilde u_L H^0h^0\cr 
        & - [{{g^2m_u^2\sin 2\alpha}\over{4M_W^2 \sin^2\beta}}
             -{1\over 2}g_Z^2s_W^2Q_u\sin 2\alpha]
            \tilde u_R^*\tilde u_R H^0h^0\cr 
        & + [{{g^2m_d^2\sin 2\alpha}\over{4M_W^2 \cos^2\beta}}
         +{1\over 2}g_Z^2(T_{3d}-s_W^2Q_d)\sin 2\alpha]
                 \tilde d_L^*\tilde d_L H^0h^0\cr 
        & + [{{g^2m_d^2\sin 2\alpha}\over{4M_W^2 \cos^2\beta}}
                  +{1\over 2}g_Z^2s_W^2Q_d\sin 2\alpha]
                 \tilde d_R^*\tilde d_R H^0h^0,\cr 
}\eqno(C.43)
$$
where (C.43) is obtained from (C.41) by the following replacement,
$$
     \cos 2\alpha \to -\sin 2\alpha, ~~~~~ \sin^2\alpha \to \sin 2\alpha,
     ~~~~~\cos^2\alpha \to -\sin 2\alpha.
\eqno(C.43a)
$$
\noindent{\bf Sfermion-Sfermion-A$^0$-A$^0$}\par
$$
\eqalign{
    {\cal L}=&+{1\over 4} g_Z^2(T_{3\nu}) \cos 2\beta \tilde \nu^*\tilde \nu (A^0)^2 \cr
    &- [{{g^2m_e^2}\over {4M_W^2}}\tan^2\beta 
        -{1\over 4}g_Z^2 (T_{3e}-s_W^2Q_e)\cos 2\beta]
        \tilde e^*_L\tilde e_L (A^0)^2\cr
    & - [{{g^2m_e^2}\over{4M_W^2}}\tan^2\beta
         -{1\over 4}g_Z^2s_W^2Q_e\cos 2\beta]
                 \tilde e_R^*\tilde e_R (A^0)^2\cr 
}$$
$$
\eqalign{
   ~~~~~~ & - [{{g^2m_u^2}\over{4M_W^2}}\cot^2\beta
           -{1\over 4}g_Z^2(T_{3u}-s_W^2Q_u)\cos 2\beta]
            \tilde u_L^*\tilde u_L (A^0)^2\cr 
        & - [{{g^2m_u^2}\over{4M_W^2}}\cot^2\beta
             -{1\over 4}g_Z^2s_W^2Q_u\cos 2\beta]
            \tilde u_R^*\tilde u_R (A^0)^2\cr 
        & - [{{g^2m_d^2}\over{4M_W^2}}\tan^2\beta
         -{1\over 4}g_Z^2(T_{3d}-s_W^2Q_d)\cos 2\beta]
                 \tilde d_L^*\tilde d_L (A^0)^2\cr 
        & - [{{g^2m_d^2}\over{4M_W^2}}\tan^2\beta
                  -{1\over 4}g_Z^2s_W^2Q_d\cos 2\beta]
                 \tilde d_R^*\tilde d_R (A^0)^2.\cr 
}\eqno(C.44)
$$
Note that (C.44) is obtained from (C.42)) by
$$
     \sin\alpha \to \sin\beta,~~~~~~~~~\cos\alpha \to \cos\beta.
\eqno(C.44a)
$$
\noindent{\bf Sfermion-Sfermion-H$^+$-H$^-$}\par
$$
\eqalign{
   {\cal L}=& -[{{g^2m_e^2}\over {2M_W^2}}\tan^2\beta 
              +{1\over 2}g_Z^2(T_{3\nu}+s_W^2Q_e)\cos 2\beta]
              \tilde \nu^*\tilde\nu H^+H^-\cr
            &-{1\over 2}g_Z^2(T_{3e})\cos 2\beta \tilde e^*_L\tilde e_L H^+H^- \cr
            &- [{{g^2m_e^2}\over{2M_W^2}}\tan^2\beta
         -{1\over 2}g_Z^2s_W^2Q_e\cos 2\beta]
                 \tilde e_R^*\tilde e_R H^+H^-\cr 
}$$
$$
\eqalign{
   ~~~~~~~~~~ & - [{{g^2m_d^2}\over{2M_W^2}}\tan^2\beta
           +{1\over 2}g_Z^2(T_{3u}+s_W^2Q_d)\cos 2\beta]
            \tilde u_L^*\tilde u_L H^+H^-\cr 
        & - [{{g^2m_u^2}\over{2M_W^2}}\cot^2\beta
             -{1\over 2}g_Z^2s_W^2Q_u\cos 2\beta]
            \tilde u_R^*\tilde u_R H^+H^-\cr 
        & - [{{g^2m_u^2}\over{2M_W^2}}\cot^2\beta
         +{1\over 2}g_Z^2(T_{3d}+s_W^2Q_u)\cos 2\beta]
                 \tilde d_L^*\tilde d_L H^+H^-\cr 
        & - [{{g^2m_d^2}\over{2M_W^2}}\tan^2\beta
                  -{1\over 2}g_Z^2s_W^2Q_d\cos 2\beta]
                 \tilde d_R^*\tilde d_R H^+H^-.\cr 
}\eqno(C.45)
$$
\noindent{\bf Sfermion-sfermion-Higgs$^0$-H$^\pm$}\par
$$
\eqalign{
    {\cal L}=&+ {{g^2}\over{2\sqrt 2 M_W^2}}
               [m_e^2{{\cos\alpha\sin\beta}\over{\cos^2\beta}}
                -M_W^2\sin(\alpha+\beta)] \tilde \nu^*\tilde e_L H^0 H^+ \cr
             &+{{g^2}\over{2\sqrt 2 M_W^2}}     
              [m_d^2{{\cos\alpha\sin\beta}\over{\cos^2\beta}}
                +m_u^2{{\sin\alpha\cos\beta}\over{\sin^2\beta}}
                -M_W^2\sin(\alpha+\beta)] 
                 \tilde u_L^*\tilde d_L H^0 H^+ \cr
             &+ g^2{{m_um_d\cos(\alpha-\beta)}\over{\sqrt 2M_W^2\sin 2\beta}}
                 \tilde u^*_R\tilde d_R H^0H^+\cr    
}$$
$$
\eqalign{
   ~~~ &- {{g^2}\over{2\sqrt 2 M_W^2}}
               [m_e^2{{\sin\alpha\sin\beta}\over{\cos^2\beta}}
                +M_W^2\cos(\alpha+\beta)] \tilde \nu^*\tilde e_L h^0 H^+ \cr
             &-{{g^2}\over{2\sqrt 2 M_W^2}}     
              [m_d^2{{\sin\alpha\sin\beta}\over{\cos^2\beta}}
                -m_u^2{{\cos\alpha\sin\beta}\over{\sin^2\beta}}
                +M_W^2\cos(\alpha+\beta)] 
                 \tilde u_L^*\tilde d_L h^0 H^+ \cr
             &- g^2{{m_um_d\sin(\alpha-\beta)}\over{\sqrt 2M_W^2\sin 2\beta}}
                 \tilde u_R^*\tilde d_R h^0H^+\cr    
}$$
$$
\eqalign{
  &+i{{g^2}\over{2\sqrt 2 M_W^2}}
          (m_e^2\tan^2\beta +M_W^2\cos 2\beta)\tilde \nu^*\tilde e_LA^0H^+ \cr
        &+i{{g^2}\over{2\sqrt 2 M_W^2}}
           (m_d^2\tan^2\beta -m_u^2\cot^2\beta +M_W^2\cos 2\beta)
           \tilde u^*_L\tilde d_L A^0H^+ ~~~~~~~~~\cr
         & + h.c.\cr
}\eqno(C.46)
$$
\noindent{\bf Sfermion-Sfermion-H$^\pm$-G$^0$}\par
$$
\eqalign{
       {\cal L} =& -i{{g^2}\over{2\sqrt 2 M_W^2}}
                  (m_e^2\tan\beta -M_W^2\sin 2\beta) 
                  \tilde \nu^*\tilde e_L H^+G^0 \cr
                 &  -i{{g^2}\over{2\sqrt 2 M_W^2}}
                  (m_d^2\tan\beta +m_u^2\cot\beta-M_W^2\sin 2\beta) 
                  \tilde u^*_L\tilde d_L H^+G^0 \cr
                 & +i{{g^2m_um_d}\over{\sqrt 2M_W^2\sin 2\beta}}
                   \tilde u^*_R\tilde d_R H^+G^0\cr
                & + h.c.  
}\eqno(C.47)
$$
\noindent{\bf Sfermion-Sfermion-H$^\pm$-G$^\mp$}\par
$$
\eqalign{
   {\cal L} =&+ ({{g^2m_e^2}\over{2M_W^2}} \tan\beta-
              {1\over 2}g_Z^2(T_{3\nu}+s_W^2Q_e)\sin 2\beta)
               \tilde\nu^*\tilde\nu H^+G^-\cr
             &-{1\over 2}g_Z^2(T_{3e})\sin 2\beta\tilde e^*_L\tilde e_L H^+G^-\cr
             &+({{g^2m_e^2}\over{2M_W^2}} \tan\beta
               +{1\over2}g_Z^2s_W^2Q_e\sin 2\beta)
                \tilde e^*_R\tilde e_R H^+G^-\cr
}$$
$$
\eqalign{
   ~~~~~~~~~~~~ &+({{g^2m_d^2}\over{2M_W^2}} \tan\beta-
           {1\over2}g_Z^2(T_{3u}+s_W^2Q_d)\sin 2\beta)
                     \tilde u^*_L\tilde u_L H^+G^-\cr
          &-({{g^2m_u^2}\over{2M_W^2}} \cot\beta
              - {1\over 2}g_Z^2s_W^2Q_u\sin 2\beta)
                \tilde u^*_R\tilde u_R H^+G^-\cr
          &-({{g^2m_u^2}\over{2M_W^2}} \cot\beta+
           {1\over 2}g_Z^2(T_{3d}+s_W^2Q_u)\sin 2\beta)
                     \tilde d^*_L\tilde d_L H^+G^-\cr
          &+({{g^2m_d^2}\over{2M_W^2}} \tan\beta
              + {1\over 2}g_Z^2s_W^2Q_d\sin 2\beta)
                \tilde d^*_R\tilde d_R H^+G^-\cr
          & + h.c.      
}\eqno(C.48)
$$
\noindent{\bf Sfermion-Sfermion-A$^0$-G$^0$}\par
$$
\eqalign{
    {\cal L}=&+{1\over 2}g_Z^2(T_{3\nu})\sin 2\beta 
                \tilde \nu^*\tilde \nu A^0G^0 ~~\cr
             &+({{g^2m_e^2}\over{2M_W^2}}\tan\beta
                 +{1\over 2}g_Z^2 (T_{3e}-s_W^2Q_e)\sin 2\beta)
                \tilde e^*_L\tilde e_L A^0G^0  ~~\cr
             &+({{g^2m_e^2}\over{2M_W^2}}\tan\beta
                +{1\over 2}g_Z^2 s_W^2Q_e\sin 2\beta)
                 \tilde e^*_R\tilde e_R A^0G^0\cr
}$$
$$
\eqalign{
 ~~~~   &-({{g^2m_u^2}\over{2M_W^2}}\cot\beta
       -{1\over 2}g_Z^2 (T_{3u}-s_W^2Q_u)\sin 2\beta)
       \tilde u^*_L\tilde u_L A^0G^0\cr
    &-({{g^2m_u^2}\over{2M_W^2}}\cot\beta
     -{1\over 2}g_Z^2 s_W^2Q_u\sin 2\beta)
       \tilde u^*_R\tilde u_R A^0G^0\cr
    &+({{g^2m_d^2}\over{2M_W^2}}\tan\beta
       +{1\over 2}g_Z^2 (T_{3d}-s_W^2Q_d)\sin 2\beta)
       \tilde d^*_L\tilde d_L A^0G^0\cr
    &+({{g^2m_d^2}\over{2M_W^2}}\tan\beta
     +{1\over 2}g_Z^2 s_W^2Q_d\sin 2\beta)
       \tilde d^*_R\tilde d_R A^0G^0.\cr
}\eqno(C.49)
$$
\noindent{\bf Sfermion-Sfermion-Higgs$^0$-G$^\pm$}\par
$$
\eqalign{
    {\cal L}=
    &-{{g^2}\over{2\sqrt 2 M_W^2}}(m_e^2{{\cos\alpha}\over{\cos\beta}}
           -M_W^2\cos(\alpha+\beta))
           \tilde\nu^*\tilde e_L H^0G^+ ~~\cr
    &-{{g^2}\over{2\sqrt 2 M_W^2}}(m_d^2{{\cos\alpha}\over{\cos\beta}}
                 -m_u^2{{\sin\alpha}\over{\sin\beta}} 
                 -M_W^2\cos(\alpha+\beta))
            \tilde u_L^*\tilde d_L H^0G^+ ~~\cr
    & -g^2{{m_um_d\sin(\alpha-\beta)}\over{\sqrt 2 M_W^2 \sin 2\beta}}
          \tilde u_R^*\tilde d_R H^0G^+ \cr
}$$
$$
\eqalign{
    &+{{g^2}\over{2\sqrt 2 M_W^2}}(m_e^2{{\sin\alpha}\over{\cos\beta}}
           -M_W^2\sin(\alpha+\beta))
           \tilde\nu^*\tilde e_L h^0G^+ \cr
    &+{{g^2}\over{2\sqrt 2 M_W^2}}(m_d^2{{\sin\alpha}\over{\cos\beta}}
                 +m_u^2{{\cos\alpha}\over{\sin\beta}} 
                 -M_W^2\sin(\alpha+\beta))
            \tilde u_L^*\tilde d_L h^0G^+ \cr
    & -g^2{{m_um_d\cos(\alpha-\beta)}\over{\sqrt 2 M_W^2 \sin 2\beta}}
          \tilde u_R^*\tilde d_R h^0G^+ \cr
}$$
$$
\eqalign{
            & -i{{g^2}\over{2\sqrt 2 M_W^2}}
                  (m_e^2\tan\beta -M_W^2\sin 2\beta) 
                  \tilde \nu^*\tilde e_L A^0G^+ \cr
                 &  -i{{g^2}\over{2\sqrt 2 M_W^2}}
                  (m_d^2\tan\beta +m_u^2\cot\beta-M_W^2\sin 2\beta) 
                  \tilde u^*_L\tilde d_L A^0G^+ ~~~\cr
                 & -i{{g^2m_um_d}\over{\sqrt 2M_W^2\sin 2\beta}}
                   \tilde u^*_R\tilde d_R A^0G^+\cr
                 & + h.c.  
}\eqno(C.50)
$$
\noindent{\bf Sfermion-Sfermion-G$^+$-G$^-$}\par
$$\eqalign{
   {\cal L}=
   &-({{g^2m_e^2}\over{2M_W^2}} -{1\over 2}g_Z^2
      (T_{3\nu}+s_W^2Q_e)\cos 2\beta) 
      \tilde \nu^*\tilde \nu G^+G^- ~~~~~~~~~~~\cr
   &+{1\over 2}g_Z^2 (T_{3e})\cos 2\beta  \tilde e_L^*\tilde e_L G^+G^-\cr
   &-({{g^2m_e^2}\over{2M_W^2}} +{1\over 2}g_Z^2 s_W^2Q_e\cos 2\beta) 
      \tilde e_R^*\tilde e_R G^+G^-\cr
}$$
$$
\eqalign{
   &-({{g^2m_d^2}\over{2M_W^2}} -{1\over 2}g_Z^2
      (T_{3u}+s_W^2Q_d)\cos 2\beta) 
      \tilde u_L^*\tilde u_L G^+G^-\cr
   &-({{g^2m_u^2}\over{2M_W^2}} +{1\over 2}g_Z^2 s_W^2Q_u\cos 2\beta) 
      \tilde u_R^*\tilde u_R G^+G^-\cr
   &-({{g^2m_u^2}\over{2M_W^2}} -{1\over 2}g_Z^2
      (T_{3d}+s_W^2Q_u)\cos 2\beta) 
      \tilde d_L^*\tilde d_L G^+G^-\cr
   &-({{g^2m_d^2}\over{2M_W^2}} +{1\over 2}g_Z^2 s_W^2Q_d\cos 2\beta) 
      \tilde d_R^*\tilde d_R G^+G^-.\cr
}\eqno(C.51)
$$
\noindent{\bf Sfermion-Sfermion-G$^0$-G$^0$}\par
$$
\eqalign{
   {\cal L}=
   &-{1\over 4}g_Z^2 (T_{3\nu})\cos 2\beta  \tilde \nu^*\tilde \nu (G^0)^2\cr
   &-({{g^2m_e^2}\over{4M_W^2}} +{1\over 4}g_Z^2
      (T_{3e}-s_W^2Q_e)\cos 2\beta) 
      \tilde e_L^*\tilde e_L(G^0)^2 ~~~~~~~~~\cr
   &-({{g^2m_e^2}\over{4M_W^2}} +{1\over 4}g_Z^2 s_W^2Q_e\cos 2\beta) 
      \tilde e_R^*\tilde e_R (G^0)^2\cr
}$$
$$
\eqalign{
   &-({{g^2m_u^2}\over{4M_W^2}} +{1\over 4}g_Z^2
      (T_{3u}-s_W^2Q_u)\cos 2\beta) 
      \tilde u_L^*\tilde u_L (G^0)^2\cr
   &-({{g^2m_u^2}\over{4M_W^2}} +{1\over 4}g_Z^2 s_W^2Q_u\cos 2\beta) 
      \tilde u_R^*\tilde u_R (G^0)^2\cr
   &-({{g^2m_d^2}\over{4M_W^2}} +{1\over 4}g_Z^2
      (T_{3d}-s_W^2Q_d)\cos 2\beta) 
      \tilde d_L^*\tilde d_L (G^0)^2\cr
   &-({{g^2m_d^2}\over{4M_W^2}} +{1\over 4}g_Z^2 s_W^2Q_d\cos 2\beta) 
      \tilde d_R^*\tilde d_R (G^0)^2.\cr
}\eqno(C.52)
$$
\noindent{\bf Sfermion-Sfermion-G$^0$-G$^\pm$}\par
$$
\eqalign{
   {\cal L}=
     &+i{{g^2}\over{2\sqrt 2 M_W^2}}(m_e^2-M_W^2\cos 2\beta)
      (\tilde\nu^*\tilde e_LG^0G^+ - \tilde e_L^*\tilde \nu G^0G^-)\cr
     &+i{{g^2}\over{2\sqrt 2 M_W^2}}(m_d^2-m_u^2-M_W^2\cos 2\beta)
      (\tilde u_L^*\tilde d_LG^0G^+ - \tilde d_L^*\tilde u G^0G^-).\cr
}\eqno(C.53)
$$
\noindent{\bf Four-slepton interactions (same generation)}\par
$$
\eqalign{
    {\cal L} =
    &-{{g_Z^2}\over 8} [\tilde\nu^*\tilde\nu\tilde\nu^*\tilde\nu
                      +2\tilde\nu^*\tilde\nu\tilde e^*_L\tilde e_L 
                      + \tilde e^*_L\tilde e_L\tilde e^*_L\tilde e_L 
       +4s_W^2 \tilde e^*_R\tilde e_R\tilde e^*_R\tilde e_R ]\cr
    & + ({1\over 2}g_Z^2 s_W^2 -{{g^2m_e^2}\over{2M_W^2\cos^2\beta}})
                     [ \tilde\nu^*\tilde\nu  \tilde e^*_R\tilde e_R 
                      +\tilde e^*_L\tilde e_L\tilde e^*_R\tilde e_R].  \cr
}\eqno(C.54)
$$
\noindent{\bf Four-slepton interactions (intergeneration)}\par
  Below I show only the interactions between the first and second 
generations.  It is  straightforward to obtain the first-third and
second-third intergeneration interactions.\par
$$
\eqalign{
     {\cal L} =
  & -{{g_Z^2}\over 4}[\tilde\nu_e^*\tilde \nu_e \tilde\nu_\mu^*\tilde \nu_\mu 
                   +  \tilde e_L^*\tilde e_L \tilde \mu_L^*\tilde \mu_L  
                   +2 \cos^2\theta_W\tilde\nu_e^*\tilde e_L \tilde \mu_L^* \tilde\nu_\mu 
                   +2 \cos^2\theta_W\tilde e_L^*\tilde\nu_e \tilde\nu_\mu^* \tilde\mu_L]\cr 
  & +{{g_Z^2}\over 2}\sin^2\theta_W
        [\tilde\nu_e^*\tilde \nu_e \tilde\mu_R^*\tilde \mu_R 
        +\tilde\nu_\mu^*\tilde \nu_\mu \tilde e_R^*\tilde e_R 
        +\tilde e_L^*\tilde e_L \tilde \mu_R^*\tilde \mu_R  
        +\tilde e_R^*\tilde e_R \tilde \mu_L^*\tilde \mu_L  
       -2\tilde e_R^*\tilde e_R \tilde \mu_R^*\tilde \mu_R]\cr
  & +g_Z^2 ({1\over 4}-{1\over 2}\sin^2\theta_W) 
        [\tilde\nu_e^*\tilde \nu_e \tilde\mu_L^*\tilde \mu_L 
        +\tilde\nu_\mu^*\tilde \nu_\mu \tilde e_L^*\tilde e_L]\cr
  & -{{g^2 m_em_\mu}\over {2M_W^2\cos^2\beta}}
        [ \tilde\nu_e^*\tilde e_R \tilde \mu_R^* \tilde\nu_\mu 
         +\tilde e_R^*\tilde\nu_e \tilde\nu_\mu^* \tilde\mu_R 
        +\tilde e_R^*\tilde e_L \tilde \mu_L^*\tilde \mu_R  
        +\tilde e_L^*\tilde e_R \tilde \mu_R^*\tilde \mu_L].\cr
}\eqno(C.55)
$$
\noindent{\bf Two-slepton-Two-squark (same generation)}\par
$$
\eqalign{
    {\cal L} =
    &-g_Z^2({1\over 4}-{1\over 2}s_W^2Q_u)
          [\tilde\nu^*\tilde\nu \tilde u_L^*\tilde u_L
          +\tilde e_L^*\tilde e_L \tilde d_L^* \tilde d_L]\cr
    &-{1\over 2}g_Z^2 s_W^2Q_u
          [\tilde\nu^*\tilde\nu \tilde u_R^*\tilde u_R
          +\tilde e_L^*\tilde e_L \tilde u_R^* \tilde u_R
          -2\tilde e_R^*\tilde e_R \tilde u_R^* \tilde u_R]\cr
    &+g_Z^2({1\over 4}+{1\over 2}s_W^2Q_d)
          [\tilde\nu^*\tilde\nu \tilde d_L^*\tilde d_L
          +\tilde e_L^*\tilde e_L \tilde u_L^* \tilde u_L]\cr
    &-{1\over 2}g_Z^2 s_W^2Q_d
          [\tilde\nu^*\tilde\nu \tilde d_R^*\tilde d_R
          +\tilde e_L^*\tilde e_L \tilde d_R^* \tilde d_R
         -2\tilde e_R^*\tilde e_R \tilde d_R^* \tilde d_R]\cr
}$$
$$
\eqalign{
  ~~~~ & -g_Z^2(Q_u-{1\over 2})s_W^2 
            \tilde e_R^* \tilde e_R\tilde u_L^*\tilde u_L
         -g_Z^2(Q_d+{1\over 2})s_W^2 
            \tilde e_R^* \tilde e_R\tilde d_L^*\tilde d_L]\cr
       & -{1\over 2} g^2[\tilde \nu^*\tilde e_L\tilde d^*_L\tilde u_L 
                       + \tilde e_L^*\tilde \nu\tilde u_L^*\tilde d_L]\cr
       &-{{g^2m_em_d}\over{2M_W^2\cos^2\beta}}
               [\tilde \nu^*\tilde e_R\tilde d^*_R\tilde u_L 
              + \tilde e_R^*\tilde \nu\tilde u_L^*\tilde d_R
              + \tilde e_L^* \tilde e_R\tilde d_R^*\tilde d_L
              + \tilde e_R^* \tilde e_L\tilde d_L^*\tilde d_R].\cr
}\eqno(C.56)
$$
\noindent{\bf Two-slepton-Two-squark (intergeneration)}\par
     Intergeneration interaction is obtained by either of the two ways:
(1) replacement in the lepton sector, $\tilde \nu_e \to \tilde \nu_\mu$,
$\tilde e \to \tilde \mu$ and $m_e \to m_\mu$;
(2) replacement in the quark sector, $\tilde u \to \tilde c$,
$\tilde d \to \tilde s$ and $m_d \to m_s$.\par
\medskip
\noindent{\bf Four-squark interactions (same generation)}\par
$$
\eqalign{
    {\cal L}=
    & -[{1\over 6}g_s^2+{1\over 2}g_Z^2({1\over 4}+s_W^2Q_uQ_d)]
            [(\tilde u_L^*\tilde u_L)^2 + (\tilde d_L^*\tilde d_L)^2]\cr
    & +[{1\over 6}g_s^2+g_Z^2({1\over 4}-(Q_uQ_d+{1\over 2})s_W^2)]
            (\tilde u_L^*\tilde u_L)(\tilde d_L^*\tilde d_L)\cr    
    & -[{1\over 6}g_s^2+{1\over 2}g_Z^2s_W^2Q_u^2]
            (\tilde u_R^*\tilde u_R)^2\cr
    & -[{1\over 6}g_s^2+{1\over 2}g_Z^2 s_W^2Q_d^2]
            (\tilde d_R^*\tilde d_R)^2\cr
}$$
$$
\eqalign{
    & -[{1\over 6}g_s^2-g_Z^2 s_W^2Q_u(Q_u-{1\over 2})]
            (\tilde u_L^*\tilde u_L)(\tilde u_R^*\tilde u_R)\cr
    & -[{1\over 6}g_s^2-g_Z^2 s_W^2Q_d(Q_d+{1\over 2})]
            (\tilde d_L^*\tilde d_L)(\tilde d_R^*\tilde d_R)\cr
    & -[{1\over 6}g_s^2-g_Z^2 s_W^2Q_d(Q_u-{1\over 2})]
            (\tilde u_L^*\tilde u_L)(\tilde d_R^*\tilde d_R)\cr
    & -[{1\over 6}g_s^2-g_Z^2 s_W^2Q_u(Q_d+{1\over 2})]
            (\tilde d_L^*\tilde d_L)(\tilde u_R^*\tilde u_R)~~~~~~~~\cr
}$$
$$
\eqalign{
  ~~~~&+[{1\over 6}g_s^2-g_Z^2s_W^2Q_uQ_d] 
         (\tilde d_R^*\tilde d_R)(\tilde u_R^*\tilde u_R)  \cr
      & +[{1\over 2}g_s^2-{{g^2m_u^2}\over {2M_W^2\sin^2\beta}}]
         [(\tilde u_R^*\tilde u_L)(\tilde u_L^*\tilde u_R)
          +(\tilde u_R^*\tilde d_L)(\tilde d_L^*\tilde u_R)]\cr
      & +[{1\over 2}g_s^2-{{g^2m_d^2}\over {2M_W^2\cos^2\beta}}]
         (\tilde d_R^*\tilde d_L)(\tilde d_L^*\tilde d_R)
        +(\tilde d_R^*\tilde u_L)(\tilde u_L^*\tilde d_R)]\cr
      & -{1\over 2}(g_s^2+g^2)
         (\tilde u_L^*\tilde d_L)(\tilde d_L^*\tilde u_L)
        -{1\over 2}g_s^2
         (\tilde u_R^*\tilde d_R)(\tilde d_R^*\tilde u_R).\cr
}\eqno(C.57)
$$
\noindent{\bf Four-quark interactions (intergeneration)}\par
$$
\eqalign{
    {\cal L}=
    & +[{1\over 6}g_s^2-g_Z^2({1\over 4}+s_W^2Q_uQ_d)]
       (\tilde u_L^*\tilde u_L)(\tilde c_L^*\tilde c_L)
      -{1\over 2}g_s^2  
       (\tilde u_L^*\tilde c_L)(\tilde c_L^*\tilde u_L) \cr
    & +[{1\over 6}g_s^2-g_Z^2 s_W^2Q_u^2]
       (\tilde u_R^*\tilde u_R)(\tilde c_R^*\tilde c_R)
      -{1\over 2}g_s^2  
       (\tilde u_R^*\tilde c_R)(\tilde c_R^*\tilde u_R) \cr
    & -[{1\over 6}g_s^2-g_Z^2 s_W^2Q_u(Q_u-{1\over 2})]
       [(\tilde u_L^*\tilde u_L)(\tilde c_R^*\tilde c_R)
       +(\tilde u_R^*\tilde u_R)(\tilde c_L^*\tilde c_L)]\cr
    & +{1\over 2}g_s^2  
       [(\tilde u_L^*\tilde c_R)(\tilde c_R^*\tilde u_L) 
       +(\tilde u_R^*\tilde c_L)(\tilde c_L^*\tilde u_R)] \cr
    & -{{g^2m_um_c}\over{2M_W^2\sin^2\beta}}
       [(\tilde u_R^*\tilde u_L)(\tilde c_L^*\tilde c_R)
       +(\tilde u_L^*\tilde u_R)(\tilde c_R^*\tilde c_L)]\cr
}$$
$$
\eqalign{
    & +[{1\over 6}g_s^2-g_Z^2({1\over 4}+s_W^2Q_uQ_d)]
       (\tilde d_L^*\tilde d_L)(\tilde s_L^*\tilde s_L)
      -{1\over 2}g_s^2  
       (\tilde d_L^*\tilde s_L)(\tilde s_L^*\tilde d_L) \cr
    & +[{1\over 6}g_s^2-g_Z^2 s_W^2Q_d^2]
       (\tilde d_R^*\tilde d_R)(\tilde s_R^*\tilde s_R)
      -{1\over 2}g_s^2  
       (\tilde d_R^*\tilde s_R)(\tilde s_R^*\tilde d_R) \cr
    & -[{1\over 6}g_s^2-g_Z^2 s_W^2Q_d(Q_d+{1\over 2})]
       [(\tilde d_L^*\tilde d_L)(\tilde s_R^*\tilde s_R)
       +(\tilde d_R^*\tilde d_R)(\tilde s_L^*\tilde s_L)]\cr
    & +{1\over 2}g_s^2  
       [(\tilde d_L^*\tilde s_R)(\tilde s_R^*\tilde d_L) 
       +(\tilde d_R^*\tilde s_L)(\tilde s_L^*\tilde d_R)] \cr
    & -{{g^2m_dm_s}\over{2M_W^2\cos^2\beta}}
       [(\tilde d_R^*\tilde d_L)(\tilde s_L^*\tilde s_R)
       +(\tilde d_L^*\tilde d_R)(\tilde s_R^*\tilde s_L)]\cr
}$$
$$
\eqalign{
~~~~~~& +[{1\over 6}g_s^2+g_Z^2({1\over 4}-(Q_uQ_d+{1\over 2})s_W^2)]
       (\tilde u_L^*\tilde u_L)(\tilde s_L^*\tilde s_L)
      -{1\over 2}g_s^2  
       (\tilde u_L^*\tilde s_L)(\tilde s_L^*\tilde u_L) \cr
    & +[{1\over 6}g_s^2-g_Z^2 s_W^2Q_uQ_d]
       (\tilde u_R^*\tilde u_R)(\tilde s_R^*\tilde s_R)
      -{1\over 2}g_s^2  
       (\tilde u_R^*\tilde s_R)(\tilde s_R^*\tilde u_R) \cr
    & -[{1\over 6}g_s^2-g_Z^2 s_W^2Q_d(Q_d+{1\over 2})]
       (\tilde u_L^*\tilde u_L)(\tilde s_R^*\tilde s_R)
      +{1\over 2}g_s^2  
       (\tilde u_L^*\tilde s_R)(\tilde s_R^*\tilde u_L) \cr
    & -[{1\over 6}g_s^2-g_Z^2 s_W^2Q_u(Q_u-{1\over 2})]
       (\tilde u_R^*\tilde u_R)(\tilde s_L^*\tilde s_L)
      +{1\over 2}g_s^2  
       (\tilde u_R^*\tilde s_L)(\tilde s_L^*\tilde u_R) \cr
}$$
$$
\eqalign{
~~~~~~& +[{1\over 6}g_s^2+g_Z^2({1\over 4}-(Q_uQ_d+{1\over 2})s_W^2)]
       (\tilde d_L^*\tilde d_L)(\tilde c_L^*\tilde c_L)
      -{1\over 2}g_s^2  
       (\tilde d_L^*\tilde c_L)(\tilde c_L^*\tilde d_L) \cr
    & +[{1\over 6}g_s^2-g_Z^2 s_W^2Q_uQ_d]
       (\tilde d_R^*\tilde d_R)(\tilde c_R^*\tilde c_R)
      -{1\over 2}g_s^2  
       (\tilde d_R^*\tilde c_R)(\tilde c_R^*\tilde d_R) \cr
    & -[{1\over 6}g_s^2-g_Z^2 s_W^2Q_u(Q_u-{1\over 2})]
       (\tilde d_L^*\tilde d_L)(\tilde c_R^*\tilde c_R)
      +{1\over 2}g_s^2  
       (\tilde d_L^*\tilde c_R)(\tilde c_R^*\tilde d_L) \cr
    & -[{1\over 6}g_s^2-g_Z^2 s_W^2Q_d(Q_d+{1\over 2})]
       (\tilde d_R^*\tilde d_R)(\tilde c_L^*\tilde c_L)
      +{1\over 2}g_s^2  
       (\tilde d_R^*\tilde c_L)(\tilde c_L^*\tilde d_R) \cr
}$$
$$
\eqalign{
       &-{1\over 2}g^2[(\tilde u_L^*\tilde d_L)(\tilde s_L^*\tilde c_L)
                      +(\tilde d_L^*\tilde u_L)(\tilde c_L^*\tilde s_L)]\cr
       &-{{g^2m_um_c}\over{2M_W^2\sin^2\beta}}               
              [(\tilde u_R^*\tilde d_L)(\tilde s_L^*\tilde c_R)
              +(\tilde d_L^*\tilde u_R)(\tilde c_R^*\tilde s_L)]
              ~~~~~~~~~~~~\cr
       &-{{g^2m_dm_s}\over{2M_W^2\cos^2\beta}}               
              [(\tilde u_L^*\tilde d_R)(\tilde s_R^*\tilde c_L)
              +(\tilde d_R^*\tilde u_L)(\tilde c_L^*\tilde s_R)].
     ~~~~~~~~~~~~~~\cr
}\eqno(C.58)
$$
\noindent{\bf Gauge-boson-Gauge-boson-Gauge-boson}\par
$$
\eqalign{
  {\cal L}=& +igc_W \lbrack
         (\partial_\mu W_\nu^--\partial_\nu W_\mu^-) W^{+\mu}Z^\nu
        - (\partial_\mu W^+_\nu-\partial_\nu W^+_\mu) W^{-\mu} Z^\nu \cr
         & ~~~~~~~~~~-(\partial_\mu Z_\nu-\partial_\nu Z_\mu)W^{+\mu} W^{-\nu}
            \rbrack \cr
         & +ie\lbrack
         (\partial_\mu W_\nu^--\partial_\nu W_\mu^-) W^{+\mu}A^\nu
        - (\partial_\mu W^+_\nu-\partial_\nu W^+_\mu) W^{-\mu} A^\nu \cr
         & ~~~~~~~~~~-(\partial_\mu A_\nu-\partial_\nu A_\mu)W^{+\mu} W^{-\nu}
            \rbrack \cr
         & +g_sf^{\alpha\beta\gamma} (\partial_\mu g_\nu^\alpha) 
                                     g_\mu^\beta g_\nu^\gamma.
}\eqno(C.59)
$$
\noindent{\bf Gauge-boson-Gauge-boson-Gauge-boson-Gauge-boson}\par
$$
\eqalign{
  {\cal L}=& +{{g^2}\over 2}[ (W_\mu^+W^{+\mu})(W_\nu^-W^{-\nu})
                -(W^+_\mu W^{-\mu})^2]\cr
           & + g^2c_W^2[W^+_\mu W^-_\nu Z^\mu Z^\nu
                               - W^+_\mu W^{-\mu} Z_\nu Z^\nu]\cr
           & + g^2c_W s_W
               [W^+_\mu W^-_\nu (Z^\mu A^\nu + Z^\nu A^\mu)
                               - 2W^+_\mu W^{-\mu} Z_\nu A^\nu]\cr
           & + e^2[W^+_\mu W^-_\nu A^\mu A^\nu
                               - W^+_\mu W^{-\mu} A_\nu A^\nu]\cr
           &-{1\over 4} g_s^2 f^{\alpha\beta\gamma}f^{\rho\sigma\gamma}
                        g_\mu^\alpha g_\nu^\beta g_\mu^\rho g_\nu^\sigma.
}\eqno(C.60)
$$
\noindent{\bf Higgs$^0$-Higgs$^0$-Higgs$^0$}\par
$$
\eqalign{
    {\cal L} =
      & -{{g_Z}\over 4}M_Z \cos 2\alpha
         [\sin(\alpha+\beta)(h^0)^3 + \cos(\alpha+\beta)(H^0)^3]\cr
      & -{{g_Z}\over 4}M_Z \cos 2\beta
         [\sin(\alpha+\beta)h^0 - \cos(\alpha+\beta)H^0](A^0)^2\cr
      & -{{g_Z}\over 4}M_Z 
         [2\sin 2\alpha\sin(\alpha+\beta)- \cos 2\alpha\cos(\alpha+\beta)]
         (h^0)^2 H^0 \cr
      & +{{g_Z}\over 4}M_Z 
         [2\sin 2\alpha\cos(\alpha+\beta)+ \cos 2\alpha\sin(\alpha+\beta)]
         (H^0)^2 h^0. \cr
}\eqno(C.61)
$$    
\noindent{\bf H$^+$-H$^-$-Higgs$^0$}\par
$$
\eqalign{
    {\cal L} =
        &-[ gM_W \cos(\alpha-\beta)
           -{{g_Z}\over 2}M_Z\cos(\alpha+\beta)\cos 2\beta] H^+H^-H^0 \cr
        &+[ gM_W \sin(\alpha-\beta)
           -{{g_Z}\over 2}M_Z\sin(\alpha+\beta)\cos 2\beta] H^+H^-h^0. \cr
}\eqno(C.62)
$$    
\noindent{\bf Higgs$^0$-Higgs$^0$-Higgs$^0$-Higgs$^0$}\par
$$
\eqalign{
    {\cal L} =
      &-{{g_Z^2}\over {32}}\cos^2 2\alpha (h^0)^4 
       -{{g_Z^2}\over {16}}\sin 4\alpha (h^0)^3H^0 \cr 
      &-{{g_Z^2}\over {16}}(3\sin^2 2\alpha -1)(h^0)^2(H^0)^2 
       +{{g_Z^2}\over {16}}\sin 4\alpha h^0(H^0)^3 \cr 
      &-{{g_Z^2}\over {32}}\cos^2 2\alpha (H^0)^4 
       -{{g_Z^2}\over {32}}\cos^2 2\beta (A^0)^4 \cr 
      &-{{g_Z^2}\over {16}}\cos 2\alpha\cos 2\beta(h^0)^2(A^0)^2 
       -{{g_Z^2}\over    8}\sin 2\alpha\cos 2\beta h^0H^0(A^0)^2 \cr 
      &+{{g_Z^2}\over {16}}\cos 2\alpha \cos 2\beta (H^0)^2(A^0)^2.  
}\eqno(C.63)
$$
\noindent{\bf H$^+$-H$^-$-Higgs$^0$-Higgs$^0$}\par
$$
\eqalign{
    {\cal L} =
    &-[  {{g^2}\over 4}\sin^2(\alpha-\beta) 
       + {{g_Z^2}\over 8} \cos 2\alpha \cos 2\beta]
          H^+H^- (h^0)^2  \cr
    &+[ {{g^2}\over 2}\sin(\alpha-\beta)\cos(\alpha-\beta) 
       -{{g_Z^2}\over 4} \sin 2\alpha \cos 2\beta] H^+H^- h^0H^0 \cr
    &-[ {{g^2}\over 4}\cos^2(\alpha-\beta) 
       -{{g_Z^2}\over 8} \cos 2\alpha \cos 2\beta] H^+H^- (H^0)^2 \cr
    & -{{g_Z^2}\over 8} \cos^2 2\beta H^+H^-(A^0)^2.
}\eqno(C.64)
$$    
\noindent{\bf H$^+$-H$^-$-H$^+$-H$^-$}\par
$$
    {\cal L} =-{{g_Z^2}\over 8} \cos^2 2\beta (H^+)^2(H^-)^2.
\eqno(C.65)
$$
\noindent{\bf Higgs-Higgs-Goldstone}\par
$$
\eqalign{
    {\cal L} =&
     +{{g_Z}\over 2} M_Z \sin 2\beta
     [\cos(\alpha+\beta)H^0 -\sin(\alpha+\beta)h^0]A^0G^0 \cr
    &+{1\over 2}(gM_W\sin(\alpha-\beta)+g_ZM_Z\cos(\alpha+\beta)\sin 2\beta)
       (H^0H^+G^- + H^0H^-G^+) \cr
    &+{1\over 2}(gM_W\cos(\alpha-\beta) -g_ZM_Z\sin(\alpha+\beta)\sin 2\beta)
       (h^0H^+G^- + h^0H^-G^+) \cr
    &+i{1\over 2}gM_W (A^0H^-G^+ - A^0H^+G^-).
}\eqno(C.66)
$$
\noindent{\bf Higgs-Goldstone-Goldstone}\par
$$
\eqalign{
    {\cal L} =
    & -  {1\over 4} g_ZM_Z\cos(\alpha+\beta)\cos 2\beta 
      [H^0(G^0)^2 + 2H^0G^+G^-]\cr
    & + {1\over 4} g_ZM_Z\sin(\alpha+\beta)\cos 2\beta 
      [h^0(G^0)^2 + 2h^0G^+G^-].\cr
}\eqno(C.67)
$$       
\noindent{\bf Higgs-Higgs-Higgs-Goldstone}\par
$$
\eqalign{
    {\cal L} =
    &+{{g_Z^2}\over 8}\sin 2\beta[\cos 2\alpha ((H^0)^2-(h^0)^2)
              -2\sin 2\alpha H^0h^0]A^0G^0 \cr
    &-{{g_Z^2}\over{16}}\sin 4\beta [(A^0)^3G^0 +2H^+H^-A^0G^0]\cr
    &+{1\over 8}(g^2\sin 2(\alpha-\beta) + g_Z^2\cos 2\alpha \sin 2\beta)
      [H^+G^- +H^-G^+][(H^0)^2-(h^0)^2 ]\cr         
    &+{1\over 4}(g^2\cos 2(\alpha-\beta) - g_Z^2\sin 2\alpha \sin 2\beta)
      [H^+G^- +H^-G^+]H^0h^0 \cr
}$$
$$
\eqalign{
    &-{{g_Z^2}\over {16}} \sin 4\beta [H^+G^- + H^-G^+](A^0)^2\cr
    &-i{{g^2}\over 4}[\cos(\alpha-\beta)H^0-\sin(\alpha-\beta)h^0]
                     [A^0H^+G^- - A^0H^-G^+]~~~~~~~~\cr
    &-{{g_Z^2}\over 8} \sin 4\beta [(H^+)^2H^-G^- + (H^-)^2H^+G^+]. 
}\eqno(C.68)
$$
\noindent{\bf Higgs-Higgs-Goldstone-Goldstone}\par
$$
\eqalign{
    {\cal L} =
     &+{{g_Z^2}\over {16}} [\cos 2\alpha \cos 2\beta ((h^0)^2- (H^0)^2)
                            +2\sin 2\alpha \cos 2\beta H^0h^0] (G^0)^2 \cr
     &-{{g_Z^2}\over {16}} (3 \sin^2 2\beta -1) (A^0)^2 (G^0)^2 \cr
     &-({{g^2}\over 4}-{{g_Z^2}\over 8}\cos^2 2\beta)H^+H^-(G^0)^2 \cr
     &-i{{g^2}\over 4}[\sin(\alpha-\beta)H^0+\cos(\alpha-\beta)h^0] 
                   [H^-G^+G^0-H^+G^0G^-]\cr
}$$
$$
\eqalign{
   ~~&+{1\over 4}(g^2-g_Z^2\sin^2 2\beta) (A^0H^-G^+G^0 + A^0H^+G^-G^0)\cr
       &-{1\over 4}(g^2\sin^2(\alpha-\beta) + {1\over 2}g_Z^2\cos 2\alpha
          \cos 2\beta)(H^0)^2G^+G^-\cr
       &-{1\over 4}(g^2\cos^2(\alpha-\beta) - {1\over 2}g_Z^2\cos 2\alpha
          \cos 2\beta)(h^0)^2G^+G^-\cr
       &-{1\over 4}(g^2\sin 2(\alpha-\beta) -g_Z^2\sin 2\alpha
          \cos 2\beta)H^0h^0G^+G^-\cr
}$$
$$
\eqalign{
   ~~~~~~&-({{g^2}\over 4}-{{g_Z^2}\over 8}\cos^2 2\beta)(A^0)^2G^+G^- 
        +{{g_Z^2}\over 4}\cos 4\beta H^+H^- G^+G^-\cr
       &-{{g_Z^2}\over 8}\sin^2 2\beta[(H^+)^2(G^-)^2 + (H^-)^2(G^+)^2].\cr
}\eqno(C.69)
$$
\noindent{\bf Higgs-Goldstone-Goldstone-Goldstone}\par
$$
\eqalign{
    {\cal L} =& {{g_Z^2}\over{16}}\sin 4\beta 
              [(G^0)^3A^0+ (G^0)^2G^+H^- + (G^0)^2G^-H^+  \cr
              &~~~~~~~~~~+ 2G^+G^-G^0A^0+2(G^-)^2G^+H^+ + 2 (G^+)^2G^-H^-].
}\eqno(C.70)
$$
\noindent{\bf Goldstone-Goldstone-Goldstone-Goldstone}\par
$$
    {\cal L} = -{{g_Z^2}\over{32}}\cos^2 2\beta
               [(G^0)^4 + 4 G^+G^-(G^0)^2 + 4(G^+)^2 (G^-)^2].
\eqno(C.71)
$$
\noindent{\bf Ghost-Ghost-Gauge-boson}\par
$$
\eqalign{
   {\cal L}=
   &~igc[\partial^\mu\bar\omega_+\omega_+ -\partial^\mu\bar\omega_-\omega_-]
     Z_\mu \cr
   &+ie[\partial^\mu\bar\omega_+\omega_+ -\partial^\mu\bar\omega_-\omega_-]
     A_\mu \cr
   &+ig[c(\partial^\mu\bar\omega_z\omega_- -\partial^\mu\bar\omega_+\omega_z)
      +s(\partial^\mu\bar\omega_\gamma\omega_- 
          -\partial^\mu\bar\omega_+\omega_\gamma)]W^+_\mu\cr
   &+ig[c(\partial^\mu\bar\omega_-\omega_z -\partial^\mu\bar\omega_z\omega_+)
       +s(\partial^\mu\bar\omega_-\omega_\gamma 
          -\partial^\mu\bar\omega_\gamma\omega_+)]W^-_\mu\cr
   &+ g_sf^{\alpha\beta\gamma}
     \partial\bar\omega^\alpha \omega^\beta g_\mu^\gamma. \cr 
}\eqno(C.72)
$$
\noindent{\bf Ghost-Ghost-Higgs}\par
$$
\eqalign{
  {\cal L}=
   &-{1\over 2}[gM_W\xi_W(\bar\omega_+\omega_+ + \bar\omega_-\omega_-)
                + g_ZM_Z\xi_Z\bar\omega_z\omega_z]\cr
   &~~~~~~[\cos(\beta-\alpha)H^0+\sin(\beta-\alpha)h^0].  \cr
}\eqno(C.73)
$$
\noindent{\bf Ghost-Ghost-Goldstone}\par
$$
\eqalign{
    {\cal L} =   
   &  -i{g\over 2}M_W\xi_W [\bar\omega_+\omega_+-\bar\omega_-\omega_-]G^0 \cr
   &  -({1\over 2}-s^2_W)g_Z M_W\xi_W[\bar\omega_+\omega_z G^+
                                          +\bar\omega_-\omega_z G^-]\cr
   &-eM_W\xi_W
       [\bar\omega_+\omega_\gamma G^+ + \bar\omega_-\omega_\gamma G^-]\cr
   &   + {g\over 2} M_Z\xi_Z[\bar\omega_z\omega_+G^- 
                           + \bar\omega_z\omega_- G^+].\cr
}\eqno(C.74)
$$

\vskip 2 truecm

\noindent{\bf Appendix D. Sfermion expansion}\par
    In Appendix C, most of the interactions of
sfermions are expressed in terms of the $f_L$ and $f_R$, 
which are not mass eigenstates.  In this Appendix, using
sfermions in the first generation, $\tilde \nu$, $\tilde e_i$, 
$\tilde u_i$ and $\tilde d_j ( i=1,2)$ as the representative, the 
product of two sfermions and of four sfermions are
expanded by the  mass eigenstates $f_1$ and $f_2$.\par
    The mass eigenstates of sfermions are defined  by (2.6).
Using the reversed expression, 
$$
    \left\lgroup\matrix{ \tilde f_L \cr \tilde f_R   }\right\rgroup =  
    \left\lgroup\matrix{ c_f  & -s_f  \cr s_f & c_f\cr}\right\rgroup
    \left\lgroup\matrix{ \tilde f_1 \cr \tilde f_2   }\right\rgroup, 
\eqno(D.1)
$$
with $c_f=\cos \theta_f$ and $s_f=\sin\theta_f$ for $f=e,u,d$,   
one finds that the products of two sfermions are decomposed as
$$
\eqalign{
   (\tilde \nu^* \tilde e_L) =&~ c_e \tilde \nu^* \tilde e_1 
                              - s_e \tilde \nu^* \tilde e_2,\cr
   (\tilde \nu^* \tilde e_R) =&~ s_e \tilde \nu^* \tilde e_1 
                              + c_e \tilde \nu^* \tilde e_2,\cr
   (\tilde u^*_L \tilde u_L) =&~ c_u^2~\tilde u^*_1 \tilde u_1 
                               -c_us_u~\tilde u^*_1 \tilde u_2 
                               -s_uc_u~\tilde u^*_2 \tilde u_1
                               +s_u^2~\tilde u^*_2 \tilde u_2, \cr
   (\tilde u^*_R \tilde u_R) =&~ s_u^2~\tilde u^*_1 \tilde u_1 
                               +s_uc_u~\tilde u^*_1 \tilde u_2 
                               +c_us_u~\tilde u^*_2 \tilde u_1
                               +c_u^2~\tilde u^*_2 \tilde u_2, \cr
   (\tilde u^*_L \tilde u_R) =&~ c_us_u~\tilde u^*_1 \tilde u_1 
                               +c_u^2~\tilde u^*_1 \tilde u_2 
                               -s_u^2~\tilde u^*_2 \tilde u_1
                               -s_uc_u~\tilde u^*_2 \tilde u_2, \cr
   (\tilde u^*_R \tilde u_L) =&~ s_uc_u~\tilde u^*_1 \tilde u_1 
                               -s_u^2~\tilde u^*_1 \tilde u_2 
                               +c_u^2~\tilde u^*_2 \tilde u_1
                               -c_us_u~\tilde u^*_2 \tilde u_2,\cr 
   (\tilde d^*_L \tilde u_L) =&~ c_dc_u~\tilde d^*_1 \tilde u_1 
                               -c_ds_u~\tilde d^*_1 \tilde u_2 
                               -s_dc_u~\tilde d^*_2 \tilde u_1
                               +s_ds_u~\tilde d^*_2 \tilde u_2, \cr
   (\tilde d^*_R \tilde u_R) =&~ s_ds_u~\tilde d^*_1 \tilde u_1 
                               +s_dc_u~\tilde d^*_1 \tilde u_2 
                               +c_ds_u~\tilde d^*_2 \tilde u_1
                               +c_dc_u~\tilde d^*_2 \tilde u_2, \cr
   (\tilde d^*_L \tilde u_R) =&~ c_ds_u~\tilde d^*_1 \tilde u_1 
                               +c_dc_u~\tilde d^*_1 \tilde u_2 
                               -s_ds_u~\tilde d^*_2 \tilde u_1
                               -s_dc_u~\tilde d^*_2 \tilde u_2, \cr
   (\tilde d^*_R \tilde u_L) =&~ s_dc_u~\tilde d^*_1 \tilde u_1 
                               -s_ds_u~\tilde d^*_1 \tilde u_2 
                               +c_dc_u~\tilde d^*_2 \tilde u_1
                               -c_ds_u~\tilde d^*_2 \tilde u_2 . 
}\eqno(D.2)
$$
\medskip
     
   The product of four sfermions are decomposed as follows.
$$
\eqalign{
    (\tilde u^*_L \tilde u_L)^2 =& 
     +c_u^4(\tilde u^*_1\tilde u_1)(\tilde u^*_1 \tilde u_1) \cr 
   & -2 c_u^3s_u (\tilde u^*_1\tilde u_1)(\tilde u^*_1 \tilde u_2)
     -2 c_u^3s_u (\tilde u^*_1\tilde u_1)(\tilde u^*_2 \tilde u_1)\cr
   & +c_u^2s_u^2 (\tilde u^*_1\tilde u_2)(\tilde u^*_1 \tilde u_2)
     +c_u^2s_u^2 (\tilde u^*_2\tilde u_1)(\tilde u^*_2 \tilde u_1)\cr
   & +2c_u^2s_u^2(\tilde u^*_1\tilde u_2)(\tilde u^*_2 \tilde u_1) 
     +2c_u^2s_u^2(\tilde u^*_1\tilde u_1)(\tilde u^*_2 \tilde u_2)\cr
   & -2c_us_u^3  (\tilde u^*_1\tilde u_2)(\tilde u^*_2 \tilde u_2)
     -2c_us_u^3  (\tilde u^*_2\tilde u_1)(\tilde u^*_2 \tilde u_2)\cr
   &  +s_u^4     (\tilde u^*_2\tilde u_2)(\tilde u^*_2 \tilde u_2). 
}\eqno(D.3)
$$
$$
\eqalign{
     (\tilde u^*_R \tilde u_R)^2 =& 
     +s_u^4      (\tilde u^*_1\tilde u_1)(\tilde u^*_1 \tilde u_1)\cr
   & +2 c_us_u^3 (\tilde u^*_1\tilde u_1)(\tilde u^*_1 \tilde u_2)
     +2 c_us_u^3 (\tilde u^*_1\tilde u_1)(\tilde u^*_2 \tilde u_1)\cr
   & +c_u^2s_u^2(\tilde u^*_1\tilde u_2)(\tilde u^*_1 \tilde u_2)
     +c_u^2s_u^2(\tilde u^*_2\tilde u_1)(\tilde u^*_2 \tilde u_1)\cr
   & +2c_u^2s_u^2(\tilde u^*_1\tilde u_2)(\tilde u^*_2 \tilde u_1) 
     +2c_u^2s_u^2(\tilde u^*_1\tilde u_1)(\tilde u^*_2 \tilde u_2)\cr
   & +2c_u^3s_u  (\tilde u^*_1\tilde u_2)(\tilde u^*_2 \tilde u_2)
     +2c_u^3s_u  (\tilde u^*_2\tilde u_1)(\tilde u^*_2 \tilde u_2)\cr
   & +c_u^4      (\tilde u^*_2\tilde u_2)(\tilde u^*_2 \tilde u_2). 
}\eqno(D.4)
$$
$$
\eqalign{
  \vert\tilde u^*_L \tilde u_R\vert^2 =& 
     + c_u^2s_u^2(\tilde u^*_1\tilde u_1)(\tilde u^*_1 \tilde u_1)\cr
   & + c_us_u(c_u^2-s_u^2)(\tilde u^*_1\tilde u_1)(\tilde u^*_1 \tilde u_2)
     + c_us_u(c_u^2-s_u^2)(\tilde u^*_1\tilde u_1)(\tilde u^*_2 \tilde u_1)
                  \cr
   &  -c_u^2s_u^2(\tilde u^*_1\tilde u_2)(\tilde u^*_1 \tilde u_2)
      -c_u^2s_u^2(\tilde u^*_2\tilde u_1)(\tilde u^*_2 \tilde u_1)\cr
   &+(c_u^4+s_u^4)(\tilde u^*_1\tilde u_2)(\tilde u^*_2 \tilde u_1) 
     -2c_u^2s_u^2(\tilde u^*_1\tilde u_1)(\tilde u^*_2 \tilde u_2)\cr
   &  -c_us_u(c_u^2-s_u^2)(\tilde u^*_1\tilde u_2)(\tilde u^*_2 \tilde u_2)
   -c_us_u(c_u^2-s_u^2)(\tilde u^*_2\tilde u_1)(\tilde u^*_2 \tilde u_2)
                  \cr
   &  +c_u^2s_u^2(\tilde u^*_2\tilde u_2)(\tilde u^*_2 \tilde u_2).
}\eqno(D.5)
$$

$$
\eqalign{
   \vert\tilde u^*_L \tilde d_L\vert^2 =& 
    +c_u^2c_d^2(\tilde u^*_1\tilde d_1)(\tilde d^*_1 \tilde u_1) \cr
 & - c_us_uc_d^2 (\tilde u^*_1\tilde d_1)(\tilde d^*_1 \tilde u_2)
   - c_u^2c_ds_d (\tilde u^*_1\tilde d_1)(\tilde d^*_2 \tilde u_1)\cr
&  - c_u^2c_ds_d (\tilde u^*_1\tilde d_2)(\tilde d^*_1 \tilde u_1)
   - c_us_uc_d^2 (\tilde u^*_2\tilde d_1)(\tilde d^*_1 \tilde u_1)\cr
 & +c_us_uc_ds_d (\tilde u^*_1\tilde d_2)(\tilde d^*_1 \tilde u_2)
   +c_us_uc_ds_d (\tilde u^*_2\tilde d_1)(\tilde d^*_2 \tilde u_1)\cr
 & +c_u^2s_d^2   (\tilde u^*_1\tilde d_2)(\tilde d^*_2 \tilde u_1) 
   +s_u^2c_d^2   (\tilde u^*_2\tilde d_1)(\tilde d^*_1 \tilde u_2)\cr
 & +c_us_uc_ds_d (\tilde u^*_1\tilde d_1)(\tilde d^*_2 \tilde u_2)
   +c_us_uc_ds_d (\tilde u^*_2\tilde d_2)(\tilde d^*_1 \tilde u_1)\cr
 & -c_us_us_d^2  (\tilde u^*_2\tilde d_2)(\tilde d^*_2 \tilde u_1) 
   -s_u^2c_ds_d  (\tilde u^*_2\tilde d_2)(\tilde d^*_1 \tilde u_2)\cr
 & -s_u^2c_ds_d  (\tilde u^*_2\tilde d_1)(\tilde d^*_2 \tilde u_2)
   -c_us_us_d^2  (\tilde u^*_1\tilde d_2)(\tilde d^*_2 \tilde u_2)\cr
 &  +s_u^2s_d^2  (\tilde u^*_2\tilde d_2)(\tilde d^*_2 \tilde u_2).
}\eqno(D.6)
$$
$$
\eqalign{
  \vert\tilde u^*_R \tilde d_R\vert^2 =& 
   + s_u^2s_d^2(\tilde u^*_1\tilde d_1)(\tilde d^*_1 \tilde u_1) \cr
 & + c_us_us_d^2 (\tilde u^*_1\tilde d_1)(\tilde d^*_1 \tilde u_2)
   + s_u^2c_ds_d (\tilde u^*_1\tilde d_1)(\tilde d^*_2 \tilde u_1)\cr
 & + s_u^2c_ds_d (\tilde u^*_1\tilde d_2)(\tilde d^*_1 \tilde u_1)
   + c_us_us_d^2 (\tilde u^*_2\tilde d_1)(\tilde d^*_1 \tilde u_1)\cr
 & +c_us_uc_ds_d (\tilde u^*_1\tilde d_2)(\tilde d^*_1 \tilde u_2)
   +c_us_uc_ds_d (\tilde u^*_2\tilde d_1)(\tilde d^*_2 \tilde u_1)\cr
 & +s_u^2c_d^2   (\tilde u^*_1\tilde d_2)(\tilde d^*_2 \tilde u_1) 
   +c_u^2s_d^2   (\tilde u^*_2\tilde d_1)(\tilde d^*_1 \tilde u_2)\cr
 & +c_us_uc_ds_d (\tilde u^*_1\tilde d_1)(\tilde d^*_2 \tilde u_2)
   +c_us_uc_ds_d (\tilde u^*_2\tilde d_2)(\tilde d^*_1 \tilde u_1)\cr
 & +c_us_uc_d^2  (\tilde u^*_2\tilde d_2)(\tilde d^*_2 \tilde u_1) 
   +c_u^2s_dc_d  (\tilde u^*_2\tilde d_2)(\tilde d^*_1 \tilde u_2)\cr
 & +c_u^2c_ds_d  (\tilde u^*_2\tilde d_1)(\tilde d^*_2 \tilde u_2)
   +c_us_uc_d^2  (\tilde u^*_1\tilde d_2)(\tilde d^*_2 \tilde u_2)\cr
 & +c_u^2c_d^2   (\tilde u^*_2\tilde d_2)(\tilde d^*_2 \tilde u_2).
}\eqno(D.7)
$$
$$
\eqalign{
  \vert\tilde u^*_L \tilde d_R\vert^2 =& 
   + c_u^2s_d^2 (\tilde u^*_1\tilde d_1)(\tilde d^*_1 \tilde u_1)\cr
 & - c_us_us_d^2 (\tilde u^*_1\tilde d_1)(\tilde d^*_1 \tilde u_2)
   + c_u^2c_ds_d (\tilde u^*_1\tilde d_1)(\tilde d^*_2 \tilde u_1)\cr
 & + c_u^2c_ds_d (\tilde u^*_1\tilde d_2)(\tilde d^*_1 \tilde u_1)
   - c_us_us_d^2 (\tilde u^*_2\tilde d_1)(\tilde d^*_1 \tilde u_1)\cr
 & -c_us_uc_ds_d (\tilde u^*_1\tilde d_2)(\tilde d^*_1 \tilde u_2)
   -c_us_uc_ds_d (\tilde u^*_2\tilde d_1)(\tilde d^*_2 \tilde u_1)\cr
 & +c_u^2c_d^2   (\tilde u^*_1\tilde d_2)(\tilde d^*_2 \tilde u_1) 
   +s_u^2s_d^2   (\tilde u^*_2\tilde d_1)(\tilde d^*_1 \tilde u_2)\cr
 & -c_us_uc_ds_d (\tilde u^*_1\tilde d_1)(\tilde d^*_2 \tilde u_2)
   -c_us_uc_ds_d (\tilde u^*_2\tilde d_2)(\tilde d^*_1 \tilde u_1)\cr
 & -c_us_uc_d^2  (\tilde u^*_2\tilde d_2)(\tilde d^*_2 \tilde u_1) 
   +s_u^2c_ds_d  (\tilde u^*_2\tilde d_2)(\tilde d^*_1 \tilde u_2)\cr
 & +s_u^2c_ds_d  (\tilde u^*_2\tilde d_1)(\tilde d^*_2 \tilde u_2)
   -c_us_uc_d^2  (\tilde u^*_1\tilde d_2)(\tilde d^*_2 \tilde u_2)\cr
 & +s_u^2c_d^2   (\tilde u^*_2\tilde d_2)(\tilde d^*_2 \tilde u_2). 
}\eqno(D.8)
$$
     The expression of $\vert\tilde u^*_R \tilde d_L\vert^2$ is obtained
from $\vert\tilde u^*_L \tilde d_R\vert^2$ by $u\leftrightarrow d$.
Explicitly it is given as     
$$
\eqalign{
  \vert\tilde u^*_R \tilde d_L\vert^2 =& 
   +s_u^2c_d^2 (\tilde u^*_1\tilde d_1)(\tilde d^*_1 \tilde u_1)\cr
 & + c_us_uc_d^2 (\tilde u^*_1\tilde d_1)(\tilde d^*_1 \tilde u_2)
   - s_u^2c_ds_d (\tilde u^*_1\tilde d_1)(\tilde d^*_2 \tilde u_1)\cr
 & - s_u^2c_ds_d (\tilde u^*_1\tilde d_2)(\tilde d^*_1 \tilde u_1)
   + c_us_uc_d^2 (\tilde u^*_2\tilde d_1)(\tilde d^*_1 \tilde u_1)\cr
 & -c_us_uc_ds_d (\tilde u^*_1\tilde d_2)(\tilde d^*_1 \tilde u_2)
   -c_us_uc_ds_d (\tilde u^*_2\tilde d_1)(\tilde d^*_2 \tilde u_1)\cr
 & +c_u^2c_d^2   (\tilde u^*_2\tilde d_1)(\tilde d^*_1 \tilde u_2) 
   +s_u^2s_d^2   (\tilde u^*_1\tilde d_2)(\tilde d^*_2 \tilde u_1)\cr
 & -c_us_uc_ds_d (\tilde u^*_1\tilde d_1)(\tilde d^*_2 \tilde u_2)
   -c_us_uc_ds_d (\tilde u^*_2\tilde d_2)(\tilde d^*_1 \tilde u_1)\cr
 & +c_us_us_d^2  (\tilde u^*_2\tilde d_2)(\tilde d^*_2 \tilde u_1)
   -c_u^2c_ds_d  (\tilde u^*_2\tilde d_2)(\tilde d^*_1 \tilde u_2)\cr
 & -c_u^2c_ds_d  (\tilde u^*_2\tilde d_1)(\tilde d^*_2 \tilde u_2) 
   +c_us_us_d^2  (\tilde u^*_1\tilde d_2)(\tilde d^*_2 \tilde u_2)\cr
 & +c_u^2s_d^2   (\tilde u^*_2\tilde d_2)(\tilde d^*_2 \tilde u_2).
}\eqno(D.9)
$$
     Setting $\tilde d=\tilde u$ in the last four equations 
(D.6)-(D.9), one recovers the first three equations (D.3)-(D.5).
$$
\eqalign{
(\tilde u^*_L \tilde u_L) (\tilde d^*_L \tilde d_L) =& 
  +c_u^2c_d^2   (\tilde u^*_1\tilde u_1)(\tilde d^*_1\tilde d_1)\cr
& -c_u^2c_ds_d  (\tilde u^*_1\tilde u_1)(\tilde d^*_1\tilde d_2)
  -c_u^2c_ds_d  (\tilde u^*_1\tilde u_1)(\tilde d^*_2\tilde d_1)\cr
& -c_us_uc_d^2  (\tilde u^*_1\tilde u_2)(\tilde d^*_1\tilde d_1)
  -c_us_uc_d^2  (\tilde u^*_2\tilde u_1)(\tilde d^*_1\tilde d_1)\cr
& +c_us_uc_ds_d (\tilde u^*_1\tilde u_2)(\tilde d^*_1\tilde d_2)
  +c_us_uc_ds_d (\tilde u^*_2\tilde u_1)(\tilde d^*_2\tilde d_1)\cr
& +c_us_uc_ds_d (\tilde u^*_1\tilde u_2)(\tilde d^*_2\tilde d_1)
  +c_us_uc_ds_d (\tilde u^*_2\tilde u_1)(\tilde d^*_1\tilde d_2)\cr
& +c_u^2s_d^2   (\tilde u^*_1\tilde u_1)(\tilde d^*_2\tilde d_2)
  +s_u^2c_d^2   (\tilde u^*_2\tilde u_2)(\tilde d^*_1\tilde d_1)\cr
& -s_u^2c_ds_d  (\tilde u^*_2\tilde u_2)(\tilde d^*_2\tilde d_1)
  -s_u^2c_ds_d  (\tilde u^*_2\tilde u_2)(\tilde d^*_1\tilde d_2)\cr
& -c_us_us_d^2  (\tilde u^*_2\tilde u_1)(\tilde d^*_2\tilde d_2)
  -c_us_us_d^2  (\tilde u^*_1\tilde u_2)(\tilde d^*_2\tilde d_2)\cr
& +s_u^2s_d^2   (\tilde u^*_2\tilde u_2)(\tilde d^*_2\tilde d_2).
}\eqno(D.10)
$$
   $(\tilde u^*_R \tilde u_R) (\tilde d^*_R \tilde d_R)$ is obtained
from $(\tilde u^*_L \tilde u_L) (\tilde d^*_L \tilde d_L)$,by 
$$
   c_u\to s_u,~~s_u\to -c_u,~~c_d\to s_d,~~s_d\to -c_d. 
\eqno(D.11)
$$
$$
\eqalign{
(\tilde u^*_R \tilde u_R) (\tilde d^*_R \tilde d_R) =& 
  +s_u^2s_d^2   (\tilde u^*_1\tilde u_1)(\tilde d^*_1\tilde d_1)\cr
& +s_u^2c_ds_d  (\tilde u^*_1\tilde u_1)(\tilde d^*_1\tilde d_2)
  +s_u^2c_ds_d  (\tilde u^*_1\tilde u_1)(\tilde d^*_2\tilde d_1)\cr
& +c_us_us_d^2  (\tilde u^*_1\tilde u_2)(\tilde d^*_1\tilde d_1)
  +c_us_us_d^2  (\tilde u^*_2\tilde u_1)(\tilde d^*_1\tilde d_1)\cr
& +c_us_uc_ds_d (\tilde u^*_1\tilde u_2)(\tilde d^*_1\tilde d_2)
  +c_us_uc_ds_d (\tilde u^*_2\tilde u_1)(\tilde d^*_2\tilde d_1)\cr
& +c_us_uc_ds_d (\tilde u^*_1\tilde u_2)(\tilde d^*_2\tilde d_1)
  +c_us_uc_ds_d (\tilde u^*_2\tilde u_1)(\tilde d^*_1\tilde d_2)\cr
& +s_u^2c_d^2   (\tilde u^*_1\tilde u_1)(\tilde d^*_2\tilde d_2)
  +c_u^2s_d^2   (\tilde u^*_2\tilde u_2)(\tilde d^*_1\tilde d_1)\cr
& +c_u^2c_ds_d  (\tilde u^*_2\tilde u_2)(\tilde d^*_2\tilde d_1)
  +c_u^2c_ds_d  (\tilde u^*_2\tilde u_2)(\tilde d^*_1\tilde d_2)\cr
& +c_us_uc_d^2  (\tilde u^*_2\tilde u_1)(\tilde d^*_2\tilde d_2)
  +c_us_uc_d^2  (\tilde u^*_1\tilde u_2)(\tilde d^*_2\tilde d_2)\cr
& +c_u^2c_d^2   (\tilde u^*_2\tilde u_2)(\tilde d^*_2\tilde d_2).
}\eqno(D.12)
$$
$$
\eqalign{
(\tilde u^*_L \tilde u_L) (\tilde d^*_R \tilde d_R) =& 
  +c_u^2s_d^2   (\tilde u^*_1\tilde u_1)(\tilde d^*_1\tilde d_1)\cr
& +c_u^2c_ds_d  (\tilde u^*_1\tilde u_1)(\tilde d^*_1\tilde d_2)
  +c_u^2c_ds_d  (\tilde u^*_1\tilde u_1)(\tilde d^*_2\tilde d_1)\cr
& -c_us_us_d^2  (\tilde u^*_1\tilde u_2)(\tilde d^*_1\tilde d_1)
  -c_us_us_d^2  (\tilde u^*_2\tilde u_1)(\tilde d^*_1\tilde d_1)\cr
& -c_us_uc_ds_d (\tilde u^*_1\tilde u_2)(\tilde d^*_1\tilde d_2)
  -c_us_uc_ds_d (\tilde u^*_2\tilde u_1)(\tilde d^*_2\tilde d_1)\cr
& -c_us_uc_ds_d (\tilde u^*_1\tilde u_2)(\tilde d^*_2\tilde d_1)
  -c_us_uc_ds_d (\tilde u^*_2\tilde u_1)(\tilde d^*_1\tilde d_2)\cr
& +c_u^2c_d^2   (\tilde u^*_1\tilde u_1)(\tilde d^*_2\tilde d_2)
  +s_u^2s_d^2   (\tilde u^*_2\tilde u_2)(\tilde d^*_1\tilde d_1)\cr
& +s_u^2c_ds_d  (\tilde u^*_2\tilde u_2)(\tilde d^*_2\tilde d_1)
  +s_u^2c_ds_d  (\tilde u^*_2\tilde u_2)(\tilde d^*_1\tilde d_2)\cr
& -c_us_uc_d^2  (\tilde u^*_2\tilde u_1)(\tilde d^*_2\tilde d_2)
  -c_us_uc_d^2  (\tilde u^*_1\tilde u_2)(\tilde d^*_2\tilde d_2)\cr
& +s_u^2c_d^2   (\tilde u^*_2\tilde u_2)(\tilde d^*_2\tilde d_2).
}\eqno(D.13)
$$

\vskip 2truecm

\noindent{\bf Appendix E.  Relation with Hikasa' convention}\par
    Here, I list the particles whose relative signs are differently defined
in ref.[4] as well as some different notation for 
parameters.
$$\matrix{ 
   {\rm this~paper} &  {\rm Hikasa~ [4]}  \cr
   \phi_1^-   &  -H_1^- \cr
   \chi_1^0   &  -\zeta_1\cr
     G^0      &  -\chi_0 ({\rm due~ to~ the~ relative~sign~ of}~ \zeta_1)\cr  
     G^\pm    &  -\chi^\pm ({\rm due~ to~ the~ relative~ sing~ of}~ H_1^-)\cr
  g_\mu^\alpha & G_\mu^\alpha \cr    
  \phi_R      & \phi_L \cr
  \phi_L      & \phi_R \cr
  \epsilon_L  & \epsilon_R  \cr
     \tilde m^2_{12}   & B\mu \cr
}\eqno(E.1)
$$
There are several trivial misprints in ref.[4], some of which 
were pointed out by T. Ishikawa.  These are
\item{$\cdot$} the first line of (6.96),  ~~~~~~~~~~~~~~~~~~~~~~~~~~~~~~~ 
                  $h^0$ should read  $H^0$.
\item{$\cdot$} the second term of the third line of (6.100),  ~~~
                  $\cos2\alpha$ should read $\sin 2\alpha$. 
\item{$\cdot$} the second line after (6.107),~~~~~~~~~~~~~~~~~~~~~~~~
                   $m_e \to m_s$ should read  $m_d \to m_s$. 
\item{$\cdot$} the first two terms of (I.5) , ~~~~~~~~~~~~~~~~~~~~~~~~
                   multiply  $g^2$.     

\vskip 2 truecm

\noindent{\bf References}\par
\item{[1]} See, for example, H.P. Nilles, Phys. Rep.110 (1984) 1.
\item{[2]} T. Ishikawa et al., The Grace manual, KEK preprint 92-19 (1993).
\item{[3]} See for example, J.F. Gunion and H.E. Haber, Nucl. Phys.
{\bf B}272(1986)1, {\it ibid}. {\bf B}278(1986) 449, 
J.F. Gunion, H.E. Haber, G. Kane and 
S. Dawson, {\it The Higgs Hunter's Guide}, 1990, Addison-Wesley Pub.Co.
\item{[4]} K. Hikasa, SUSY manuscript version July 5,1995, 
unpublished.
\item{[5]} SUSY23 v2.0, J. Fujimoto et al., Comp. Phys. Commun. 111(1998) 185.
\item{[6]} LEP2 working group report 
(http://pdg.lbl.gov/monte$\_$carlo$\_$num.html).
\item{[7]} S. Katsanevas and P. Morawitz, Physics at LEP2, CERN 96-01 vol.2 
(1996) 328. 
\item{[8]} M. Kuroda, Meiji Gakuin Ronsou, 594 (1997) 1 and KEK CP-057.
\item{[9]} K.-I. Aoki et al., Prog. Theor. Phys. Suppl.73(1982) 1.

\vskip 2 truecm
\end


\end